\documentclass[twocolumn,prb,nofootinbib,notitlepage,showpacs,amsmath,amstex,amssymb,citeautoscript,longbibliography]{revtex4-2}
\pdfoutput=1
\usepackage[english]{babel}
\usepackage{letltxmacro}
\usepackage{latexsym}
\LetLtxMacro{\ORIGselectlanguage}{\selectlanguage}
\makeatletter
\DeclareRobustCommand{\selectlanguage}[1]{%
  \@ifundefined{alias@\string#1}
    {\ORIGselectlanguage{#1}}
    {\begingroup\edef\x{\endgroup
       \noexpand\ORIGselectlanguage{\@nameuse{alias@#1}}}\x}%
}
\newcommand{\definelanguagealias}[2]{%
  \@namedef{alias@#1}{#2}%
}
\makeatother
\definelanguagealias{en}{english}
\definelanguagealias{English}{english}
\usepackage{graphicx}
\usepackage{amsmath}
\usepackage{amsfonts}
\usepackage{amssymb}
\usepackage{color}
\usepackage{soul} 
\usepackage{amssymb}
\usepackage{wasysym}
\usepackage{comment}
\usepackage{dsfont}
\usepackage{float}
\usepackage{braket}
\usepackage{hyperref}
\hypersetup{
    bookmarks=false,         
    unicode=false,          
    pdftoolbar=false,        
    pdfmenubar=true,        
    pdffitwindow=false,     
    pdfstartview={FitH},    
    pdftitle={Localization of a mobile impurity interacting with an Anderson insulator},    
    pdfauthor={P. Brighi et al.},     
    pdfsubject={},   
    pdfcreator={},   
    pdfproducer={}, 
    pdfkeywords={many-body localization} {DMRG} {disordered systems}, 
    pdfnewwindow=true,      
    colorlinks=true,       
    linkcolor=black,          
    citecolor=blue,        
    filecolor=magenta,      
    urlcolor=blue           
}

\begin{document}
\title{Localization of a mobile impurity interacting with an Anderson insulator}
\author{Pietro Brighi$^1$}
\author{Alexios A. Michailidis$^{1,2}$}
\author{Kristina Kirova$^1$}
\author{Dmitry A. Abanin$^2$}
\author{Maksym Serbyn$^1$}
\affiliation{$^1$IST Austria, Am Campus 1, 3400 Klosterneuburg, Austria}
\affiliation{$^2$Department of Theoretical Physics, University of Geneva, 24 quai Ernest-Ansermet, 1211 Geneva, Switzerland}
\date{\today}

\begin{abstract}
Thermalizing and localized many-body quantum systems present two distinct dynamical phases of matter. Recently, the fate of a localized system coupled to a thermalizing system viewed as a quantum bath received significant theoretical and experimental attention. In this work, we study a mobile impurity, representing a small quantum bath, that interacts locally with an Anderson insulator with a finite density of localized particles. Using static Hartree approximation to obtain an effective disorder strength, we formulate an analytic criterion for the perturbative stability of the localization. Next, we use an approximate dynamical Hartree method and the quasi-exact time-evolved block decimation (TEBD) algorithm to study the dynamics of the system. We find that the dynamical Hartree approach which completely ignores entanglement between the impurity and localized particles predicts the delocalization of the system. In contrast, the full numerical simulation of the unitary dynamics with TEBD suggests the stability of localization on numerically accessible timescales. Finally, using an extension of the density matrix renormalization group algorithm to excited states (DMRG-X), we approximate the highly excited eigenstates of the system. We find that the impurity remains localized in the eigenstates and entanglement is enhanced in a finite region around the position of the impurity, confirming the dynamical predictions. Dynamics and the DMRG-X results provide compelling evidence for the stability of localization. 
\end{abstract}
\maketitle

\section{Introduction\label{Sec:Intro}}
Many body localization (MBL) is an example of a stable dynamical phase of matter that avoids thermal equilibrium. The existence of the MBL phase may be understood from the stability of localization in the Anderson Hamiltonian~\cite{Anderson1958} with respect to weak but finite interactions~\cite{Gornyi2005a,Basko2006,Abanin2019}. Recently a number of dynamical properties of MBL~\cite{Bardarson2012,Serbyn2013a} were explained via the existence of an extensive number of quasilocal conserved quantities~\cite{Serbyn2013,Huse2014,Imbrie2016}. At the same time, experiments provided strong support for the stability of the MBL phase on long timescales, verified a number of theoretical predictions, and started exploring new regimes where the theoretical understanding is incomplete~\cite{Schreiber2015,Bloch2017a,Greiner2019a,Greiner2019b}. 
In particular, experiments suggested the existence of MBL in higher dimension~\cite{Choi2016,Bloch2017c} and also probe the so-called many-body mobility edges~\cite{Guo2019}, although theoretically their existence is subject to debate~\cite{DeRoeck2016,Luitz2015,Brighi2020,Yao20,Pomata,DeRoeck2017,Altman2019,Pal2019,Mirlin2020,Eisert2020}. 

Another open question concerns the stability of localization in an MBL system coupled to a so-called quantum bath, represented by another quantum system that thermalizes in the absence of the coupling. In the case of a thermodynamically large bath, where the back-action of the localized system can be neglected, the system-bath coupling is expected to result in delocalization. However, considering a bath whose dimension is comparable to the localized system, or smaller, can yield distinct outcomes, especially if the back-action on the bath is taken into account. The MBL degrees of freedom can localize the bath -- a phenomenon dubbed ``MBL proximity effect'' by~\textcite{Nandkishore2015a}. Alternatively, the bath can thermalize the formerly localized system~\cite{Luitz2017}.

Inspired by the avalanche mechanism~\cite{DeRoeck2017} for the delocalization transition, a number of works studied the effect of a ``thermal grain'' coupled to a localized system~\cite{Huse2015,Luitz2017,Pollmann2019}. These works described the bath by a random matrix type Hamiltonian and account for the quasi-local nature of integrals of motion while considering coupling between the bath and localized system. 
In these models the bath lacks spatial structure, thus it cannot be localized by the MBL system, excluding \emph{a priori} the MBL proximity effect. 
In order to keep the microscopic structure of the bath, Refs.~\cite{Nandkishore2015a,Hyatt2017,Goihl2019} represented it by a set of thermalizing particles. In this framework, thermal and localized degrees of freedom coexist and are coupled through local interaction. It has been numerically shown~\cite{Hyatt2017} that localization can globally persist, if the bandwidth of the thermal particles is small. However, the fingerprints of localization of the bath under the influence of the disordered degrees of freedom was not studied in detail and still remains an open question.

All the studies discussed so far relied on the use of exact diagonalization (ED), which dramatically limits the system sizes available. In contrast, recent experimental studies~\cite{Rubio-Abadal2019,Leonard2020} have addressed this problem on bosonic quantum simulators, enabling the study of large systems. In Ref.~\cite{Rubio-Abadal2019} a ``global" setup was used, where the thermal degrees of freedom are homogeneously distributed through the $2$d lattice. Varying the number of thermal particles, the experiment showed evidence of the stability of localization when the thermalizing bosons are a small fraction of the total. On the other hand, the authors of Ref.~\cite{Leonard2020} used a different approach. There, a $1$d chain is split into a disorder-free segment, that represents a bath, and is connected to a disordered segment. The experiment investigated the stability of localization while changing the size of the disorder-free segment. While localization is stable for small thermal chains, signs of delocalization were observed when the bath constitutes half of the whole system.

In this work, inspired by Ref.~\cite{Rubio-Abadal2019}, we study a one-dimensional system of two hard-core bosonic species interacting with one another. The disordered particles form an Anderson insulator and are coupled locally to a single disorder-free boson representing the smallest possible bath. Analyzing the model under a mean-field type approximation, we formulate a perturbative criterion for stability of localization, that suggests that localization remains robust at strong interaction and disorder. This is in contrast with the results discussed in Ref.~\cite{Krause2021}, where ergodicity is introduced in the system by a stable doublon, that can freely move in the strong interaction limit, therefore facilitating delocalization in the case when single-particle localization length is long enough.

In an accompanying paper~\cite{Brighi2021a}, we studied the dynamics of the model discussed here in large systems using matrix product state (MPS)~\cite{Verstraete2006} techniques. There, at strong interactions and disorder we observed the localization of the bath through the back-action of the Anderson insulator. Here, besides analytic results supporting localization in such a regime, we provide a thorough comparison of fully interacting dynamics with an approximate time-dependent Hartree method, which neglects entanglement and quantum correlations among the two particle species. This latter technique reveals delocalization at long times, supported by the diffusive spreading of the particle constituting  the bath and decaying memory of the initial state of the Anderson insulator. This result highlights the fundamental role played by entanglement, which we study in detail in the present work.

Finally, we use density-matrix renormalization group for excited states method (DMRG-X)~\cite{Serbyn2016,DMRG-X,Pollman2016}, to study highly excited eigenstates of the Hamiltonian in large chains. This effectively allows us to probe localization at infinite times. Analyzing the expectation value of density in eigenstates, we observe localization of the small bath due to the interaction with the Anderson insulator. Furthermore, we find that eigenstates show area-law entanglement, thus  providing complementary support for the persistence of localization at strong interactions.

The structure of this paper is as follows. In Section~\ref{Sec:Model} we introduce the model and describe the typical initial state chosen for the dynamics. In Section~\ref{Sec:Hartree} we analyze the Hartree limit and study perturbatively the effect of the interaction. Following that, in Section~\ref{Sec:Dynamics}, we introduce the time-dependent Hartree approximation for the dynamics, showing its results and comparing them with quasi-exact TEBD results. Finally, in Section~\ref{Sec:DMRG-X} we present our DMRG-X study.

\section{Model\label{Sec:Model}}
In this work, we study two bosonic species of particles, interacting through an on-site potential. The \textit{disordered} bosons ($d$-bosons) are subject to a random potential drawn from a uniform distribution $\epsilon_i\in[-W,W]$ and are governed by the Hamiltonian $\hat{H}_d$
\begin{equation}
\label{Eq:Hd}
\hat{H}_d = t_d\sum_{i=1}^{L-1}(\hat{d}^\dagger_i\hat{d}_{i+1} + \text{h.c.}) + \sum_{i=1}^L\epsilon_i\hat{n}_{d,i},
\end{equation}
where $\hat{d_i}$ is the annihilation operator for the $d$-bosons, $\hat{n}_{d,i} = \hat{d}^\dagger_i \hat{d}_i$ is their density operator, and $t_d$ is the hopping strength. These bosons realize an Anderson insulator~\cite{Anderson1958}, that is localized at any density of bosons thereby providing a specific system that avoids thermalization.  

The small quantum bath will be represented by the \textit{clean} bosons ($c$-bosons) described by $\hat{H}_c$, 
\begin{equation}
\label{Eq:Hc} 
\hat{H}_c = t_c\sum_{i=1}^{L-1}(\hat{c}^\dagger_i\hat{c}_{i+1} + \text{h.c.}),
\end{equation}
that are characterized by a single hopping parameter $t_c$ and are not subject to disordered potential. Finally, the two boson species are coupled via the on-site Hubbard interaction,
\begin{align}
\label{Eq:Hint}
\hat{H}_\text{int} = U\sum_{i=1}^L\hat{n}_{c,i}\hat{n}_{d,i},
\end{align}
where $U$ is the interaction strength and  $\hat{n}_{c,i} = \hat{c}^\dagger_i \hat{c}_i$ is the number operator of $c$-bosons. 

The full interacting Hamiltonian then reads:
\begin{equation}
\label{Eq:H}
\hat{H} = \hat{H}_d+\hat{H}_c+\hat{H}_\text{int}.
\end{equation}
The system described by the Hamiltonian~(\ref{Eq:H}) has $U(1)\times U(1)$ symmetry, as the number of both types of particles, $\hat{N}_{c/d} = \sum_i\hat{n}_{c/d,i}$, is conserved. In what follows we restrict to a sector of finite $d$-bosons density and to the case $t_d=t_c=1$. In particular, density of $\nu_d = N_d/L =1/3$ would be used unless specified otherwise. At the same time, throughout this work we consider the presence of a \emph{single} $c$-boson, that realizes the smallest possible quantum bath with non-trivial spatial structure and local coupling to a localized system. In this respect, our setting resembles the experimental setup of Ref.~\cite{Rubio-Abadal2019} which was, however, performed on a two dimensional lattice and considered various densities of clean particles. 

Although the boson density is sufficient to specify a particular sector of the Hilbert space, we further restrict ourselves to states where $d$-bosons have a globally homogeneous distribution. In particular, we will study initial states corresponding to a $d$-bosons density wave, with the single $c$-boson located on the central site, as exemplified by $\ket{\psi_0}$ below on a system of $L=18$ sites and with $\nu_d=1/3$
\begin{equation}
\label{Eq:psi0}
\ket{\psi_0} = |{\bullet}{\circ}{\circ}{\bullet}{\circ}{\circ}{\bullet}{\circ}{\color{red}\bullet}{\bullet}{\circ}{\circ} {\bullet}{\circ}{\circ} {\bullet}{\circ}{\circ}\rangle,
\end{equation}
where empty circles represent empty sites, and black (red) circles are sites occupied by $d$- and $c$-boson respectively. Such initial state resembles the configurations used in experiments and can be characterized by the so-called imbalance~\cite{Bloch2017a,Bloch2017b}, quantifying the memory of the initial density-wave configuration in the system after a quench.

While a strictly periodic arrangement of $d$-bosons akin to the state~(\ref{Eq:psi0}) is not required, we assume that the density of  $d$-bosons is on average distributed uniformly on a scale that is larger than a few lattice spacings. This assumption is important since the presence of large empty/occupied regions in the chain would imply the effective absence of disorder for $c$-boson in that region. While such configurations could be used to imitate another experimental study of MBL-bath coupling~\cite{Leonard2020}, states where extensive regions of the chain are fully empty or occupied by $d$-bosons are far from typical initial product states. Moreover, we expect that if the localization of $d$-bosons persists, such states are nearly decoupled from spatially homogenous initial states of the type~(\ref{Eq:psi0}). Indeed, in order to connect the highly inhomogeneous state such as $\ket{\psi_0} = |{\bullet}{\bullet}{\bullet}{\bullet}{\bullet}{\bullet}{\circ}{\circ}{\circ}{\circ}{\circ}{\color{red}\bullet}{\circ}{\circ} {\circ}{\circ}{\circ}{\circ}\rangle$ to the density-wave state in Eq.~(\ref{Eq:psi0}), the tunneling of an extensive number of $d$-boson over long distances is required. 

\begin{figure*}[t]
\includegraphics[width=1.95\columnwidth]{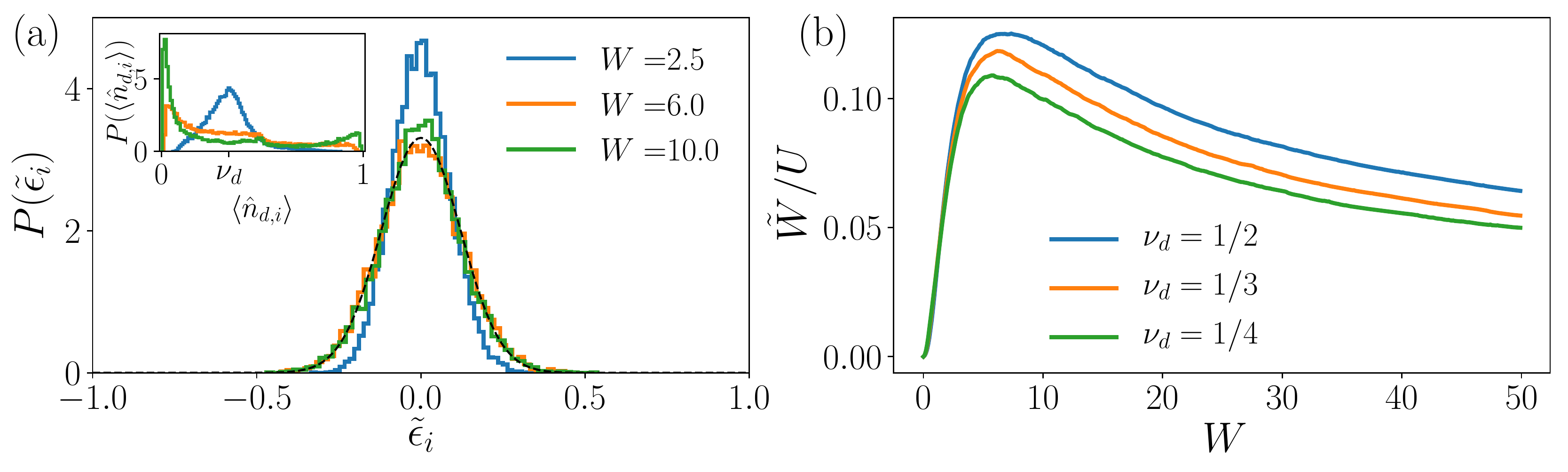}
\caption{\label{Fig:effective disorder}
(a) Subtracting deterministic components from the distribution of $\langle \hat{n}_{d,i}\rangle$ (inset) results in an approximately Gaussian distribution of the effective disorder potential $\tilde{\epsilon}_i$, shown here for $\nu_d=1/3$.  Standard deviation of this distribution, $\sigma$, is used to define the effective disorder strength, $\tilde{W}/U=\sigma$. 
(b) The effective disorder strength $\tilde{W}$  shows a non-monotonic behavior as a function of $W$ with a maximum at $W^*\approx 5$. The data shown are obtained for $L=500$ sites and averaged over $50$ disorder realizations.
}
\end{figure*}

\section{Hartree approximation and criterion for stability of localization\label{Sec:Hartree}}
First, we formulate the Hartree approximation and use it to study the Hamiltonian~(\ref{Eq:H}). This facilitates the choice of parameters in the Hamiltonian, and allows to formulate an analytic criterion for stability of localization with respect to two-particle tunneling processes.

\subsection{Effective Disorder\label{Sec:Disorder}}
The Hartree approximation adopted in this section consists of replacing the density operator $\hat{n}_{d,i}$ in Eq.~(\ref{Eq:H}) with its infinite-time average $\langle\hat{n}_{d,i}\rangle$. This approximation would be fully justified in the case of instantaneous relaxation of $d$-bosons with respect to $c$-boson dynamics, i.e.\ if $t_d,W\gg t_c$. Moreover, as in an Anderson insulator the fluctuations of the expectation value $\langle\hat{n}_{d,i}\rangle$ are finite at all times, the Hartree approximation overestimates localization. In spite of these shortcomings, the Hartree approximation will assist us with the choice of model parameters and will also allow defining an effective disorder strength thus quantifying the analytic criterion for stability of localization.

The infinite-time average of the $d$-bosons density is given by the diagonal ensemble generated by the eigenstates of $\hat{H}_d$:
\begin{equation}
\label{Eq:nd-inf}
\begin{split}
\langle \hat{n}_{d,i} \rangle &= \lim_{T\to\infty}\frac{1}{T}\int_0^Tdt\,\bra{\psi(t)}\hat{n}_{d,i}\ket{\psi(t)}\\
&=  \sum_n |c_n|^2\bra{E_n}\hat{n}_{d,i}\ket{E_n},
\end{split}
\end{equation}
where $\{\ket{E_m}\}$ is the eigenbasis of $\hat{H}_d$, $c_n = \langle\psi_0\ket{E_n}$ and $\ket{\psi(t)} = \hat{T}e^{-\imath\int_0^t \hat{H}_d(t') dt'}\ket{\psi_0}=e^{-\imath \hat{H}_d t}\ket{\psi_0} $. The initial state $\ket{\psi_0}$ is taken to be a density wave, see Eq.~(\ref{Eq:psi0}). Due to the random potential in the Anderson Hamiltonian, the $d$-bosons occupation at infinite times acquires a quasi-random nature and is distributed according to $P(\langle \hat{n}_{d,i})\rangle$ in the range $[0,1]$. 

Thanks to the noninteracting nature of the problem in the Hartree approximation, the eigenstates in Eq.~(\ref{Eq:nd-inf}) can be written as product states of single particle orbitals $\ket{\ell}$. Then, by going into eigenbasis one can decouple the sum over ${L}\choose{N_d}$ terms in Eq.~(\ref{Eq:nd-inf}) into a double sum over the $N_d$ occupied sites in the initial state and the $L$ single particle orbitals. The problem of finding the infinite time occupation values then reduces to finding eigenstates of a single-particle Hamiltonian, whose Hilbert space scales as $L$, thus greatly reducing the complexity of the problem. The infinite-time average of the $d$-bosons density, hence, can be carried out through Eq.~(\ref{Eq:nd-inf}) for very large systems, $L=500$, where boundary effects on its probability distribution are negligible, and at various disorder strengths.

The infinite-time occupation of the $d$-bosons plays the role of an effective random chemical potential experienced by the $c$-boson in the Hartree approximation. As shown in the inset of Figure~\ref{Fig:effective disorder}(a), when the disorder is strong compared to the hopping parameter, $t_d=1$, $P(\langle \hat{n}_{d,i})\rangle$ resembles a bimodal distribution, with two distinct peaks at $\langle \hat{n}_{d,i}\rangle\sim0,1$. This behavior can be understood as a consequence of the strong localization of $d$-bosons for disorder $W=10$. Thus, even at infinite time, the $d$-bosons remain close to their initial position and expectation value of  $\langle \hat{n}_{d,i}\rangle$ remains close  to its initial value, i.e.~either $0$ or $1$, depending on the considered site. As disorder is decreased, the two peak structure gradually disappears, and the distribution acquires a single peak around the average density $\nu_d$. 

In order to define the effective disorder $\tilde{\epsilon}_i$ generated by the $d$-bosons, we need to isolate the random part of the distribution of $\langle \hat{n}_{d,i}\rangle$. To this end, we subtract from $\langle \hat{n}_{d,i}\rangle$ the uniform and the period-$1/\nu_d$ contribution. In Fourier space this corresponds to modifying the Fourier harmonics of density, $\tilde{n}_d(k) = \sum_j \langle\hat{n}_{d,j}\rangle e^{-\imath k j}$, as follows:
\begin{equation}
\label{Eq:effective disorder}
\tilde{\epsilon}(k) = \tilde{n}_d(k) - (1-\alpha)L\nu_d\delta_{k,0}-\alpha\sum_{j=1}^L f(j,\nu_d) e^{-\imath kj},
\end{equation}
where $\alpha = \sum_j \langle\hat{n}_{d,j}\rangle e^{-\imath\pi\nu_d j}$ corresponds to the weight of the $1/\nu_d$ harmonic in the particle distribution and $f(j,\nu_d) = \sum_{n=0}^{N_d-1}\delta_{j,n/\nu_d+1}$ is a functional representation of the initial density wave configuration. Transforming $\tilde{\epsilon}(k)$ back to the real space, we obtain the effective random potential $\tilde{\epsilon}_i$. Its distribution differs from the one of the infinite-time density, as it can be seen in Figure~\ref{Fig:effective disorder}(a). In particular, $P(\tilde{\epsilon}_i)$ is centered around zero for all values of disorder, and has an approximately Gaussian shape. Thus, we use the standard deviation of this distribution as the effective disorder strength $\tilde{W} = U \mathop{\rm std} P(\tilde{\epsilon}_i)$ experienced by $c$-boson in the Hartree approximation.

The effective disorder strength measured in units of $U$, $\tilde W/U$ is shown as a function of the disorder strength experienced by $d$-bosons, $W$,  in Figure~\ref{Fig:effective disorder}(b). First, we note a non-monotonic dependence of $\tilde W$ on the $d$-boson disorder strength, $W$.  The effective disorder strength $\tilde W$ presents a maximum around $W^*\approx 5$, whose position depends weakly on the density of $d$-bosons. This behavior can be naturally explained by considering two opposite limits: at weak $W$ the localization length of $d$-bosons is much larger and the initial period-2/3/4 density wave configuration is washed out at late times resulting in weak effective disorder $\tilde W$. In the opposite limit of very strong $W$, the $d$-bosons remain frozen close to their initial positions resulting in a nearly perfect periodic potential experienced by $c$-boson. However such periodic potential is unable to localize $c$-boson and is subtracted in Eq.~(\ref{Eq:effective disorder}), thus again resulting in a weak effective disorder. Given that $\tilde W$ is expected to decrease for very large and small $W$, we expect it to achieve a maximal value at some intermediate disorder $W$.  
Finally, we study in Fig.~\ref{Fig:effective disorder}(b) the dependence of the effective disorder on the $d$-bosons density $\nu_d$. As $\nu_d$ is increased, $\tilde W$ increases accordingly due to the fact that the effective random potential $\tilde{\epsilon}_i$ is generated by the $d$-bosons density. We note, however, that for $\nu_d>1/2$ the effective disorder would decrease again, because of the hard-core nature of the bosons.

\subsection{Localization length of the $c$-boson in Hartree approximation\label{Sec:LocHartree}}

We demonstrated that in the Hartree approximation the $c$-boson experiences an approximately Gaussian-distributed random potential with disorder strength $\tilde{W}$ that depends on the initial state and disorder experienced by $d$-bosons. Since an arbitrary weak disorder potential suffices to localize a single particle in a one-dimensional lattice, the $c$-boson in the Hartree approximation is always localized.  We proceed with the calculation of its localization length $\xi_c$ that provides a characteristic lengthscale of localization and can be compared with the localization length of the $d$-bosons, $\xi_d$.

To obtain the localization length for the $c$-boson we use the weak disorder approximation, justified at weak values of $U$ as the random potentials $\tilde{\epsilon}_i$ are restricted to $[-1,1]$. Perturbing around the tight-binding limit with the transfer matrix method~\cite{Thouless1972}, we obtain $\xi_c(k)\approx {8t^2_c \sin^2(k)}/{\langle\tilde{\epsilon}^2_i\rangle}$ that depends on the momentum $k$ that determines single particle energy in absence of disorder, $E(k) = -2t_c\cos(k)$. The average localization length is calculated by performing the integral over the complete band
\begin{equation}
\label{Eq:xic ave}
\xi_c\approx \frac{8t^2_c}{\langle\tilde{\epsilon}^2_i\rangle} \int_{-2t_c}^{2t_c} d\varepsilon \left[1-\left(\frac{\varepsilon}{2t_c}\right)^2\right]\rho(\varepsilon),
\end{equation}
where $\rho(\varepsilon)$ is the usual density of states of the tight-binding Hamiltonian. Carrying out the integral we realize that it contributes only to a numerical factor as the $c$-boson hopping terms cancel out. Recalling that, since $\langle\tilde{\epsilon}_i\rangle=0$, the variance $\langle \tilde{\epsilon}^2_i\rangle$ corresponds to the definition of the effective disorder strength squared, we obtain
\begin{equation}
\label{Eq:xic final}
\xi_c\approx \frac{4t_c^2}{a\tilde{W}^2},
\end{equation}
where $a=1$ is the lattice spacing.
\begin{figure}[t]
\includegraphics[width=0.95\columnwidth]{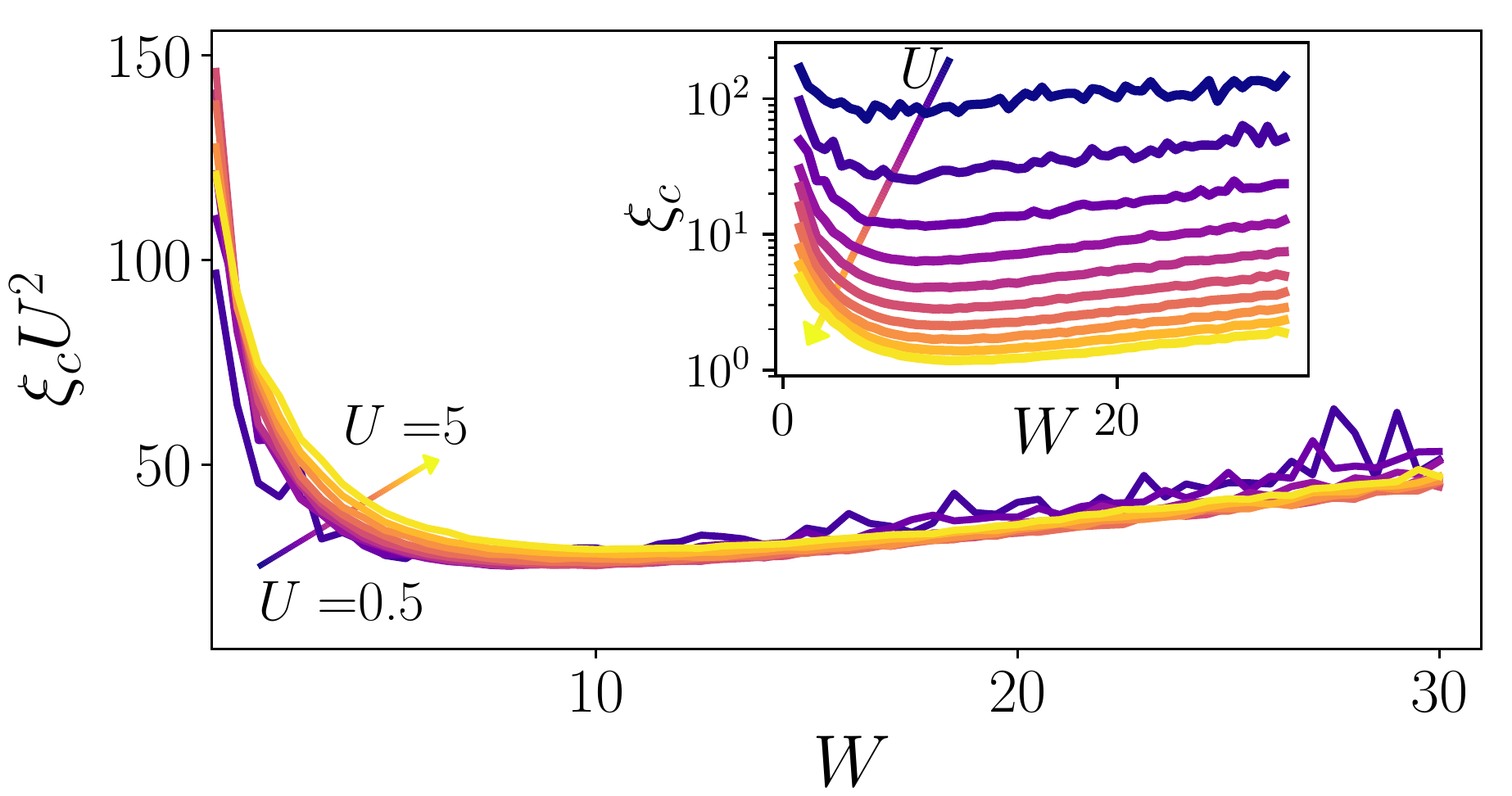}
\caption{\label{Fig:xi Hartree}
The localization length of the $c$-boson from the Hartree approximation for initial period-3 density wave state of $d$-bosons. Inset shows bare data, for different values of $U$ ranging from $U=0.5$ (dark blue) to $U=5$ (yellow). For a broad range of interaction strength, $\xi_c$ has a minimum at $W^*\approx5$. The main plot shows the collapse of data, confirming the scaling $\xi_c\sim U^{-2}$.
}
\end{figure}

\begin{figure*}[t]
\includegraphics[width=1.95\columnwidth]{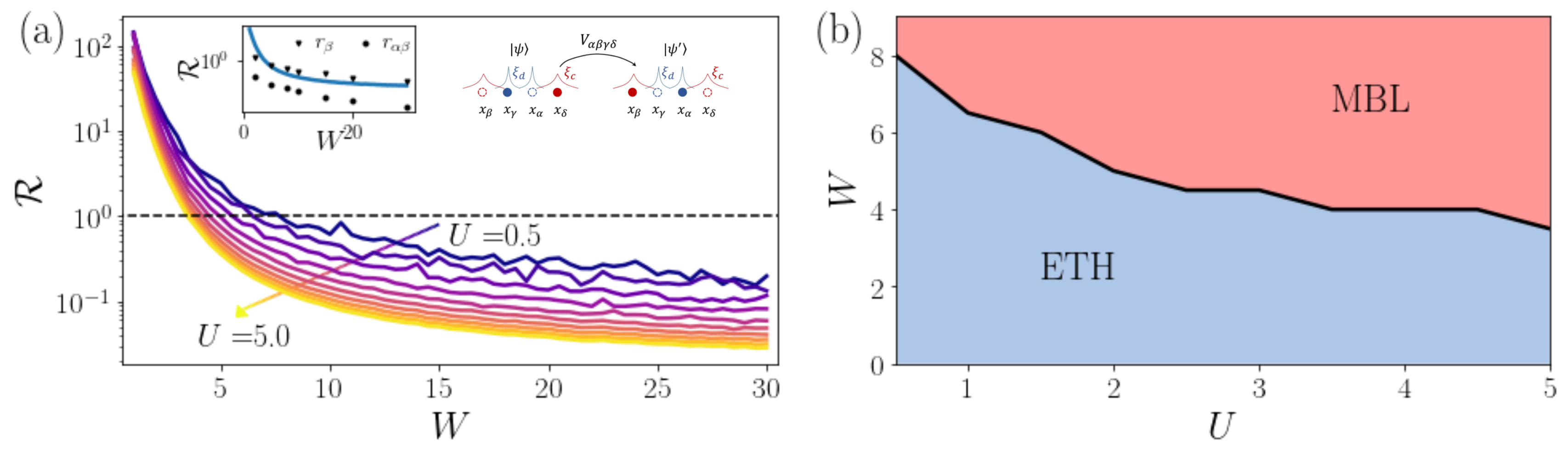}
\caption{\label{Fig:Resonances}
(a) The typical ratio between matrix element and level spacing as a function of disorder for different coupling strengths $U$ rapidly decreases as the interaction strength increases. In the left inset we compare analytic expression for $\mathcal{R}$ with its numerical estimate. The right inset shows the tunneling process induced by $\hat{H}_\text{int}$. 
(b) Phase diagram resulting from the criterion ${\cal R}=1$ reveals a broad region where the localization is perturbatively stable, which increases with increasing $U$ and $W$. 
}
\end{figure*}

The resulting simple expression for $\xi_c$, Eq.~(\ref{Eq:xic final}), has two main consequences. First, we expect that $\xi_c$ inherits the non-monotonic dependence on $W$ and has a minimum approximately when the effective disorder strengths is maximal in Fig.~\ref{Fig:effective disorder}. Second, since $\tilde{W}\propto U$, we expect that localization length diverges as $\xi_c\propto 1/U^2$ at small values of $U$ where weak disorder approximation is controllable. To confirm these predictions, we extract the localization length of $c$-bosons from numerical simulations. Using the effective disorder obtained from the eigenstates of the $d$-boson Hamiltonian, we use exact diagonalization to calculate the single-particle wave functions of $c$-bosons resulting from the Hamiltonian $H_c^\text{Hartree} = H_c+\sum_i U \langle \hat{n}_{d,i}\rangle \hat{n}_{c,i}$. Each eigenstate $\ket{\phi_\varepsilon}$ of this Hamiltonian can be characterized by an (energy-dependent) localization length $\xi_c(\varepsilon)$. After obtaining $\xi_c(\varepsilon)$ for each eigenstate through an exponential fit of its probability distribution in the real space, $|\bra{i}\phi_\varepsilon\rangle|^2$, we average the localization length over all eigenstates and further over $50$ disorder realizations.

The localization length resulting from the numerical simulation is shown in Fig.~\ref{Fig:xi Hartree}, where the non-monotonic behavior of $\xi_c$ and its scaling with the interaction strength becomes apparent. We note that for the adopted choice of hoppings $t_d=t_c=1$, $\xi_c$ always exceeds the localization length of $d$-bosons and is tunable by the interaction strength in a broad range. In particular, for disorder strength $W=6.5$, used in the accompanying  paper~\cite{Brighi2021a}, $\xi_d\approx 0.5$ is smaller than one lattice spacing, while $\xi_c$ takes values from $1.5$ to $100$ lattice spacings, as the interaction strength is decreased from $U=5$ to $U=0.5$ respectively. This result also assists in the choice of disorder strength: in order to facilitate the numerical studies, we fix disorder at $W=6.5$ so that the localization length of $c$-bosons is close to its minimum.

\subsection{Analytic criterion for stability of localization\label{Sec:Stability}}
The drawback of the Hartree approximation presented above is that it ignores fluctuations of $d$-bosons density, thus strongly favoring localization.  However, it provides a useful starting point to address the perturbative stability of such localized system. To this end, we use the basis of Anderson localized orbitals provided by Hartree approximation to address the stability of the system with respect to interactions between clean and disordered bosons. 

We transform the interaction Hamiltonian~(\ref{Eq:Hint}) to the basis of localized orbitals via the relation $\hat{d}_\alpha = \sum_i \psi_{d\alpha}(i)\hat{d}_i$, where $\hat{d}_\alpha$ is the annihilation operator of an Anderson orbital and $\psi_{d\alpha}(i)=\langle{\alpha}|i\rangle$ is the corresponding wave function and similar expressions hold for $c$-bosons. This leads to 
\begin{equation}
\label{Eq:Interaction Anderson basis}
\hat{H}_\text{int} = \sum_{\alpha\beta\gamma\delta}V_{\alpha\beta\gamma\delta} \hat{d}^\dagger_\alpha \hat{c}^\dagger_\beta\hat{d}_\gamma\hat{c}_\delta.
\end{equation}
As shown in the schematic representation in the inset of Figure~\ref{Fig:Resonances}, the matrix element $V_{\alpha\beta\gamma\delta}$ corresponds to a correlated hopping process with its value being given by the overlap of the wave functions of the corresponding localized orbitals, 
\begin{equation}
\label{Eq:V}
V_{\alpha\beta\gamma\delta}= U\sum_{i=1}^L\psi^*_{d\alpha}(i)\psi^*_{c\beta}(i)\psi_{d\gamma}(i)\psi_{c\delta}(i),
\end{equation}  
where the envelope of the wave functions decays on the scale of the corresponding localization length, $|\psi_{d\alpha} (i)|\sim e^{-|i-x_\alpha|/(2\xi_d)}/{\sqrt{2\xi_d}}$, where $x_\alpha$ is the site around which the orbital is localized. In order to address the stability of localization with respect to such correlated hoppings triggered by interaction, we compare the matrix element of this process to the corresponding level spacing.  

Keeping only the leading behavior of the wave functions, i.e.~the exponential decay, and neglecting their oscillatory behavior results in an upper bound for the matrix element in Eq.~(\ref{Eq:V}) and thus favors delocalization. In this case, the matrix element can be easily estimated as 
\begin{equation}
\label{Eq:V1}
V_{\alpha\beta\gamma\delta} \approx \frac{U}{4\xi_c\xi_d}\sum_{i=1}^L e^{-\frac{|x_\alpha-i|+|x_\gamma-i|}{\xi_d}}
 e^{-\frac{|x_\beta-i|+|x_\delta-i|}{\xi_c}},
\end{equation}
from where one can immediately realize that it is exponentially suppressed whenever any of terms in the exponents $|x-i|/\xi_{c/d}>1$ exceeds one. This restricts the localized orbitals that can efficiently participate in the hopping process:
\begin{equation}
\label{Eq:loc centers constraint}
\begin{cases}
|x_\alpha-x_\gamma|&\lesssim\xi_d,\quad |x_\alpha-x_\beta|\lesssim\max(\xi_c,\xi_d)\\
|x_\beta-x_\delta|&\lesssim\xi_c,\quad |x_\gamma-x_\delta|\lesssim\max(\xi_c,\xi_d).
\end{cases}
\end{equation}
Using the fact that $\xi_c\geq \xi_d$, the summation cancels the $\xi_d$ in the denominator resulting in a simple estimate
\begin{equation}
\label{Eq:V res}
V_{\alpha\beta\gamma\delta}\approx \frac{U}{4\xi_c}.
\end{equation}

The second element needed to understand if such tunneling processes are resonant is the typical level spacing $\delta_{cd}$, obtained as the ratio of the typical energy difference $\Delta E$ to the number of states $\mathcal{N}$ connected by the interaction. Since the two states differ for the position of the two bosons, the typical energy difference is given by the sum of the two disorder strengths: $\Delta E\approx W+\tilde{W}$. To account for the number of states $(\ket{\psi},\ket{\psi'})$ connected by the interaction we need to consider the constraints~(\ref{Eq:loc centers constraint}), which restrict the possible configurations. First, in the initial state the two bosons must lie within a distance $\xi_c$ from one another, as $\xi_c\geq \xi_d$, thus contributing $2\xi_c$ possible configurations. Due to the localized nature of wave functions, the position of the bosons in the final state must lie within a distance $\xi_c$ and $\xi_d$ for the $c$-boson and the $d$-bosons respectively. A na{\"i}ve computation would then give $\mathcal{N} = (2\xi_c)(2\xi_c)(2\xi_d) = 8\xi^2_c\xi_d$. However, one must be careful as also in the final state the two bosons must be separated by at most $\xi_c$ for the matrix element not to vanish. This reduces the total number of states, as not all the  moves increasing the distance among the bosons are allowed. A careful computation reveals that for a given $d$-boson hopping there are only $(3/2)\xi_c$ possible choices, thus reducing the total number of states. Finally, one has to take into account the finite density of the $d$-bosons, together with their hard-core nature, that requires that an initially occupied site must be left empty after the hopping and vice versa, leading to
\begin{equation}
\label{Eq:number of states}
\mathcal{N} \approx 6\xi^2_c\xi_d\nu_d(1-\nu_d), \quad \delta_{cd} = (W+\tilde W)/\mathcal{N} .
\end{equation}

Gathering the results of Eqs.~(\ref{Eq:V res})-(\ref{Eq:number of states}) we obtain the expression for the ratio of level spacing to the matrix element,
\begin{equation}
\label{Eq:Res}
\mathcal{R} = \frac{3}{2}U\frac{\xi_c\xi_d\nu_d(1-\nu_d)}{W+\tilde{W}}.
\end{equation}
Condition $\mathcal{R}<1$ provides a criterion for stability of localization. 
Recalling that the $c$-boson localization length depends on the interaction strength $\xi_c\approx\xi_c^0/U^2$, the criterion $\mathcal{R}<1$ can be rearranged as 
\begin{equation}
\label{Eq:Res Criterion}
(W+\tilde{W})U>\frac{3}{2}\xi_c^0\xi_d\nu_d(1-\nu_d),
\end{equation}
suggesting that localization remains stable at strong interactions and large disorder.

The typical probability of resonance $\mathcal{R}$ can also be evaluated numerically and compared with the prediction of Eq.~(\ref{Eq:Res}).  To this end, after diagonalizing the Hamiltonian in the Hartree approximation, for each $c$-boson eigenstate we select states lying within the localization lengths $\xi_c$ and $\xi_d$ from one another and evaluate for all of them the matrix element $V_{\alpha\beta\gamma\delta}$ and the energy difference. Then we define the resonance value $\mathit{r}_\beta$ fixing the initial state of the $c$-boson, e.g. $\beta$, and maximizing the ratio of the matrix element to the energy difference over all other available states. In a similar way we define the resonance $\mathit{r}_{\alpha\beta}$, where both the initial state of the $c$-boson ($\beta$) and of the $d$-boson ($\alpha$) are fixed. Being a more constrained version of $\mathit{r}_\beta$, the following inequality holds
\begin{equation}
\label{Eq:maxima Res}
\mathit{r}_\beta = \max_{\alpha\gamma\delta}\Bigr[\frac{V_{\alpha\beta\gamma\delta}}{\Delta E}\Bigr]\geq \mathit{r}_{\alpha\beta} = \max_{\gamma\delta}\Bigr[\frac{V_{\alpha\beta\gamma\delta}}{\Delta E}\Bigr].
\end{equation}
Finally, we obtain the typical resonance probability by taking the median of the distribution of $\mathit{r}_\beta$ and $\mathit{r}_{\alpha\beta}$. The results of this numerical evaluation are shown in the inset of Figure~\ref{Fig:Resonances}(a) together with the analytic prediction~(\ref{Eq:Res}) and show an overall agreement. In particular, we notice that $\mathit{r}_{\alpha\beta}$ is always smaller than $\mathit{r}_\beta$.

From Figure~\ref{Fig:Resonances}(a) we observe that the disorder $W$ at which the condition~(\ref{Eq:Res Criterion}) is satisfied decreases as the interaction strength $U$ increases. Taking disorder value where ${\cal R}= 1$ as a critical point, we obtain an estimate for the phase diagram shown in Figure~\ref{Fig:Resonances}(b), which predicts localization for strong interaction and strong disorder. Note, that our considerations assume a homogeneous initial distribution of $d$-bosons, as discussed at the end of Sec.~\ref{Sec:Model}. Figure~\ref{Fig:Resonances}(b) suggests that at disorder $W=6.5$ system is expected to be localized for sufficiently large $U\gtrapprox 1.5$. In what follows we explore the properties of the localized system for $U=12$, a point that is located deep in the localized regime according to our stability condition.

\section{Quench dynamics as a probe of localization\label{Sec:Dynamics}}

The static Hartree approximation introduced above allowed to formulate a criterion for the stability of localization. Below we proceed with probing the dynamics of the full model in the localized regime. First, we extend the Hartree approximation attempting to include dynamical effects in the system. More importantly, we study the full quantum dynamics using TEBD. In this section we fix $U=12$ and $W=6.5$ and investigate the dynamics of a quench from the state~(\ref{Eq:psi0}) with density $\nu_d=1/3$.

\subsection{Time-dependent Hartree approximation\label{Sec:TDHF}}

The static Hartree approximation relied on the assumption of quick equilibration of the $d$-boson density. This overestimated localization in the system, as we shall demonstrate below. Here, we adopt a different approximation: we assume that the full many-body wave function can be decomposed into a product of $c$- and $d$-bosons wave functions respectively. Such representation completely ignores entanglement between two boson species. However, it allows performing fast and efficient simulation of dynamics on long timescales for large systems. We note here that similar approaches to the approximation of dynamics in MBL system have already been used~\cite{Bera2021}, showing results qualitatively similar to what we present below. 

The product state structure of the wave function of the full system allows to reduce the full many-body Schr\"odinger equation to the simultaneous evolution of two non-interacting but time dependent Hamiltonians. Specifically, the time evolution of $c$- and $d$-bosons is governed by time-dependent Hamiltonians $\hat{H}^\text{dH}_c(t)$ and $\hat{H}^\text{dH}_d(t)$ respectively:
\begin{equation}
\label{Eq:Hd(t)}
\begin{split}
\hat{H}^\text{dH}_d(t) &= t_d\sum_{i=1}^{L-1}(\hat{d}^\dagger_{i+1}\hat{d}_i + \text{h.c.}) + \sum_{i=1}^L V_{d,i}(t)\hat{n}_{d,i},
\\ &V_{d,i}(t) = \epsilon_i + U\langle\hat{n}_{c,i}(t)\rangle 
\end{split}
\end{equation}
\begin{equation}
\label{Eq:Hc(t)}
\begin{split}
\hat{H}^\text{dH}_c(t) &= t_c\sum_{i=1}^{L-1}(\hat{c}^\dagger_{i+1}\hat{c}_i + \text{h.c.}) + \sum_{i=1}^L V_{c,i}(t)\hat{n}_{c,i},
\\ &V_{c,i}(t) =  U\langle\hat{n}_{d,i}(t)\rangle,
\end{split}
\end{equation}
where the expectation value of densities of corresponding boson species reads $\langle\hat{n}_{c/d,i}(t)\rangle = \bra{\psi_{c/d}(t)}\hat{n}_{c/d}\ket{\psi_{c/d}(t)}$, and $\ket{\psi_{c/d}(t)}$ is the wave function obtained from the time-evolution of the initial state~(\ref{Eq:psi0}) with the correspondent time-dependent Hamiltonian:
\begin{equation}
\label{Eq:psi_cd(t)}
\ket{\psi_{c/d}(t)} = \hat{T}\exp\Bigr(-\imath\int_0^t dt'\hat{H}^\text{dH}_{c/d}(t')\Bigr)\ket{\psi^0_{c/d}}.
\end{equation}
Although the $d$-bosons formally are described by a many-body wave function due to their finite density, one can exploit their non-interacting nature to perform an efficient simulation of dynamics. In fact, as explained after Eq.~(\ref{Eq:nd-inf}), both the initial state and the eigenstates can be written as product states. After rotating the orbitals into the real space basis and back, the time-evolution reduces to a double sum over initially occupied sites and orbitals. The $d$-bosons can, then, be treated separately as single particles, and their overall wave-function can be reconstructed from the single-particle evolved states, thus reducing the complexity from exponential to linear in $L$. 

\begin{figure}[b]
\includegraphics[width=.95\columnwidth]{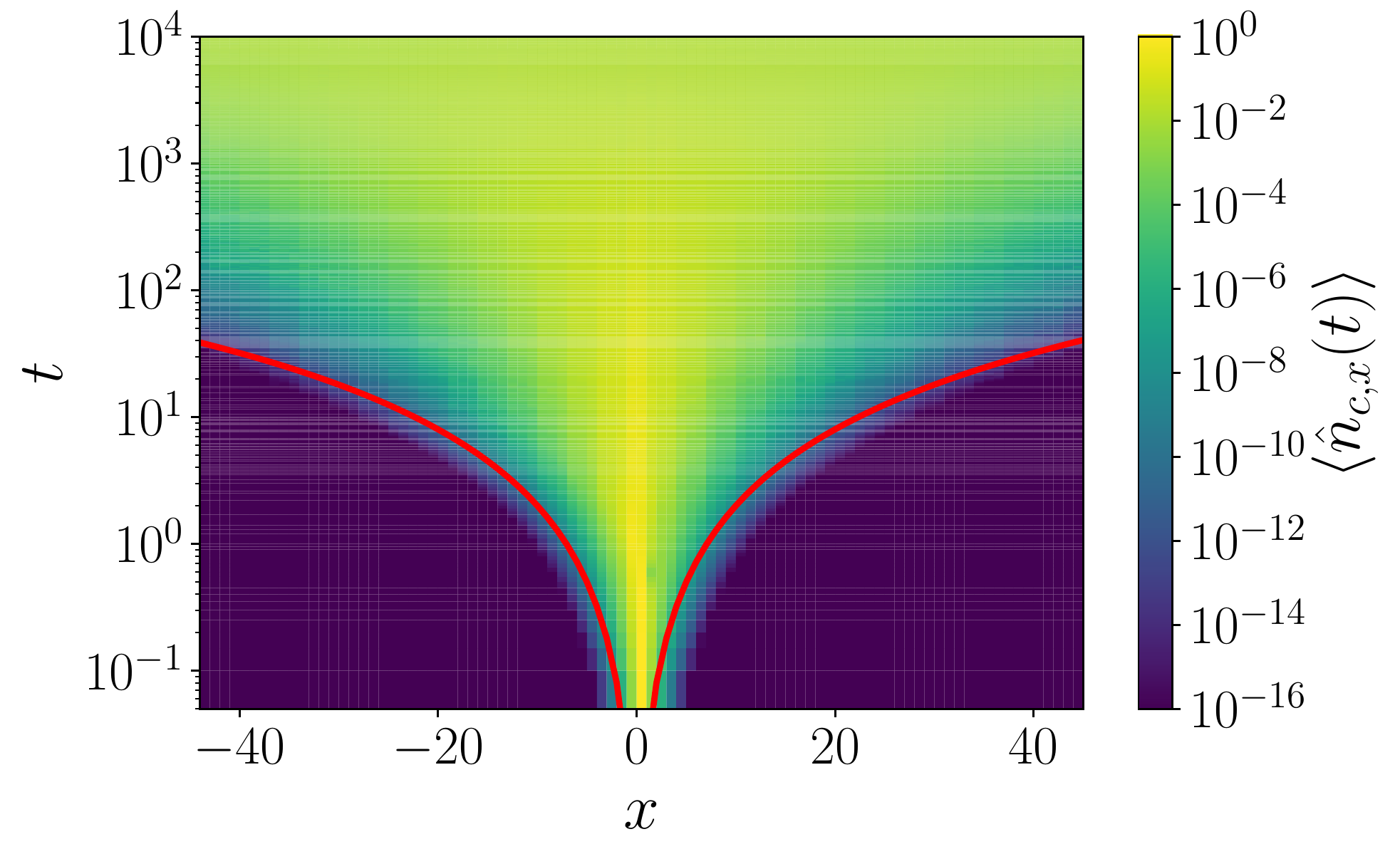}
\caption{\label{Fig:nct tdHF}
Evolution of the density profile of the clean boson in the dynamic Hartree approximation has a clear diffusive behavior. The red line corresponds to the diffusive ``lightcone'', $\langle\hat{n}_{c,x}(t)\rangle\sim\sqrt{t}$.  The data is averaged over $400$ disorder realizations, $L=90$.}
\end{figure}

First, we study the dynamics of the $c$-boson, illustrated in Figure~\ref{Fig:nct tdHF}, where we plot the density profile of the clean boson $\langle \hat{n}_{c,x}(t)\rangle$ as a function of distance from its initial location, $x=i-L/2$, and time. We observe that in contrast to the case of static Hartree approximation, where the $c$-boson spreads within an exponentially localized envelope (see Appendix~\ref{App:Hartree-Dynamics}), the dynamic Hartree approximation results in a diffusive spreading of $c$-boson and its complete delocalization over the entire system on a timescale $t\sim L^2$. Once the $c$-boson spreads over the whole chain, it does not reach a steady state, which can be attributed to the fact that this approximation oversimplifies the true many-body character of the problem. 

\begin{figure}[t]
\includegraphics[width=.95\columnwidth]{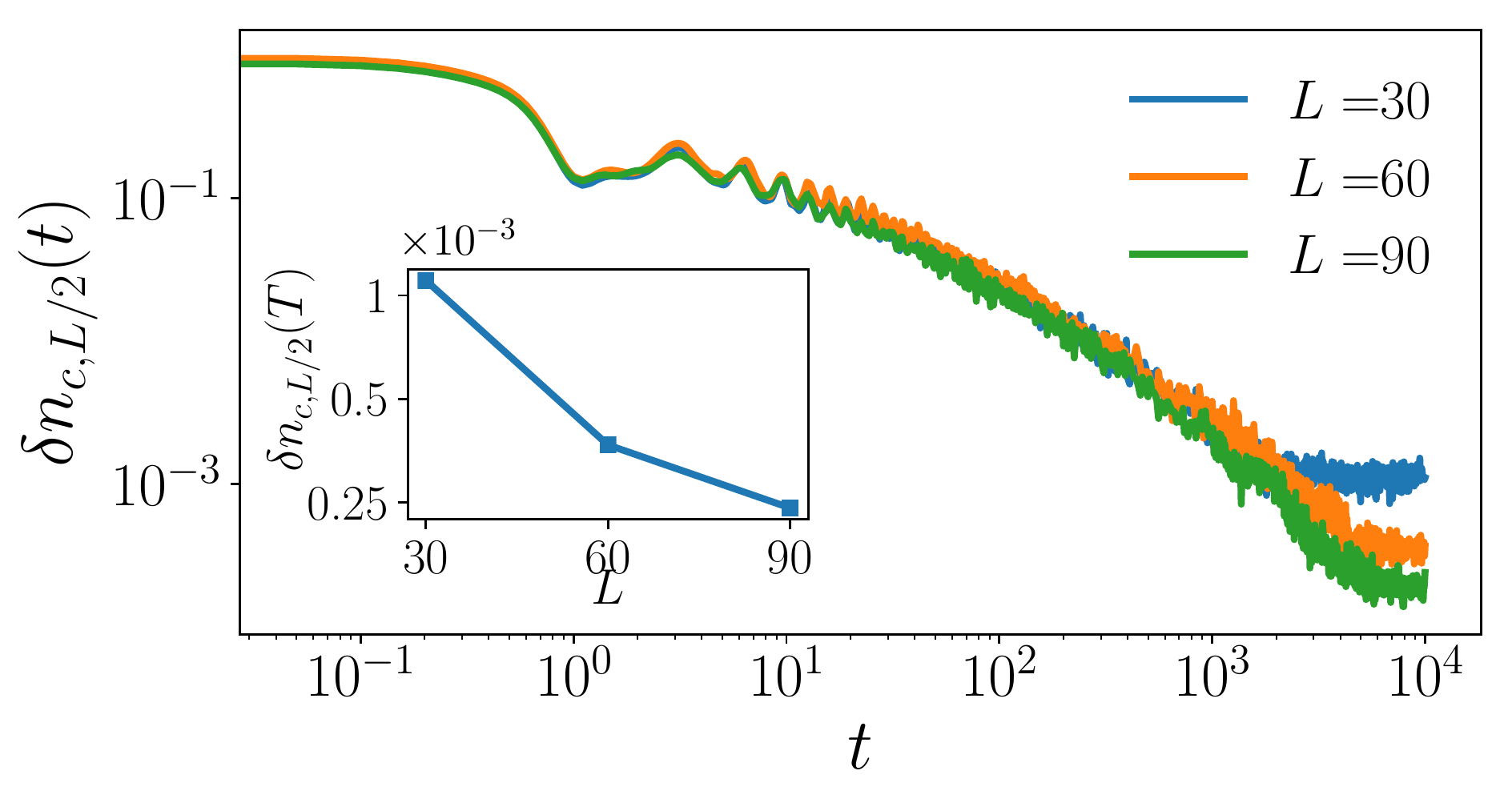}
\caption{\label{Fig:fluct c}
Fluctuations of the $c$-boson density decay slightly faster than polynomially in time. Inset shows that the saturation value of density fluctuations reached at late times decreases monotonically with the system size as $1/L^{b}$, with $b\approx1.6$. This data is obtained averaging over $400$ disorder realizations.}
\end{figure}

The fluctuations of the $c$-boson density result in a time-dependent potential acting on disordered bosons, see Eq.~(\ref{Eq:Hd(t)}). In order to quantify the density fluctuations of the clean boson, we average the absolute value of the deviation of its density from the mean value, $\delta n_{c,i}(t) = |\langle\hat{n}_{c,i}(t)\rangle-\overline{\langle\hat{n}_{c,i}\rangle}|$,
where $\overline{\langle\hat{n}_{c,i}\rangle}$ corresponds to the average over the interval $t\in[9.9\times 10^3,10^4]$.
As shown in Figure~\ref{Fig:fluct c}, the density fluctuations around the initial position of the $c$-boson decay in time, until they eventually reach a plateau. Note the long timescale involved in reaching the saturation value, even for  a relatively small system of $L=30$ this happens at times $t\gtrsim 10^3$.

The dynamics of the $d$-bosons can be intuitively understood as resulting from an Anderson insulator globally coupled to a weak, non-Markovian, local noise. While it is known~\cite{Knap2017,Marino2018,Gefen1984,Logan1987,Scardicchio2021} that Anderson localization is unstable with respect to global noise, a recent work~\cite{BarLev2021} has demonstrated that coupling of an Anderson insulator to a local Markovian white noise leads to a logarithmically slow particle transport and entanglement growth.  Our dynamics differs from that of Ref.~\cite{BarLev2021} in that $d$-bosons are coupled to fluctuations of the  $c$-boson density globally throughout the entire length of the chain. Another important difference is that the density fluctuations of the clean boson produce a temporally and spatially correlated noise. 

\begin{figure}[t]
\includegraphics[width=.95\columnwidth]{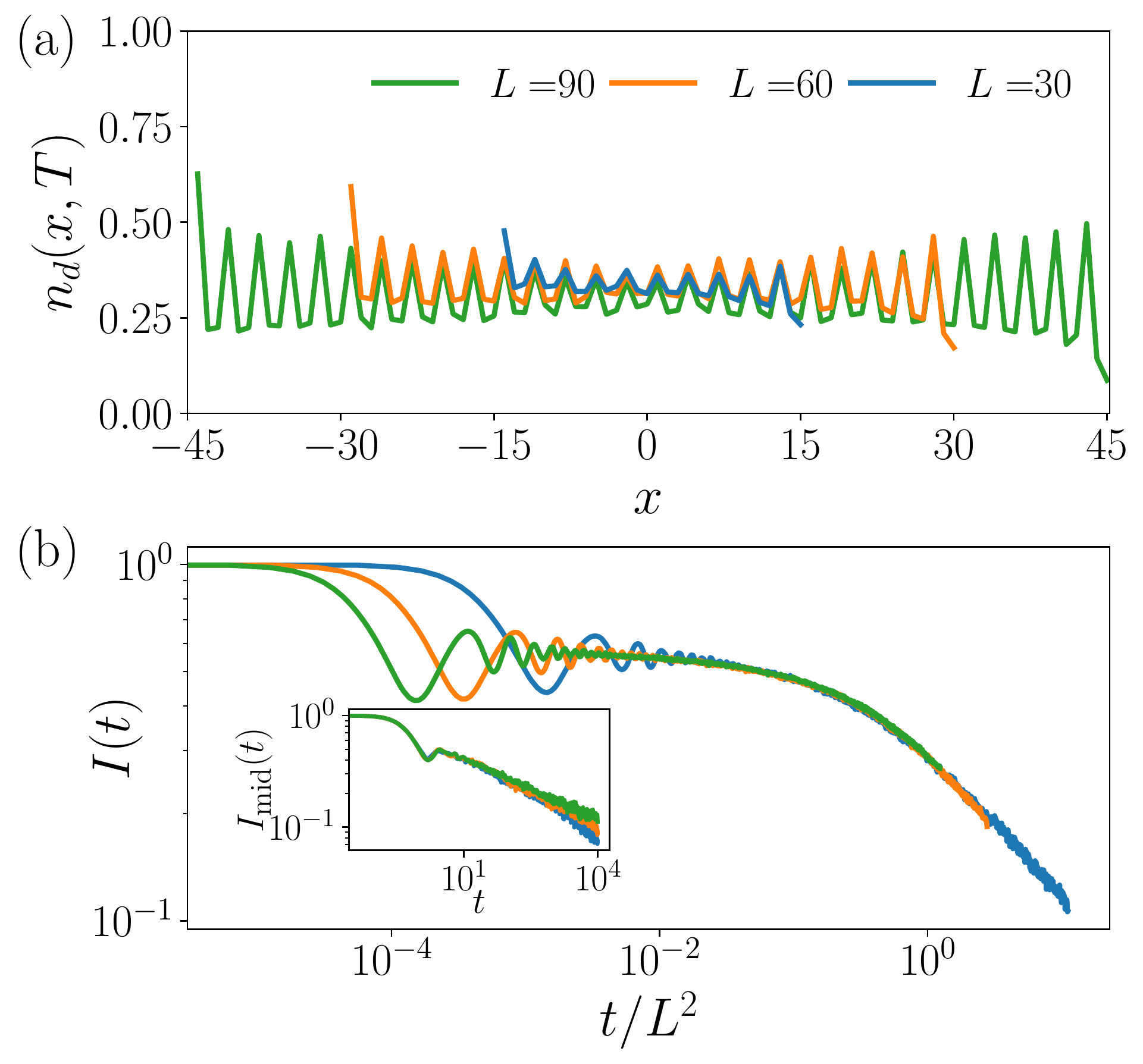}
\caption{\label{Fig:nd Id tdHF}
(a) The density profile of $d-$bosons at late time $T=10^4$ retains memory of the initial state. Comparing data for different system sizes, we observe that larger systems keep stronger memory of the initial state.
(b) Inset shows that the imbalance in the middle of the chain follows an approximately power-law relaxation without any signs of saturation at late times. The main panel reveals that imbalance dynamics collapses for different system sizes after rescaling of the time axis by $1/L^2$.  The averaging is performed over $400$ disorder realizations.
}
\end{figure}

In Figure~\ref{Fig:nd Id tdHF}(a) we show the density profile of $d$-bosons at late times, $T=10^4$, for different system sizes $L=30$, $60$, and $90$. Although the relaxation is strongest in the middle of the chain, the memory of initial density wave configuration survives even at late times. 
To understand the dynamics of relaxation of density profile, we consider the imbalance $I(t)$~\cite{Schreiber2015}. Imbalance quantifies the memory of the initial state using the difference in occupation among initially occupied and empty sites ($N_o$ and $N_e$ respectively), 
\begin{equation}
\label{Eq:Id}
I(t) = \frac{N_o(t)-N_e(t)}{N_o(t)+N_e(t)}.
\end{equation}
For the initial period-3 density wave state considered here, the explicit form of $N_{o/e}$ reads:
\begin{equation}
\label{Eq:No Ne}
N_o =  \sum_{i=1}^{L/3}\hat{n}_{d,3i-2}\;,\quad N_e = \frac{1}{2}\sum_{i=1}^{L/3}\Bigr(\hat{n}_{d,3i}+\hat{n}_{d,3i-1}\Bigr).
\end{equation}
Figure~\ref{Fig:nd Id tdHF}(b) reveals that $I(t)$ decays without any signs of saturation even at times $T=10^4$. After rescaling the time axis with a factor of $1/L^2$ we observe the collapse of the data, which further supports the intuition that although the imbalance is not a conserved quantity, its relaxation will happen on timescale that scales as $t\propto L^2$ with the system size.  The decay of imbalance, after a plateau extending up to $t\approx 0.1L^2$ follows a power-law scaling, $I(t)\sim \left({L^2}/{t}\right)^{\beta}$, with $\beta\approx 0.4$ as shown in the inset of Fig.~\ref{Fig:I comparison}(a). 

In summary, the dynamic Hartree approximation suggests a complete delocalization of the clean particle, that is followed by the melting of the imbalance of disordered bosons. This does not agree with our expectations from the criterion for stability of localization obtained in Sec.~\ref{Sec:Stability}, and also disagrees with the results of the TEBD dynamics presented below. It is natural to attribute the delocalization observed here to the nature of dynamic Hartree approximation, that completely neglects the many-body character of the problem and discards the correlations between the two species of bosons.  The inclusion of such correlations is expected to lead to a more efficient relaxation of density fluctuations and may stop the diffusive spreading of $c$-boson as we illustrate below.

\subsection{Stability of localization from TEBD dynamics}

We now turn our attention to the fully interacting system, thus reintroducing all the effects neglected in the Hartree approximation (both static and time-dependent). In order to simulate dynamics of large systems we use the time-evolving block decimation (TEBD) method~\cite{Vidal2003} with time-step $dt=0.05$, truncation $\varepsilon=10^{-9}$ and maximum bond dimension $\chi=3000$, delegating all details of numerical implementation to Appendix~\ref{App:TEBD}.

\subsubsection{Stability of localization of clean and disordered bosons}
Studying the dynamics of the $c$-boson in the Hartree limit, see Appendix~\ref{App:Hartree-Dynamics}, we realized that it presents different behaviors according to the distance from the central site. Close to the center, within a time-dependent radius $R(t)$, the density is exponentially suppressed with a decay length $\ell_c(t)$. On the other hand, farther from the middle, it shows rather a Gaussian expansion. In the accompanying work, Ref.~\cite{Brighi2021a}, we analyzed the dynamical localization of the $c$-boson observed in the TEBD dynamics. Observing features similar to the static Hartree case, there we  proposed the following ansatz for the dynamics of clean boson density, 
\begin{equation}
\label{Eq:nc(x,t)}
n_c(x,t) = \mathcal{N}_c(t)\exp\Bigr(-\frac{|x|}{\ell_c(t)\tanh\bigr(R(t)/|x|\bigr)}\Bigr),
\end{equation}
where $\mathcal{N}_c(x,t)$ is a normalization factor. Eq.~(\ref{Eq:nc(x,t)}) smoothly interpolates between the two behaviors observed, as $|x|$ runs across $R(t)$. Based on phenomenological observations, we suggested the following form for the decay length
\begin{equation}
\label{Eq:ell_c(t)}
\ell_c(t) = \xi_c\frac{\log\bigr(1+t/T_0\bigr)}{1+\log\bigr(1+t/T_0\bigr)},
\end{equation}
which implies saturation of the decay length to a constant localization length $\xi_c$ on timescales $t\gg T_0$.

\begin{figure}[t]
\includegraphics[width=.95\columnwidth]{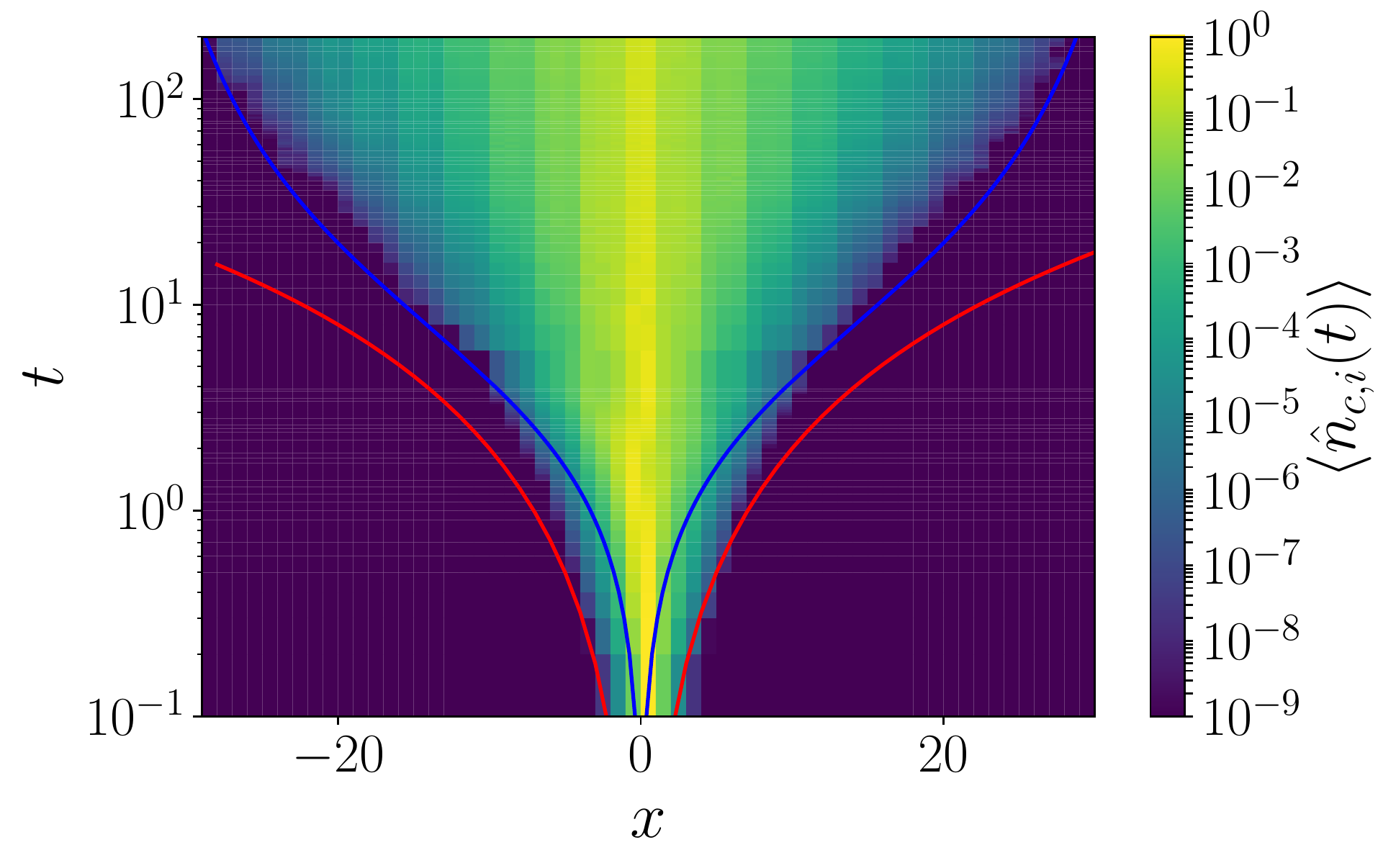}
\caption{\label{Fig:nct TEBD}
The spreading of the $c$-boson density in the $L=60$ system is well described by the ansatz of Eq.~(\ref{Eq:nc(x,t)}) (blue curve), while it deviates from the diffusive behavior (red curve). }
\end{figure}

Figure~\ref{Fig:nct TEBD} shows the evolution of the density profile of the clean boson with time. While at early times it approximately agrees with a diffusive profile extracted from time-dependent Hartree approximation (red line and Fig.~\ref{Fig:nct tdHF}), at times of order one, the density spreading slows down. To describe the envelope of the density profile, we use the fact that when $R(t)$ saturates to $L/2$, the ansatz~(\ref{Eq:nc(x,t)}) reduces to $n_c(x,t)\approx e^{-|x|/\ell_c(t)}$. By imposing the condition $n_c(x,t)=\text{const}$, we obtain the scaling $|x|\approx \ell_c(t)$ shown in Fig.~\ref{Fig:nct TEBD} by the blue line, that provides better description for the envelope of the density profile.

\begin{figure}[b]
\includegraphics[width=0.95\columnwidth]{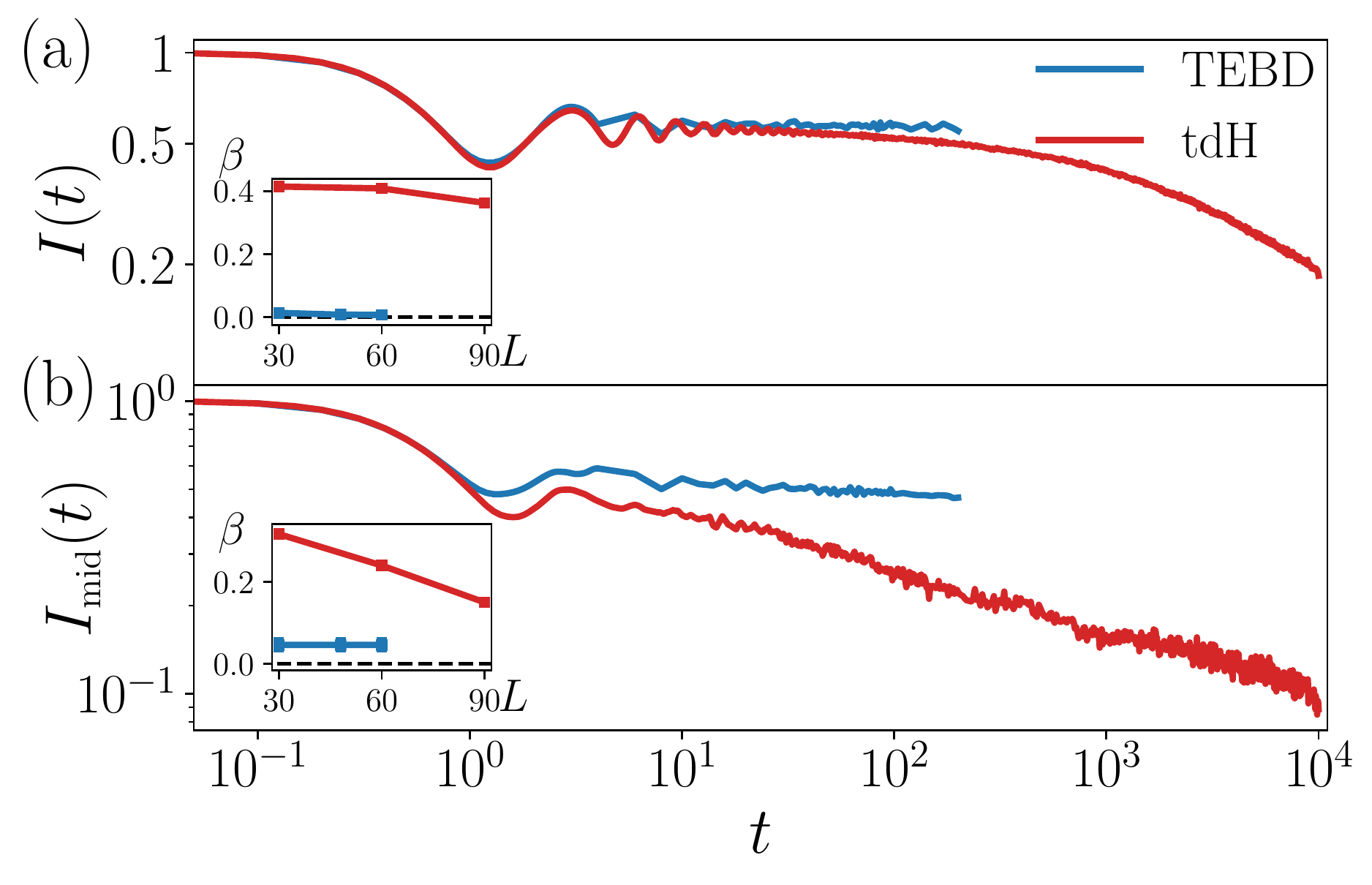}
\caption{\label{Fig:I comparison}
(a) TEBD dynamics (blue line) reveals a broad plateau in imbalance at late times, whereas time-dependent Hartree approximation (red line) predicts a power-law decay.
(b) The imbalance of central six sites in TEBD dynamics reveals a weak power-law decay. The stronger deviation between TEBD and approximate time-dependent Hartree data highlights the importance of entanglement for the accurate description of the dynamics. Data is shown for $L=60$ and averaged over $400$ and $50$ disorder realizations for time-dependent Hartree and TEBD respectively.
}
\end{figure}

Next, we consider the dynamics of imbalance of $d$-bosons, defined in Eq.~(\ref{Eq:Id}), that is shown in Fig.~\ref{Fig:I comparison}(a).  While TEBD and time-dependent Hartree approximation approximately agree at times up to $t\approx 10$,  at later times the imbalance obtained from TEBD remains approximately constant, while the time-dependent Hartree approximation predicts a power-law decay of the imbalance. This is further highlighted by the exponent $\beta$, shown in the inset, that is compatible within error bars with value $\beta=0$ for TEBD dynamics, which corresponds to non-decaying imbalance. 

In order to quantify the stronger relaxation close to the center of the chain, triggered by the presence of the clean boson, we also study the local imbalance in the middle of the chain, $I_\text{mid}(t)$ defined according to Eq.~(\ref{Eq:Id}) but restraining the sum in Eq.~(\ref{Eq:No Ne}) to the six central sites. The quantity  $I_\text{mid}(t)$ shows a stronger decay in Fig.~\ref{Fig:I comparison}(b) as compared to its global counterpart in Fig.~\ref{Fig:I comparison}(a). However, again the TEBD dynamics reveals a much weaker effect of the $c$-boson compared to the Hartree approximation. Fitting the decay to the power-law form, the resulting exponent remains small and it does not change with system size (inset). Hence, we expect the effect on the central region to be system size independent, thus leaving the boundaries unaffected as $L\to\infty$. 

\subsubsection{Entanglement dynamics}

After demonstrating the stability of localization in TEBD dynamics, we present additional details for the picture of entanglement spreading provided in Ref.~\cite{Brighi2021a}.  Thanks to $U(1)$ conservation, we can write the reduced density matrix of the first $i$ sites from the left, $\rho_i$, in block-diagonal form, $\rho_i = \sum_n p_i^{(n)}\rho_i^{(n)}$, where $p_i^{(n)}$ corresponds to the weight of the $n$-particles sector and $\mathrm{tr}\rho_i^{(n)}=1$. It is then convenient to separate the entanglement entropy, $S(i) = -\mathrm{tr}\rho_i\log\rho_i$, in two contributions: the configuration entanglement $S_C(i)$ and the particle entanglement $S_n(i)$~\cite{Lukin2019}
\begin{align}
\label{Eq:S}
S(i) = &S_n(i)+S_C(i) \\
\label{Eq:Sc}
S_C(i) = -&\sum_n p_i^{(n)} \mathrm{tr}\rho_i^{(n)}\log\rho_i^{(n)}\\
\label{Eq:Sn}
S_n(i) = &-\sum_n p_i^{(n)}\log p_i^{(n)}.
\end{align}

\begin{figure}[b]
\includegraphics[width=.95\columnwidth]{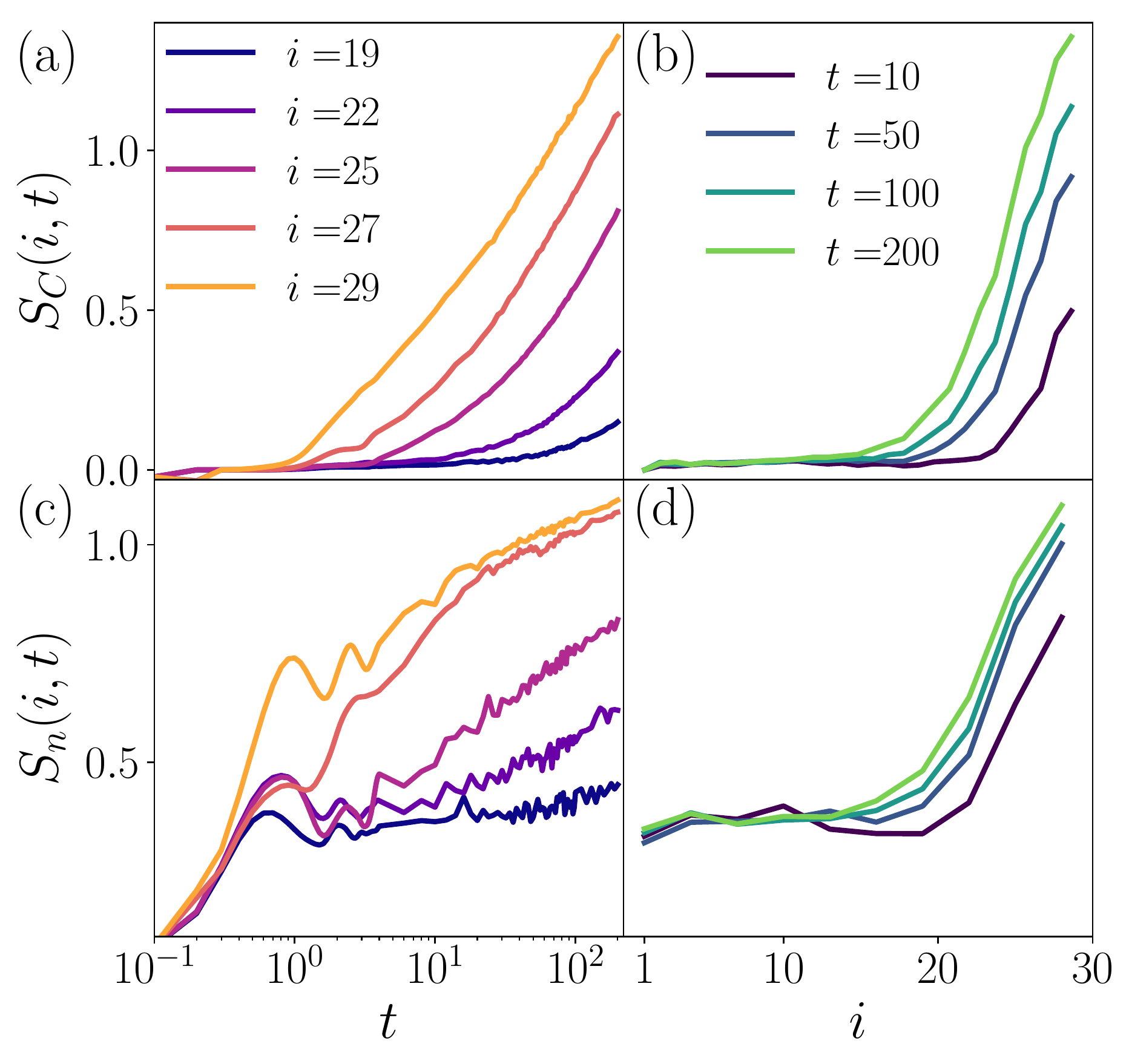}
\caption{\label{Fig:S(t)}
(a) The configuration entropy shows a steady logarithmic growth, whose start is delayed as the cut gets farther from the center of the chain, due to the weaker effect of the interaction at the boundaries. 
(b) The real space profile of the configuration entropy is highly non-uniform, with entropy being maximal in the center of the chain, where the $c$-boson is localized. Far from the center, $S_C(i)$ is close to zero, indicating the absence of interactions in the boundaries up to late times.
(c) The growth of number entropy is similarly delayed with increasing distance from the center of the chain. It shows slow growth, compatible with the weak relaxation of $d$-bosons observed in the imbalance dynamics.
(d) The number entropy also has a non-uniform profile, with its values far from the center being consistent with the particle entropy of the Anderson insulator.
This data refer to a system of $L=60$ sites, averaged over $50$ disorder realizations.
}
\end{figure}

\begin{figure}[t]
\includegraphics[width=.95\columnwidth]{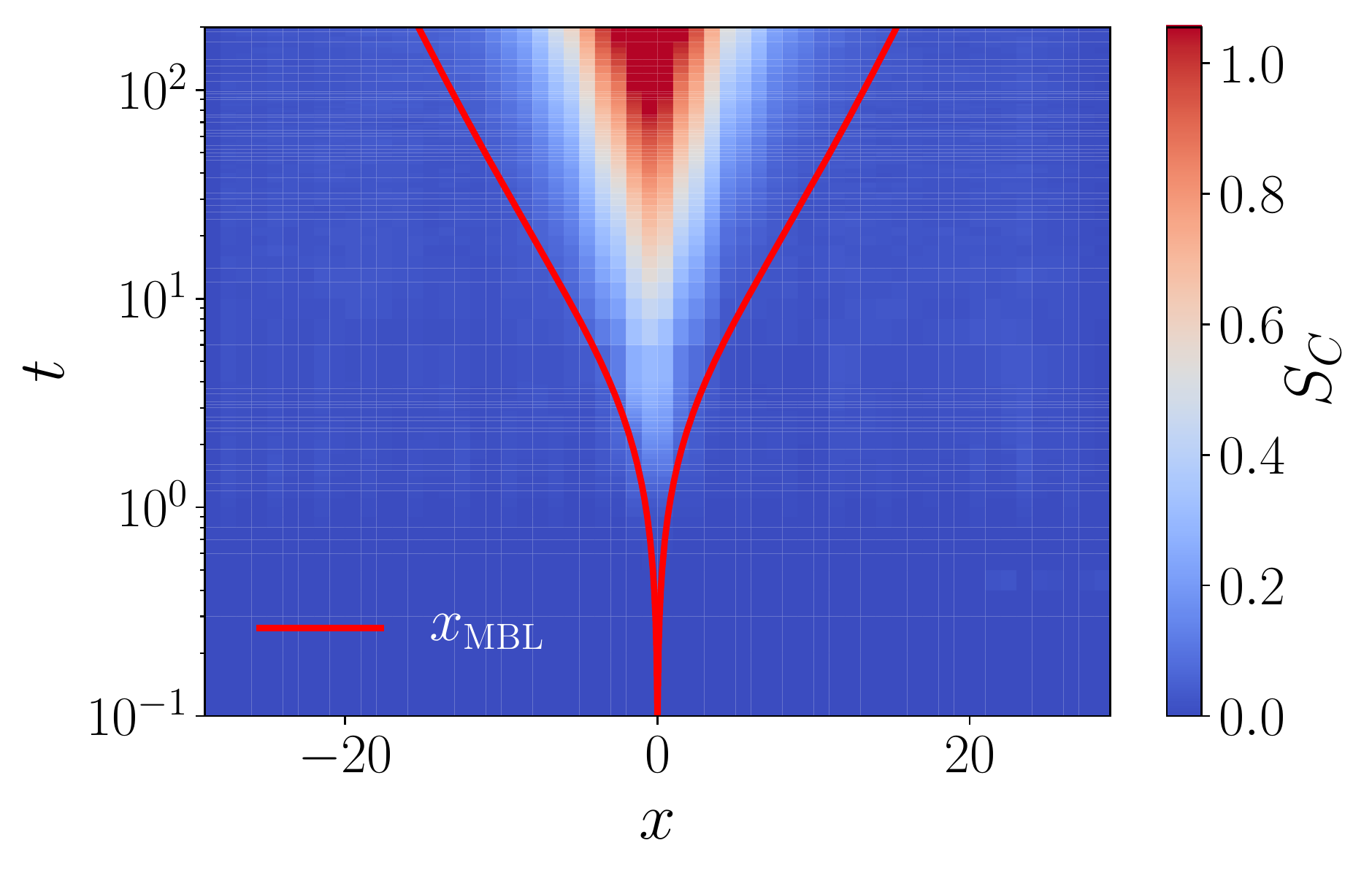}
\caption{\label{Fig:S_C spread}
The configuration entropy $S_C(i,t)$ at $U=12$ and $L=60$, characterizing the interacting nature of the system, presents an extremely inhomogeneous growth. In agreement with the picture of propagation of MBL, its growth is constrained within the ``MBL light-front'' $x_\text{MBL}(t)$ shown by the red line. When $x\gg x_\text{MBL}$, the configuration entropy $S_C$ is close to zero, suggesting that system remains effectively non-interacting until late times. 
}
\end{figure}

The configuration entropy $S_C(i,t)$ arises from the superposition of different product states in $\ket{\psi(t)}$ and measures the correlation between particle configurations in the two subsystems. The growths of configuration entropy can be attributed to the interacting nature of the system. Thus configuration entropy does not show interesting dynamics in Anderson insulator, while displaying a logarithmic growth in a many-body localized phase~\cite{Lukin2019} providing the main contribution to the entanglement growth~\cite{Serbyn2013a}. Particle number entanglement entropy $S_n(i)$ arises from the occupation of different sub-sectors of the density matrix $\rho_i$ and hence accounts for the particle transport in the system. Recent studies~\cite{Kiefer-Emmanouilidis2020,Kiefer-Emmanouilidis2021} suggested an extremely slow, albeit finite, growth $S_n(i,t)\sim\log\log t$, which they attributed to slow particle transport. Subsequent work by \textcite{Luitz2020} however suggested absence of such transport and saturating particle number entanglement.  

The dynamics of the two contributions to the entanglement in our model is presented in Figure~\ref{Fig:S(t)}, where the first row shows the configuration entanglement growth for different cuts in the chain~(a) and its profile at different times~(b). In the second row, we show the dynamics of particle number entanglement. Similarly to the configuration entropy, $S_n(i,t)$ has a non-homogeneous growth across the system shown in  Fig.~\ref{Fig:S(t)}(c)-(d).

Separating entanglement $S(i)$ into configurational and particle part, it is clear that the logarithmic growth is due to the configuration entropy, $S_C(i)\sim \xi_S\log(t)$, as clearly shown in Figure~\ref{Fig:S(t)}(a).  The configuration entanglement growth has a non-homogeneous behavior in the system, explained by the propagation of MBL phenomenology discussed in Ref.~\cite{Brighi2021a}. Introducing an effective interaction $U_\text{eff}(x,t) = Un_c(x,t)$, the dynamics of $S_C(i)$ can be described by the ansatz $S_C(i,t)\approx\xi_S\log(1+U_\text{eff}(i,t)t)$. This explains the delay of onset of growth of configurational entropy in Fig.~\ref{Fig:S(t)}(a) away from the center of the chain. In addition, this phenomenology  predicts an linearly decreasing  entanglement profile away from the center of the chain. Indeed, using an exponential profile for the density of the clean boson at late times we obtain the following effective interaction strength, $U_\text{eff}\approx Ue^{-|i-L/2|/\ell_c(t)}$. This gives the following asymptotic behavior of configurational entanglement, 
\begin{equation}\label{Eq:}
S_C(i,t\gg 1)\sim \text{const}-\frac{\xi_S}{\ell_c(t)}|i-L/2|,
\end{equation}
that decays linearly away from the center of the chain with slope given by $1/\xi_S$. This prediction is confirmed by the numerical results shown in Fig.~\ref{Fig:S(t)}(b), yielding the value of $\xi_S\approx0.31$.

According to the effective interaction picture describing the propagation of MBL, the interacting nature spreads through the system producing a many-body localization lightcone. At late times, when $n_c(i,t)\approx\mathcal{N}_c(t)e^{-|i-L/2|/\ell_c(t)}$, this can be captured analytically by solving $S_C(x,t)=\text{const}$ for $x$ and defines the MBL ``light-front",
\begin{equation}
\label{Eq:xMBL}
x_\text{MBL}(t) = \ell_c(t)\log\bigr(\mathcal{N}_c(t)Ut\bigr).
\end{equation}
As shown in Figure~\ref{Fig:S_C spread}, the prediction of the MBL lightcone (red curve) provides an accurate description of the actual behavior of the spread of configuration entropy, thus testing the genuine propagation of the many-body nature through the chain.

\section{Probing eigenstates with DMRG-X\label{Sec:DMRG-X}}

Although TEBD enabled simulation of dynamics for large systems, the maximal time remained limited by entanglement growth.  Below we use the DMRG-X algorithm~\cite{Serbyn2016,DMRG-X,Pollman2016} to probe eigenstates of large systems, providing an effective insight into infinite time behavior.

\begin{figure}[b]
\includegraphics[width=.95\columnwidth]{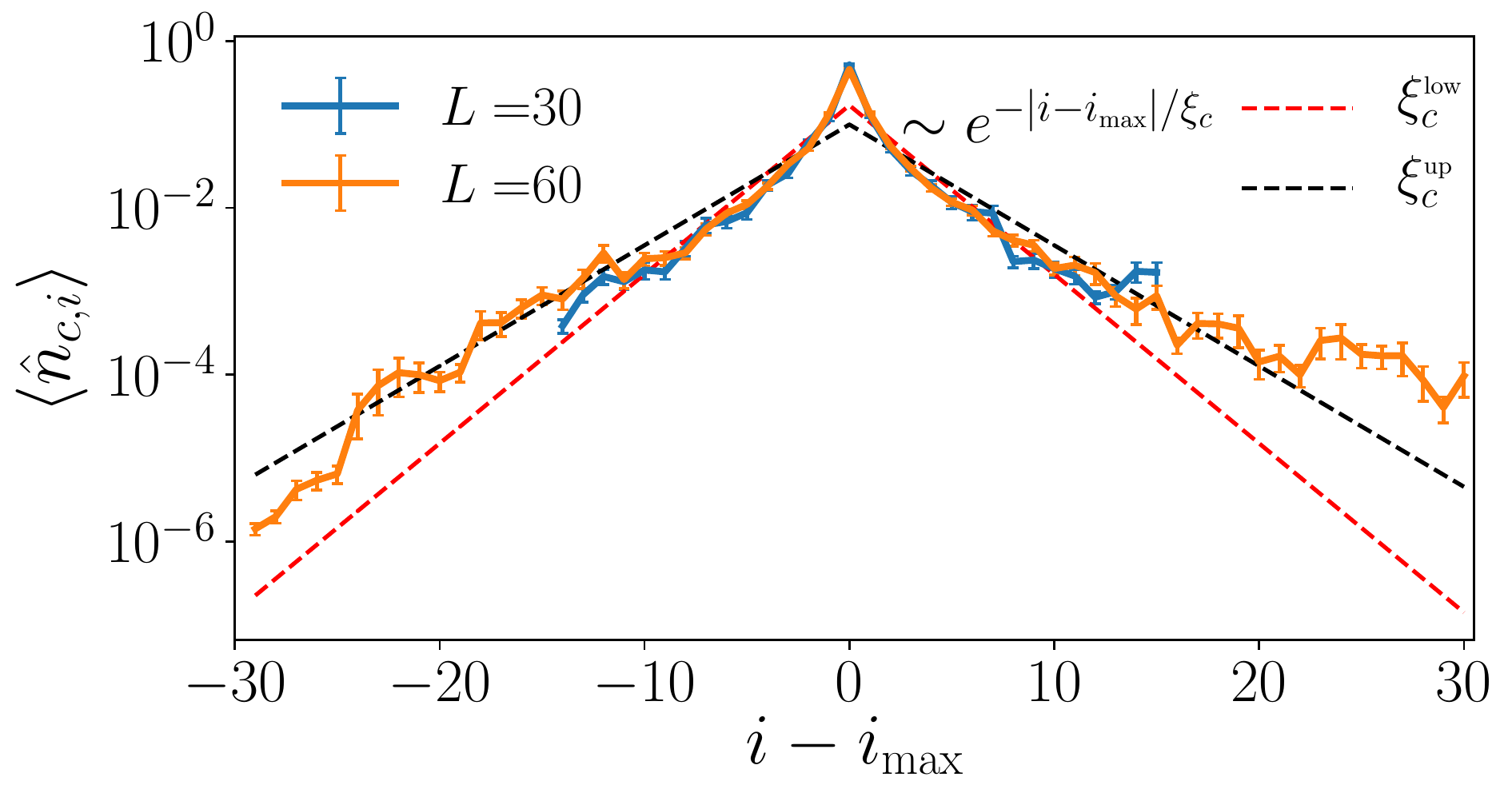}
\caption{\label{Fig:nc DMRG}
The density profile of the $c$-boson averaged over different eigenstates as a function of the distance from $i_\text{max}$ shows an exponential decay for both the system sizes considered. Through a fit to the density in different regions we obtain an upper and lower bound for the localization length, with $\xi^\text{low}_c\approx 2.1$ (red dashed line) corresponding to the fit excluding the central site and $\xi^\text{up}_c\approx 3$ (black dashed line) obtained from the fit that exclude 12 central sites.}
\end{figure}

\subsection{$c$- and $d$-boson localization}

We extract highly excited eigenstates of the Hamiltonian~(\ref{Eq:H}) with $L=30$ and $60$ sites, $U=12$ and 1/3 filling of $d$-bosons. We obtain in total $250$ (500) eigenstates in the middle of the spectrum by targeting $25$ different energies for each of the $10$ ($20$) disorder realizations considered for $L=30$ (60) chain respectively.~\footnote{We note, that the first 10 disorder realizations for $L=60$ system are chosen in such a way that the values of the random energies of the $30$ central sites agree with those for $L=30$ system.} The states are obtained as a result of $100$ sweeps where the targeted eigenstate is variationally approximated by an MPS of maximum bond dimension $\chi=500$ and $250$ for systems of size $30$ and $60$, respectively. The algorithm is initialized with different copies of the initial state~(\ref{Eq:psi0}), where the $c$-boson initial position is chosen among empty sites in the central region. We refer the reader to Appendix~\ref{App:DMRG-X} for details on the implementation of the algorithm as well as its performance metrics.

For each of the eigenstates we extract the peak position of the $c$-boson density, $i_\text{max}$. The average $c$-boson density plotted as a function of the distance from its peak presents an exponential profile, as shown in Fig.~\ref{Fig:nc DMRG}. We notice that the density profile around the peak is enhanced with respect to the tails, which can be potentially attributed to the large value of $U$ producing stable doublons. The exponentially decaying density confirms the localization of the clean particle and allows us to extract the upper and lower bound on the localization length, $\xi^\text{low}_c \approx 2.1$ and $\xi^\text{up}_c \approx 3$  resulting from fitting its profile close to the center (excluding the central site) and at the edges of the chain. The two bounds  quantitatively agree with the lengthscale $\ell_c(t\to\infty) = \xi_c = 2.5$ \emph{extrapolated} from TEBD dynamics in Ref.~\cite{Brighi2021a}, suggesting that the localization of the $c$-boson observed in dynamics is not a transient effect, but it is a property of the system, provided the correct sector of the Hilbert space is explored.

\begin{figure}[t]
\includegraphics[width=.95\columnwidth]{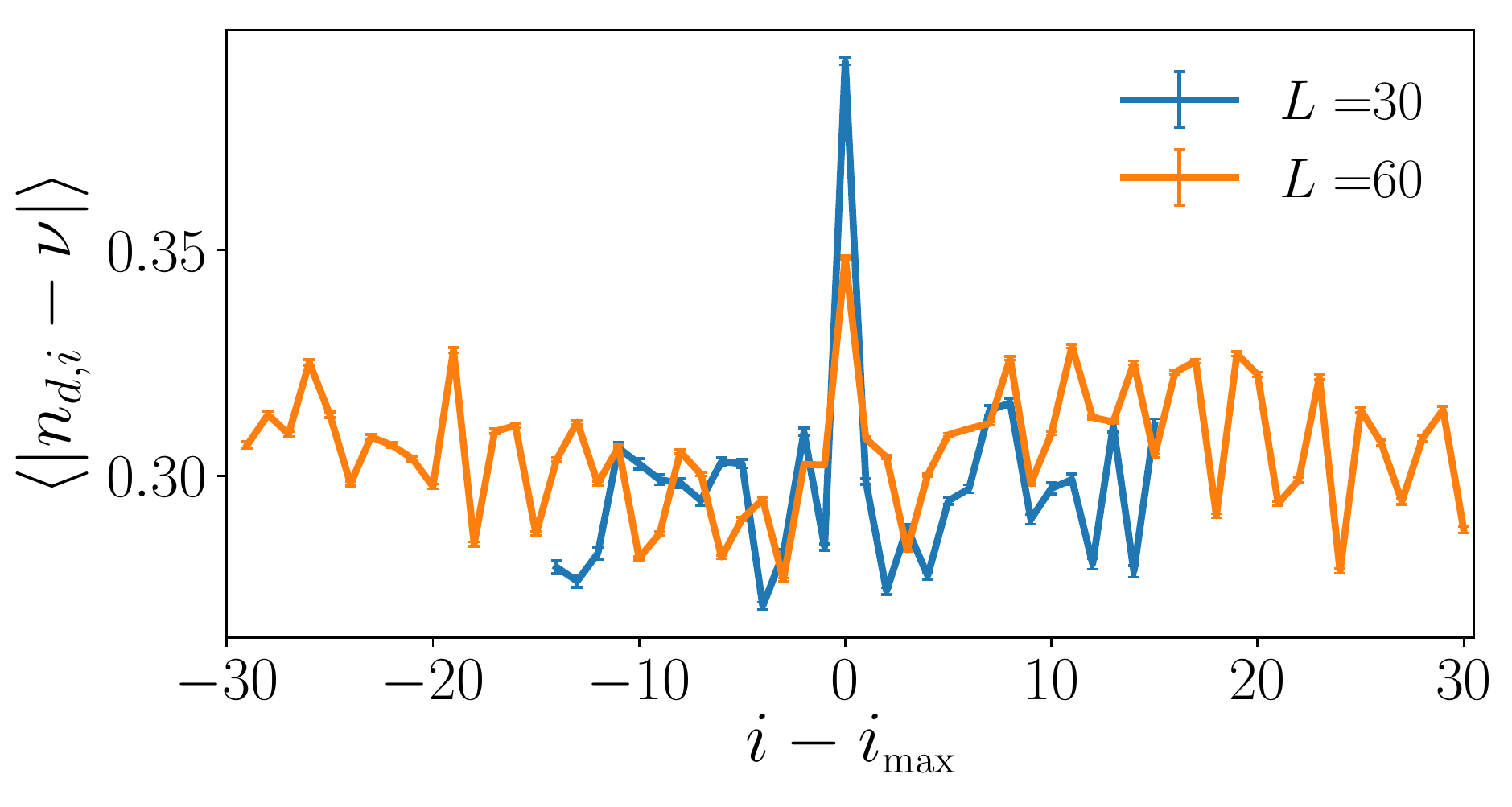}
\caption{\label{Fig:nd fluct DMRG}
Strong deviation of the $d$-bosons density away from the thermal value $\nu_d=1/3$ consistent between different system sizes  indicates the breakdown of thermalization in eigenstates. }
\end{figure}

Probing potential localization of $d$-bosons in eigenstates is more complicated: the finite particle density does not allow to define a localization center.  Therefore, we consider the average deviations of the $d$-bosons density from the thermal value given by their average density $\nu_d = 1/3$ as a function of the distance from $i_\text{max}$.  In Figure~\ref{Fig:nd fluct DMRG}, we show the averaged absolute value of the deviation of the density expectation value from $\nu_d$, which is limited to the range of values $0<\langle| \hat{n}_{d,i}-\nu|\rangle<1-\nu_d$. While in a thermalizing system we expect this quantity to be small and decrease exponentially with the system size, the eigenstates of our problem have on average large expectation value of $\langle| \hat{n}_{d,i}-\nu|\rangle$. In addition,  increasing the system size does not lead to a decrease of the average distance from the thermal value, thus suggesting that the $d$-bosons density strongly fluctuates around $\nu_d$ in eigenstates due to their localization. Interestingly, we notice a slight peak of this quantity at $i=i_\text{max}$, which is in agreement with the enhancement of the $c$-boson density observed in Fig.~\ref{Fig:nc DMRG} and may be attributed to the effect of doublons. In the region around the center, however, we observe a slight weakening of localization, corresponding to smaller values of $\langle| \hat{n}_{d,i}-\nu|\rangle$. This can be attributed to the fact that in the vicinity of the peak of the $c$-boson, the interactions are effectively stronger, leading to an increased relaxation of the $d$-bosons.

\begin{figure}[b]
\includegraphics[width=.95\columnwidth]{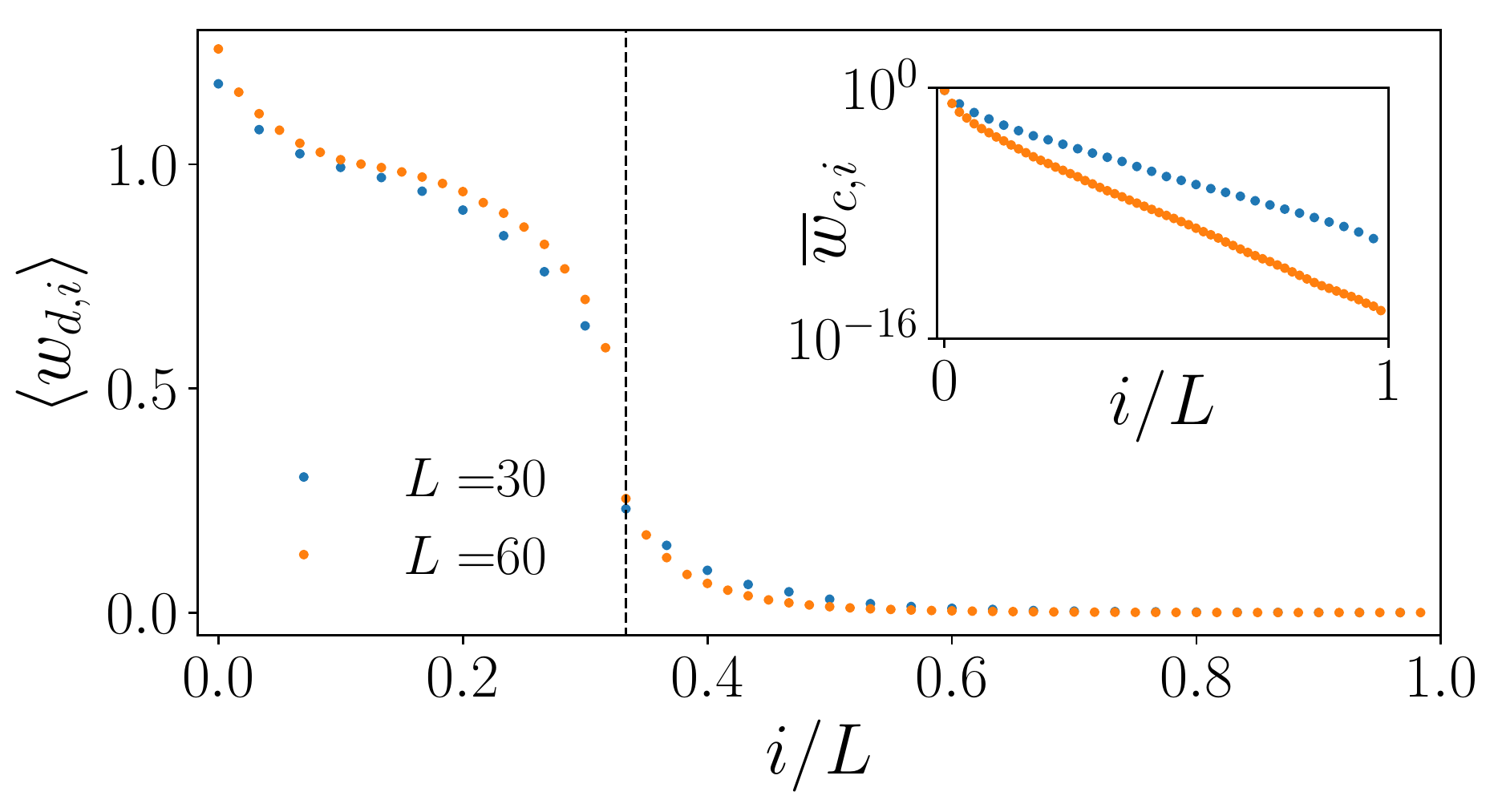}
\caption{\label{Fig:spDM spectrum}
The spectrum of $\rho_d$ shows a step behavior, dropping quickly to small values at $i/L=\nu_d=1/3$ (dashed line). This corresponds to the presence of $N_d$ occupied single-particle orbitals, thus confirming localization. We note that $w_{d,i}>1$ is due to the bosonic nature of these particles. The inset shows the log-averaged spectrum of $\rho_c$ on a logarithmic scale, revealing that there is a single significantly occupied orbital, while all the others have an exponentially decaying weight.
}
\end{figure}

\begin{figure}[b]
\includegraphics[width=.95\columnwidth]{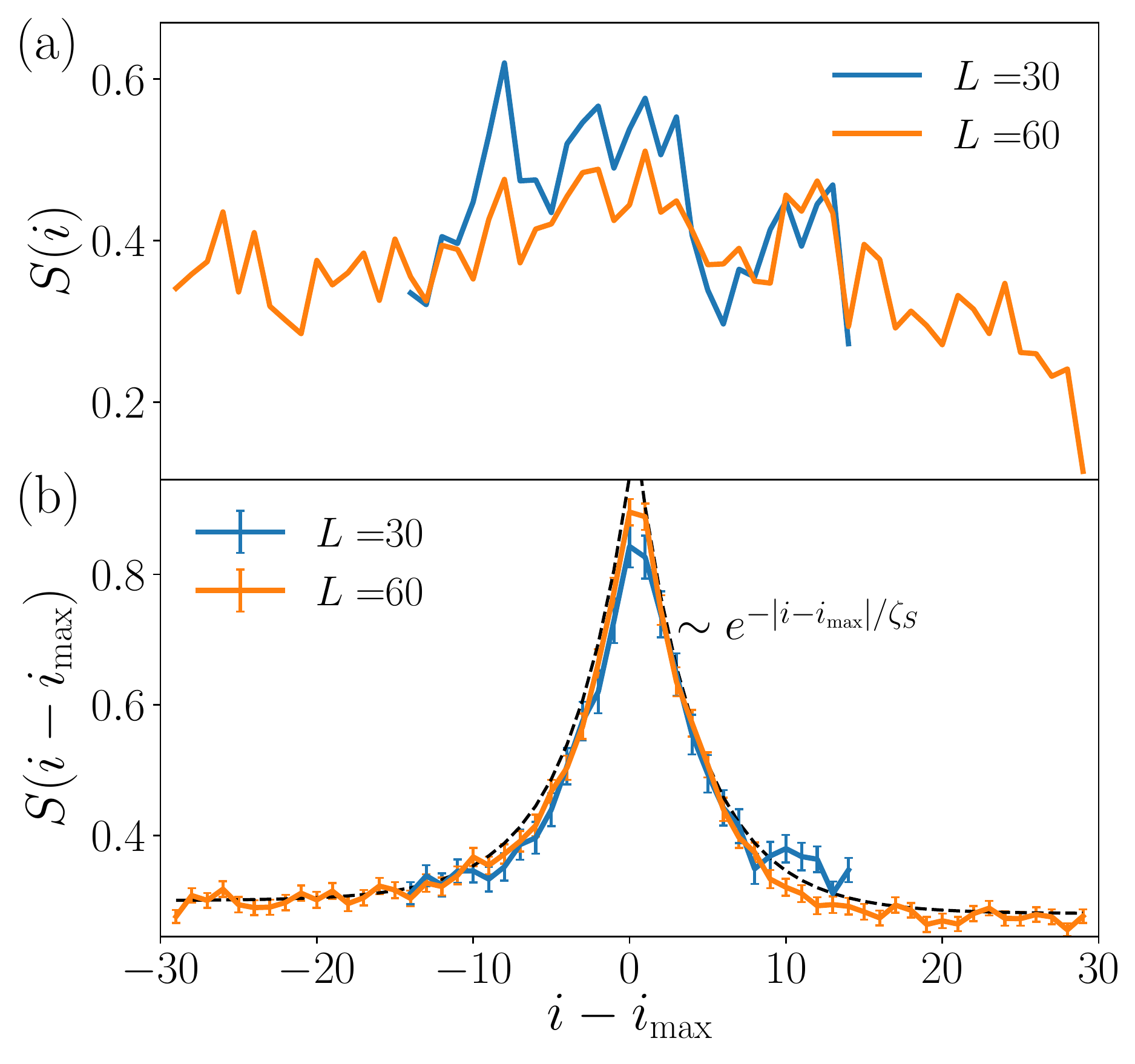}
\caption{\label{Fig:S DMRG}
(a) Average entanglement as a function of the cut position results in a featureless profile. The agreement in entanglement between different system sizes where they overlap is due to particular disorder choice.
(b) Averaging the entanglement profile as a function of the distance from the center of localization of $c$-boson, $i_\text{max}$, shows  enhancement in entanglement caused by the presence of the clean boson. The dashed line shows the comparison to an exponential fit. The collapse of entanglement profile between different system sizes suggests area-law entanglement scaling. 
}
\end{figure}

Further evidence of  localization for both types of particles is found in the spectrum of the single-particle density matrices $\rho_{c}=\langle \hat{c}^\dagger_i\hat{c}_j\rangle$ ($\rho_{d}=\langle \hat{d}^\dagger_i\hat{d}_j\rangle$)~\cite{Bardarson2015,Bardarson2017}. In the particular case treated here, the von Neumann entropy of $\rho_c$ corresponds to the intra-species entanglement $S_{cd} = -\operatorname{tr}\rho_c\log\rho_c$. We study its distribution and average value among eigenstates, obtaining average entropy of $ S_{cd}\approx 0.85$ for both system sizes. Furthermore, the eigenvalues of $\rho_{c/d}$, $w_{c/d,i}$, sorted by decreasing value, can be interpreted as occupation numbers of single-particle orbitals~\cite{Bardarson2015,Bardarson2017}. In the case of an MBL system, it is expected that particles sit on almost localized sites, hence only the first $N_{c/d}$ orbitals should be significantly occupied. In Figure~\ref{Fig:spDM spectrum} we show the average, $\langle \cdot\rangle$, and log-average, $\overline{x}  = e^{\langle\ln x\rangle}$, value of the ordered eigenvalues for $\rho_d$ and $\rho_c$ respectively. The scaling of $\langle w_{d,i}\rangle$ shown in the inset presents a step behavior, consistent with the localization of the $d$-bosons. Similarly, the typical value $\overline{w}_{c,i}$ presents a single large eigenvalue, while the rest decay exponentially.

\subsection{Entanglement structure of eigenstates}

In Figure~\ref{Fig:S DMRG}(a) we show the bipartite entanglement profile  averaged as a function of the position of the cut $i$. While the entanglement profile of eigenstates suggests an area-law entanglement scaling characteristic of many-body localization~\cite{Serbyn2013,Bauer2013}, this representation of data does not show any effect from the presence of the clean boson. To observe the influence of the $c$-boson, we average the entanglement profile defined with respect to the distance from the localization site of the $c$-boson, $i-i_\text{max}$. Figure~\ref{Fig:S DMRG}(b) reveals a peak in $S(i-i_\text{max})$ around zero, indicating that the clean boson is responsible for additional entanglement in eigenstates. Away from $i_\text{max}$ entanglement saturates to a constant value, revealing a clear area-law scaling. Fitting the decay of entanglement away from the center with an exponential of the type $S_0+ce^{-|i-i_\text{max}|/\zeta_S}$ allows to extract the value of $\zeta_S\approx4$, suggesting that the single $c$-boson is capable of generating non-trivial entanglement patterns on a scale larger than its localization length. 

The increase in entanglement of eigenstates that decays exponentially with the distance away from the location of $c$-boson provides further support to the localized nature of eigenstates. Moreover, this provides a complementary view on the picture of the inhomogeneous effective interaction triggered by the presence of $c$-boson and resulting in the dynamical ``propagation of MBL'' presented in~\cite{Brighi2021a}.

\section{Discussion}

We investigated the effect of the coupling of a small local bath represented by a single clean boson to a one dimensional  Anderson insulator with finite particle density. First, we used a static Hartree approximation that overestimates localization to obtain the parameters range most favorable for localization. In addition, we analyzed the effect of two-particle resonances triggered by the interaction, estimating the region of parameters where localization is perturbatively stable with respect to such resonances.   

Focusing on the parameter values where the perturbative criterion predicts localization, we studied the dynamics in the coupled system.  We demonstrated that the time-dependent Hartree approximation, which neglects the entanglement between the localized particles and the degree of freedom representing the bath, predicts a slow delocalization of the system consistent with diffusion. In contrast, the TEBD numerical simulation, which takes into account all quantum correlations among the two bosonic species, shows evidence of localization within the experimentally relevant timescales achieved. In particular, the imbalance of disordered particles saturates, suggesting that memory of the initial density-wave configuration is retained even at long times. In addition simulations reveal logarithmic growth of configuration entanglement, in agreement with the MBL phenomenology, and extremely slow growth of particle number entanglement, ascribed to the weak relaxation of the disordered particles due to the interaction with the small bath. In agreement with Ref.~\cite{Brighi2021a}, we observed an enhanced effect of the bath on the $d$-bosons in the center of the system, while far from it the difference with the Anderson case is negligible. 

Finally, in order to probe the fate of localization at infinite times, we used DMRG-X method to extract highly excited eigenstates of the Hamiltonian. We observe that the clean boson remains localized in eigenstates, confirming the intuition gained from the study of dynamics. We further notice area-law scaling of the entanglement entropy of eigenstates, with an enhancement in the vicinity of the localization site of the $c$-boson, $i_\text{max}$. These results  provide evidence for the effective stability of localization even at infinite times and further support for the picture proposed in the accompanying work~\cite{Brighi2021a}.

In summary, our work provides evidence for MBL proximity effect and a phenomenological picture of dynamics~\cite{Brighi2021a} for large systems that are beyond the reach of numerical exact diagonalization. Our predictions are readily verifiable in state of the art experimental setups~\cite{Rubio-Abadal2019,Leonard2020} that are capable of probing the particle dynamics with a single-site resolution. 

However, there remain many open questions. Describing a more ``powerful'' thermal bath requires a finite $c$-bosons density. The persistence of the phenomenology proposed here in such a scenario is non-trivial and its study can lead to a better understanding of the MBL proximity effect. In the same direction, a careful investigation of the role of the interaction strength is in order for an accurate description of MBL-bath systems. In particular, in the opposite limit of weak interaction the $c$-boson is expected to delocalize quickly yielding a very weak and uniform effective coupling. This regime can provide a different point of view on the MBL-bath model. A further interesting direction would be to address the stability of our predictions to inter-species interactions, i.e.~replacing the Anderson insulator Hamiltonian $\hat{H}_d$ with an MBL Hamiltonian. Although we expect the qualitative picture discussed above to hold, a different entanglement dynamics is supposed to arise in this case. Finally, replacing the Hamiltonian evolution with a periodic Floquet drive may allow the investigation of the long-time behavior of the model, since it would avoid the large number of singular value decompositions required for reaching long times with TEBD algorithm. In this framework, the saturation of the measured quantities could be observed, thus providing insight into the infinite-time fate of the system.

\begin{acknowledgments}
We thank M.~Ljubotina for insightful discussions. P.~B., A.~M.\ and M.~S.\ acknowledge support by the European Research Council (ERC) under the European Union's Horizon 2020 research and innovation program (Grant Agreement No.~850899). D.~A.~was supported by the Swiss National Science Foundation and by the European Research Council (ERC) under the European Union's Horizon 2020 research and innovation program (Grant Agreement No.~864597). The development of parallel TEBD code was supported by S.~Elefante from the Scientific Computing (SciComp) that is part of Scientific Service Units (SSU) of IST Austria. Some of the computations were performed on the Baobab cluster of the University of Geneva.
\end{acknowledgments}

\appendix 

\section{Clean boson dynamics in static Hartree disorder\label{App:Hartree-Dynamics}}
\begin{figure}[b]
\includegraphics[width=0.95\columnwidth]{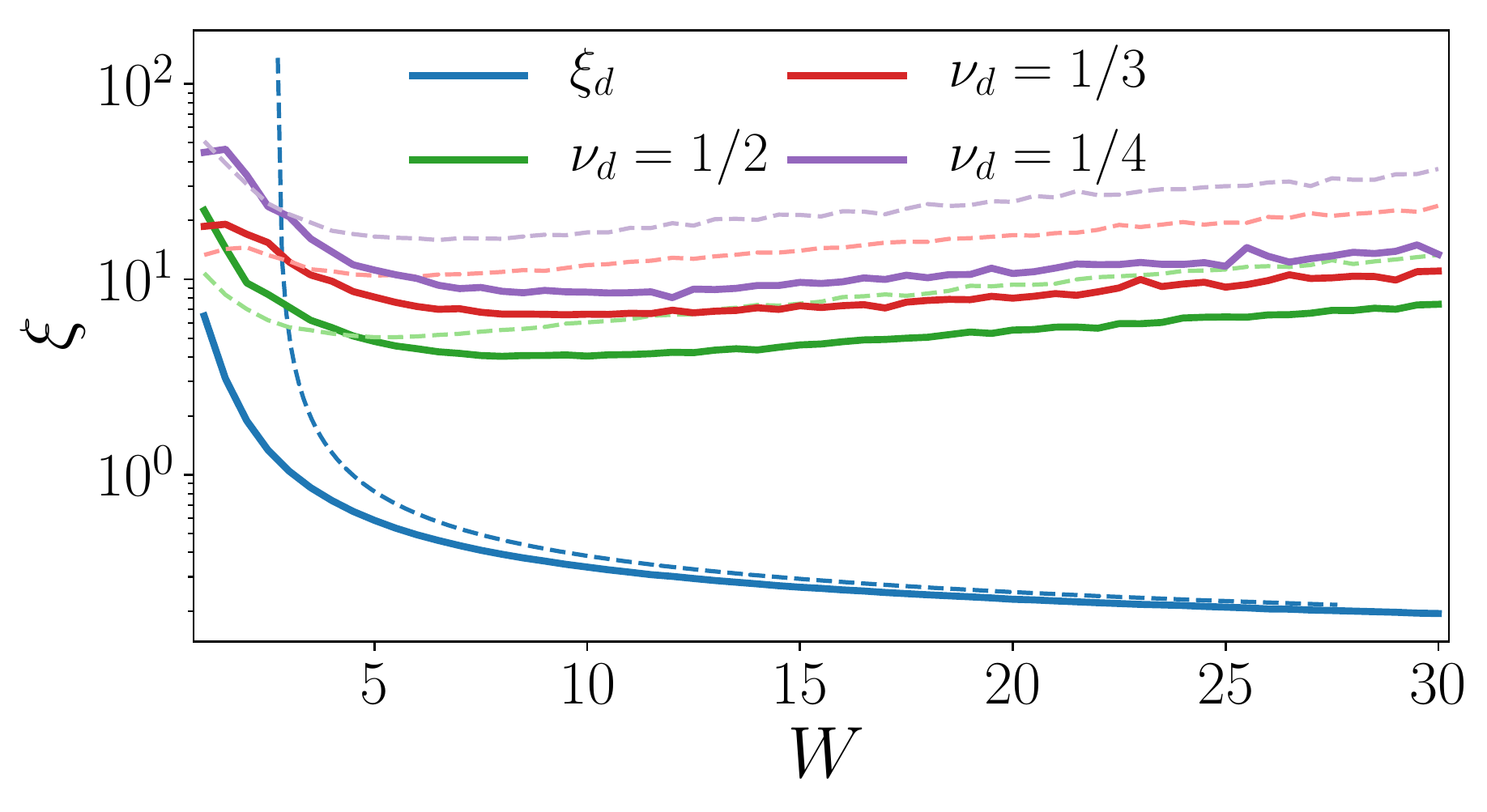}
\caption{\label{Fig:xi_c Hartree nud}
Hartree induced localization length compared with the analytic result of Eq.(\ref{Eq:xic final}), dashed lines, reveals good agreement at weak disorder, while at larger $W$ the deviation of $P(\tilde{\epsilon}_i)$ from Gaussian leads to a worse agreement. The localization length $\xi_c$ reflects the behavior of the effective disorder, leading to larger $\xi_c$ for systems far from half filling. The weaker nature of the effective disorder results in $\xi_c\gg\xi_d$ for all the parameters studied, thus highlighting the much weaker localization of the $c$-boson.}
\end{figure}

In the main part of the current paper, we discussed the \textit{static} Hartree approximation, yielding an effective disorder acting on the $c$-boson and consequently its localization. As the effective disorder changes with the filling fraction of the $d$-bosons, we expect also the localization length to change accordingly. This should lead to a weaker localization whenever the $d$-bosons density, $\nu_d$, deviates from half filling. This is confirmed by the numerical results shown in Figure~\ref{Fig:xi_c Hartree nud}, where we observe stronger localization for $\nu_d=1/2$. Additionally, we compare the data with the prediction of Eq.~(\ref{Eq:xic final}) in the main text (dashed lines).  Although qualitatively similar, the quantitative agreement fails as the disorder strength $W$ increases. We attribute this discrepancy to the deviation of the distribution of effective disorder from Gaussian discussed in the main text.

\begin{figure*}
\includegraphics[width=1.95\columnwidth]{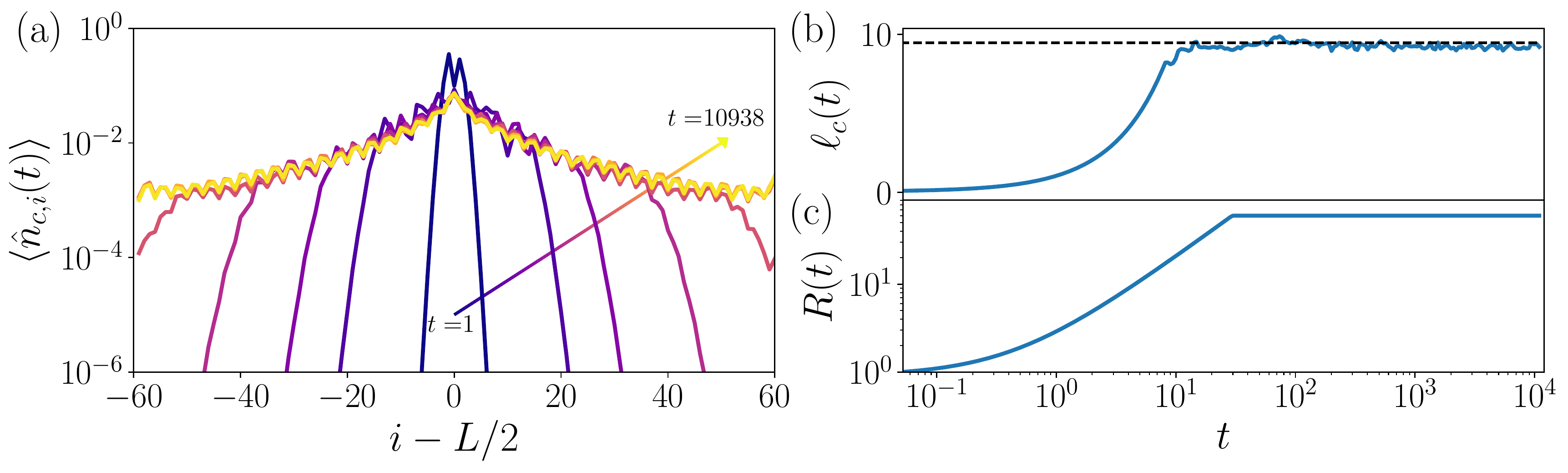}
\caption{\label{Fig:static Hartree dynamics}
The density profiles~(a) of the single particle in the static Hartree disorder at $U=2.5$ show an exponential decay within a range $R(t)$, outside of which they rather have a Gaussian behavior. The exponential decay is characterized by a decay length $\ell_c(t)$~(b), which after an initial transient is constant in time and saturates to the value obtained in Fig.~\ref{Fig:xi_c Hartree nud}. The extent of the exponentially suppressed region, $R(t)$, instead grows in a power-law fashion until it reaches the boundaries of the system~(c).
}
\end{figure*}

Next, we consider the $c$-boson dynamics in the static Hartree approximation fixing the disorder strength to $W=6.5$, close to the minimum of its localization length. The dynamics in this setup can be obtained numerically for large systems without resorting to large computational resources, as the single-particle Hilbert space dimension only scales as the system size $L$. Therefore, we study the density dynamics of a $c$-boson initialized in the center of a $L=120$ chain evolving the initial state up to $t\approx10^4$. The density profiles obtained in this way in Figure~\ref{Fig:static Hartree dynamics}(a) reveal the presence of two distinct behaviors, depending on the distance from the center. Close to the initial position of the particle, the density in Fig.~\ref{Fig:static Hartree dynamics}(a) decays exponentially, consistently with the expectation for a localized particle. However, further away from the initial location of the particle, the density dependence is consistent with a Gaussian profile. While the Gaussian profile is expected from the diffusive spreading of the particle that could take place at early times for weak disorder (weak localization), the coherent backscattering from static disorder leads to the localization of the particle. 

The two distinct regimes in the density profile can be described by introducing two separate lengthscales: the distance $R(t)$ corresponds to crossover from exponential to diffusive behavior, and $\ell_c(t)$ describes the slope of exponential decay. Using these lengthscales, the dynamics of the density can be captured by the following ansatz, 
\begin{equation}
\label{Eq:n ansatz}
n(x,t) \propto \exp\Bigr(-\frac{|x|}{\ell_c(t)\tanh\bigr(R(t)/|x|\bigr)}\Bigr),
\end{equation}
that smoothly interpolates between exponential and Gaussian decays for $x<R(t)$ and $x>R(t)$ respectively. 

The ansatz~(\ref{Eq:n ansatz}) can then be used to fit the numerical data and to obtain the behavior of the two parameters $\ell_c(t)$ and $R(t)$, shown in Figure~\ref{Fig:static Hartree dynamics}(b)-(c). We observe that for the static Hartree approximation, the decay length after a transient quickly saturates to the corresponding localization length $\xi_c$ shown in Figure~\ref{Fig:xi_c Hartree nud} by a dashed line. The extent of the exponential decay region around the center, $R(t)$, instead increases as a power-law, before its saturation to half system size, meaning that the localized region extends over the whole chain. In the companion paper~\cite{Brighi2021a} and in the main text we use the same ansatz to study the dynamics in the fully interacting system, obtaining similar results, although interactions result in much more intricate dynamics.

\section{Details and benchmarks of TEBD implementation \label{App:TEBD}}

\begin{figure}[b]
\includegraphics[width=.95\columnwidth]{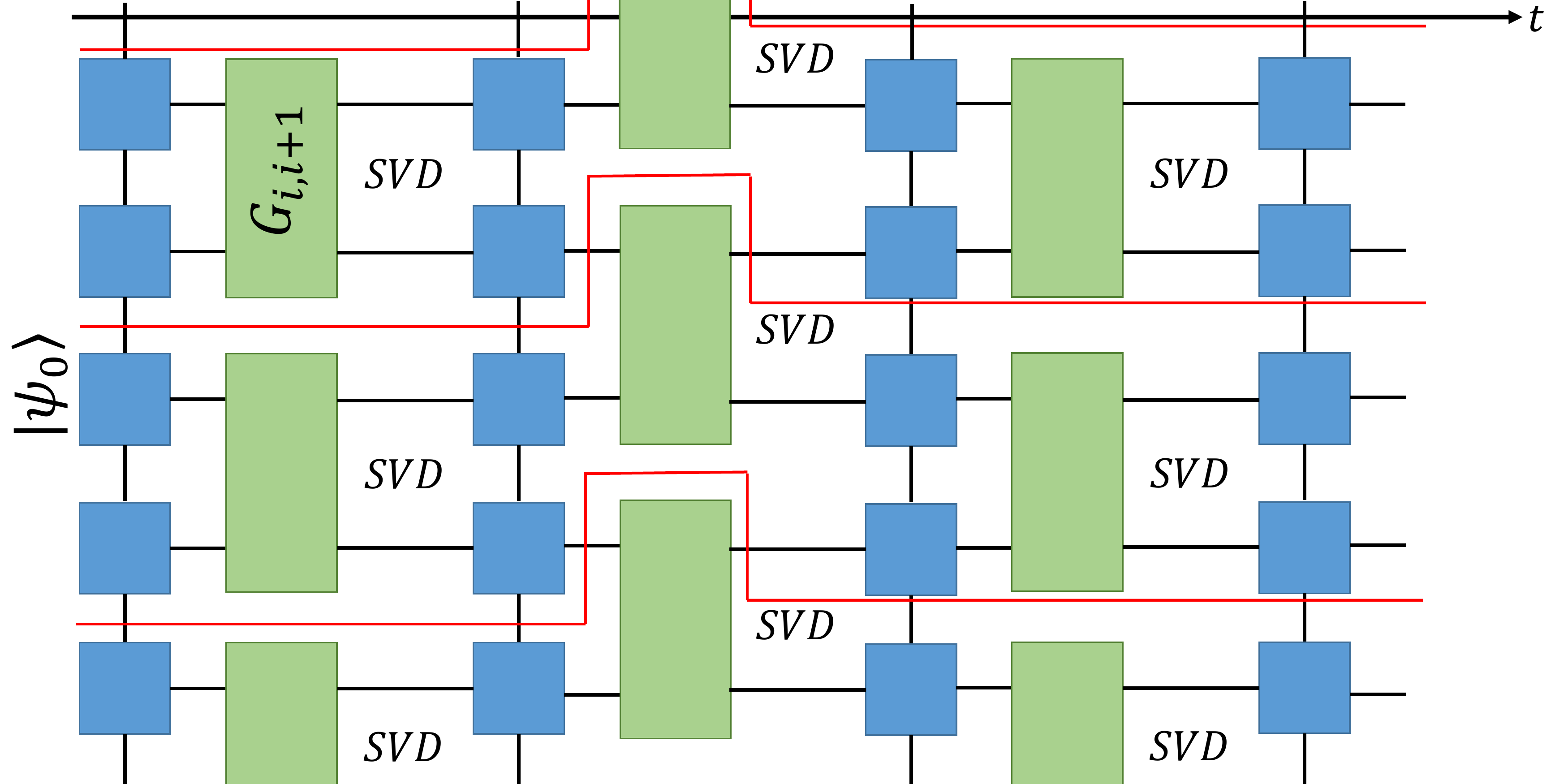}
\caption{\label{Fig:TEBD}
Cartoon representation of the TEBD algorithm. The time evolution can be naturally parallelized, by sharing each of the operations contoured in red to a different core.
}
\end{figure}

We use the TEBD algorithm~\cite{Vidal2003} to perform the time evolution of the wave function of the large system represented as an MPS. The algorithm splits the unitary evolution into time steps (Trotterization), $dt$, and further divides them into even and odd gates, $\hat{G}_{i,i+1} = \exp{\bigr(-\imath dt\hat{h}_{i,i+1}\bigr)}$, where $\hat{h}_{i,i+1}$ is the Hamiltonian density on sites $i$ and $i+1$. Since even(odd) gates do not overlap with one another, all even (odd) gates can be applied independently to the state. This allows a natural parallel implementation of the algorithm, where all even (odd) gates are applied simultaneously.

The TEBD algorithm has two main sources of error: the finite value of time-step used in Trotterization and the truncation error in the singular-value decomposition (SVD). The time step error na\"ively scales as $O(dt^2)$, but it can be reduced to $O(dt^4)$ with an appropriate higher order Suzuki-Trotter decomposition. By choosing sufficiently small time steps, then, the accuracy of the simulation remains reliable for long timescales. The accuracy of the simulation can be controlled by the degree of energy conservation violation, i.e. by tracking $\Delta E(t) = |\bra{\psi_0}\hat{H}\ket{\psi_0}-\bra{\psi(t)}\hat{H}\ket{\psi(t)}$. In our simulations, we fixed $dt=0.05$. In spite of a slower execution time, such a small time-step ensures very accurate results, as evidenced by $\Delta E(T)<10^{-4}$ at the final simulation time $T=200$.

The second source of error, i.e. the singular values truncation $\varepsilon$, arises from the singular-value decomposition performed after each gates application in order to restore the MPS form. In order to avoid the exponential growth of the MPS bond dimension $\chi$, one needs to select the most relevant singular values by neglecting all those which are smaller than a certain threshold $\varepsilon$. The truncation then corresponds to neglecting some weight of the wave-function, and the error can be estimated by the sum of the discarded singular values $\sum_{\lambda_i<\varepsilon}\lambda_i^2$. A further limit on the number of singular values is imposed by the maximal bond dimension, $\chi$. If during the time evolution the number of singular values smaller than $\varepsilon$ becomes larger than $\chi$, i.e. the bond dimension is saturated, then the control on the accuracy of the results is lost, as there is no guarantee that large singular values will be discarded in the future. In our simulations, we fix $\epsilon=10^{-9}$ and $\chi=3000$. This set of parameters guarantees the reliability of the results as the maximum bond dimension is saturated only in the last 10 time-steps (in hopping parameter units) of less than $1/10$ of the central bonds. Furthermore, we can ensure accuracy by comparison to exact diagonalization results on smaller systems, which for the chosen simulation parameters show discrepancies in local observables and entanglement typically of the order $O(10^{-4})$.

The choice of these parameters, however, makes the simulations extremely demanding. However, the structure of the algorithm is naturally suited for parallelization, as shown in Figure~\ref{Fig:TEBD}, and the use of a parallel implementation of the code allowed us to study the dynamics even in the high entanglement regime, where most of the tensors have a large bond dimension of order $1000$.

\section{Details and benchmarks of DMRG-X implementation\label{App:DMRG-X}}
The DMRG-X algorithm~\cite{Serbyn2016,DMRG-X,Pollman2016} used to obtain the highly excited eigenstates of the Hamiltonian (\ref{Eq:H}) relies on the shift-invert method applied to MPS states. The variant of the algorithm used in this work is described in detail in the supplementary material of~\cite{Serbyn2016}. In this section, we will give a brief overview of the algorithm and then explore the dependence of the quality of the results with respect to various tuning parameters such as bond dimension, system size, etc.

The DMRG-X algorithm used can be understood as a simple modification of the standard two-site DMRG~\cite{SCHOLLWOCK201196} algorithm which is extensively used to target ground states of one-dimensional systems. We describe the two differences compared to the ground state algorithm.

First, the ground state DMRG uses the matrix product operator (MPO) form of the Hamiltonian, $H$, to perform the variational optimization. The DMRG-X algorithm, in contrary, employs $\mathcal{H} = (H - E_{0})^2$, which is still efficiently represented by a MPO with a slightly larger bond dimension. The energy $E_{0}$ is the target energy, i.e. the energy of the eigenstates we are aiming to calculate. We square the operator in order to make it semi-positive definite. This is done in order to use iterative inversion algorithms (see below) which require such property and can be avoided if a different iterative algorithm is chosen. We have numerically observed that different iterative algorithms are less stable and converge slower for most cases.

\begin{figure}[t]
\includegraphics[width=.95\columnwidth]{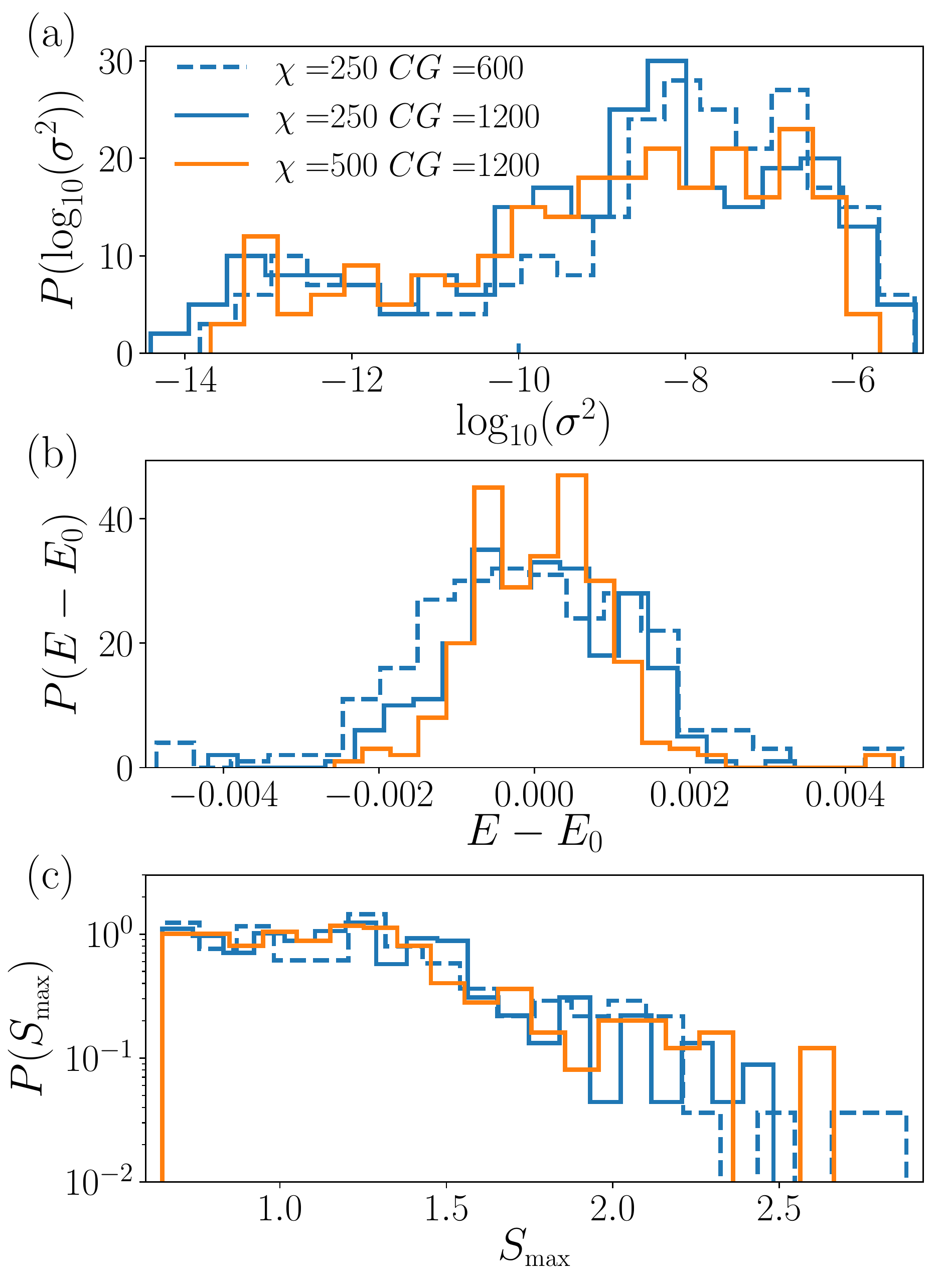}
\caption{\label{Fig:distributions DMRG}
(a) Increasing the bond dimension $\chi$ leads to a better distribution of the variance of the energy of the states $\sigma^2$, shifting it to lower values. Similarly, the increase of the number of conjugate gradient iterations at fixed $\chi$ improves the dataset quality.
(b) The distribution of the distance of the final energy $E$ from the target energy $E_0$. At fixed bond dimension $\chi=250$ doubling the number of conjugate gradient iterations produces a dramatic improvement of the final energy distribution. Further increasing the bond dimension to $\chi=500$ then gives an additional refinement of the distribution.
(c) The entanglement distribution $P(S)$ for the bonds closest to $i_\text{max}$ becomes narrower as the number of conjugate gradient iterations is increased. However, as the bond dimension is increased at fixed $CG$ the MPS states can host larger entanglement, as is confirmed by the broader $P(S_{max})$ at $\chi=500$. 
}
\end{figure}

Second, in the DMRG-X algorithm, we are not aiming to find the eigenstate of $\mathcal{H} $ with the largest magnitude eigenvalue. Instead, the goal is to find the eigenstate corresponding to the lowest magnitude eigenvalue, i.e. the one which corresponds to the eigenstate closest to energy $E_{0}$. This is taken into account in the local optimization step, present in DMRG-type algorithms, i.e.\ when two adjacent local tensors are optimized. In this step, one generates an effective operator by tracing the left and right environments, $\mathcal{H}_\text{eff}$~\cite{SCHOLLWOCK201196}. Calculating the lowest magnitude eigenvalue of $\mathcal{H}_\text{eff}$ is equivalent to calculating the largest eigenvalue of $\mathcal{H}_\text{eff}^{-1}$. We use a power-method to converge to the largest magnitude eigenvalue geometrically. Instead of applying $\mathcal{H}_\text{eff}^{-1}$ at each iteration, we solve the linear system $\mathcal{H}_\text{eff}\textbf{x}_{i+1} = \textbf{x}_{i} $ using a conjugate gradient algorithm. Due to the sweeping nature of the DMRG-type algorithms we have observed that a single iteration of the local power method leads to a faster convergence of the global wavefunction (for fixed runtime), indicating that increasing the number of sweeps is a more important factor. The iterative algorithm scales as $O(CG \chi^3)$, where $CG$ is the number of conjugate gradient iterations, instead of the $O(\chi^6)$ scaling of the exact diagonalization of $\mathcal{H}_\text{eff}$. This scaling, together with the explicit conservation of both particle numbers, allow us to reach bond dimensions up to $\chi = 500$.

For the simulations presented in this work we used the following parameters: In the global level we restricted to 100 DMRG sweeps. However, we do not always keep the final state but the state with the least energy variance as measured at the end of every sweep for the last 50 sweeps. This is because the algorithm tends to switch between different eigenstates as it converges towards the target energy and during the switching period it is not always accurately converged. This behaviour is not present in the standard DMRG algorithm and is attributed to the exponentially large density of states around the target energy. For the solution
of each local optimization problem, we use a single power method iteration which is performed using the conjugate gradient algorithm. For the singular value decomposition which is used to extract the updated local tensors we have used a very low probability truncation cutoff $\epsilon = 10^{-28}$, in order to avoid any statistical bias towards eigenstates of low entanglement. We found that the most important tuning parameters are the bond dimension $\chi$ and the number of conjugate gradient iterations $CG$. Below we show various benchmarks for different parameters and system sizes.

We compare the quality of the resulting eigenstates for $\chi=250,500$ and $CG=600,1200$. A first check is obtained by comparison of the energy variance 
\begin{equation}
\label{Eq:var}
\sigma^2 = \langle \hat{H}^2\rangle -\langle\hat{H}\rangle^2
\end{equation}
of the resulting states. As true eigenstates have exactly $0$ energy variance, a small $\sigma^2$ is an indicator of good convergence. In Figure~\ref{Fig:distributions DMRG}(a), we compare the variance distribution for the different range of parameters, observing, as expected, that increasing both the number of iterations and the bond dimension leads to a better quality result.

The low-valued distribution of the variance suggests that the final states obtained with the DMRG-X algorithm are actually extremely close to exact eigenstates of $\hat{H}$. As a next step, one must check whether the obtained states are indeed highly excited. We thus study the distribution of the energy difference among the obtained state and the target energy $E_0$. As shown in Figure~\ref{Fig:distributions DMRG}(b), increasing $\chi$ and $CG$ yields a narrower distribution close to $0$.

\begin{figure}[t]
\includegraphics[width=.95\columnwidth]{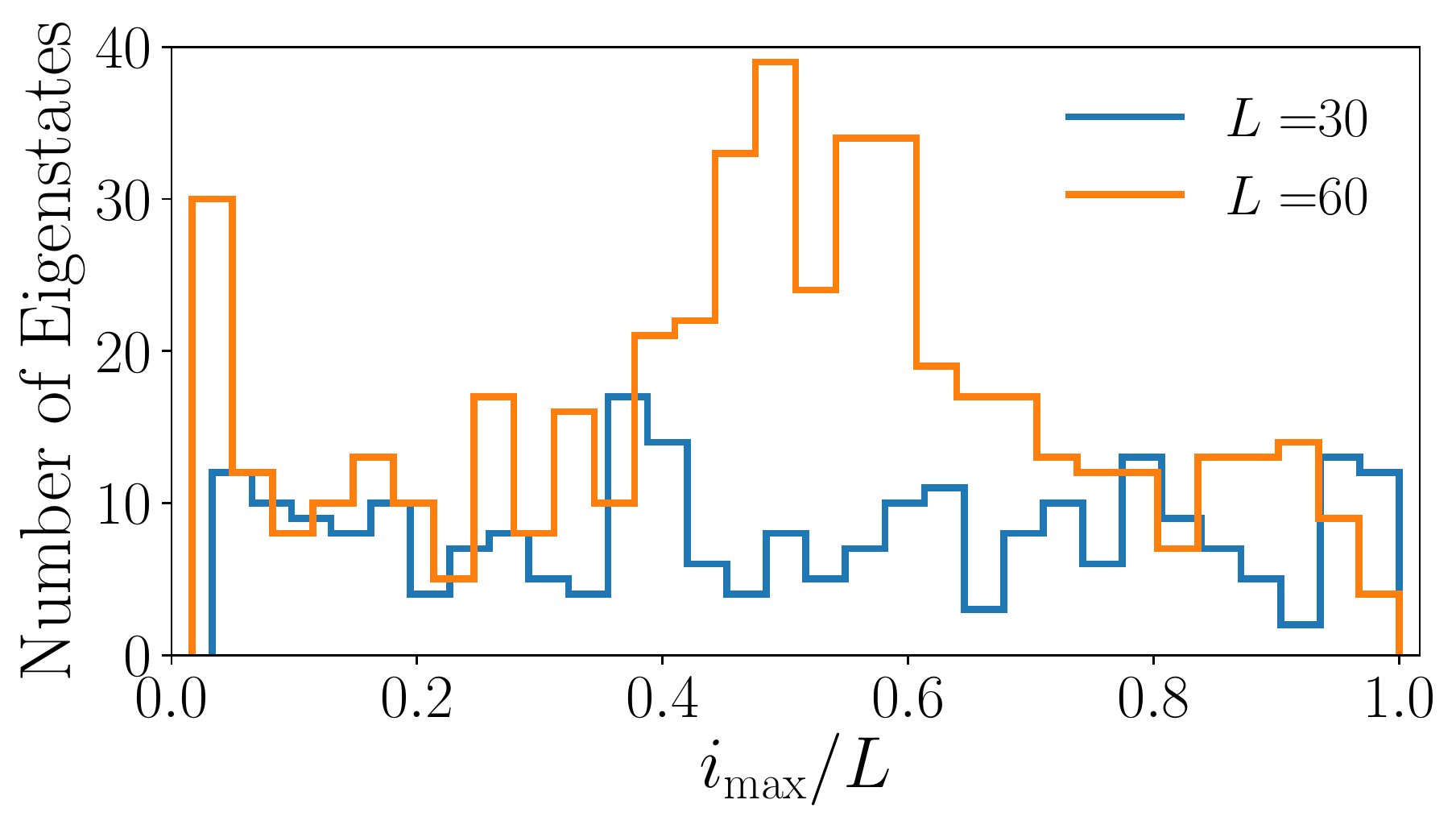}
\caption{\label{Fig:Imax DMRG}
The distribution of the peak of the $c$-boson is almost homogeneous through the whole system in the $L=30$ case, but as the system size is doubled the typical position of $i_\text{max}$ is more restrained to the center of the chain. The broad distribution of $i_\text{max}$ is expected, as the local optimization performed in the algorithm can indeed update the position of the $c$-boson from its initial value and ``move'' it through the system.
}
\end{figure}

Since the amount of entanglement entropy an MPS state can host is strictly related to its bond dimension, it is important to observe how the entanglement varies as we increase $\chi$. We thus study the distribution of the entanglement entropy on the $10$ bonds closer to the peak of the $c$-boson, $i_\text{max}$, where entanglement is supposed to be largest. The results are shown in Figure~\ref{Fig:distributions DMRG}(c) and show that at fixed bond dimension, a larger number of iterations leads to a narrower distribution, in agreement with the small entanglement of exact eigenvalues of MBL Hamiltonians. On the other hand, fixing $CG=1200$ and increasing the bond dimension gives access to states with larger entanglement, as verified by the broader distribution of $S$ at $\chi=500$.

Finally, we study the distribution of the position of the peak of the $c$-boson. As initial state in the DMRG-X algorithm we always used a version of $\ket{\psi_0}$, Eq.~(\ref{Eq:psi0}), with the $c$-boson shifted of a few sites from the center, provided the site is not already occupied by a $d$-boson to avoid stable doublons. However, the local optimization performed by the algorithm can easily ``move" the $c$-boson, as can be seen by the almost even distribution of $i_\text{max}$ shown in Figure~\ref{Fig:Imax DMRG}, for $L=30$. This behavior is consistent in all the datasets studied and does not seem to have a particular relation with the quality of the eigenstates. As the system size is doubled, the distribution of the peak of the $c$-boson becomes more centered around $L/2$, probably due to the larger density of state which increases the possibility of finding an eigenstate close to the target energy for smaller displacement of the $c$-boson.


\begin{thebibliography}{58}%
\makeatletter
\providecommand \@ifxundefined [1]{%
 \@ifx{#1\undefined}
}%
\providecommand \@ifnum [1]{%
 \ifnum #1\expandafter \@firstoftwo
 \else \expandafter \@secondoftwo
 \fi
}%
\providecommand \@ifx [1]{%
 \ifx #1\expandafter \@firstoftwo
 \else \expandafter \@secondoftwo
 \fi
}%
\providecommand \natexlab [1]{#1}%
\providecommand \enquote  [1]{``#1''}%
\providecommand \bibnamefont  [1]{#1}%
\providecommand \bibfnamefont [1]{#1}%
\providecommand \citenamefont [1]{#1}%
\providecommand \href@noop [0]{\@secondoftwo}%
\providecommand \href [0]{\begingroup \@sanitize@url \@href}%
\providecommand \@href[1]{\@@startlink{#1}\@@href}%
\providecommand \@@href[1]{\endgroup#1\@@endlink}%
\providecommand \@sanitize@url [0]{\catcode `\\12\catcode `\$12\catcode
  `\&12\catcode `\#12\catcode `\^12\catcode `\_12\catcode `\%12\relax}%
\providecommand \@@startlink[1]{}%
\providecommand \@@endlink[0]{}%
\providecommand \url  [0]{\begingroup\@sanitize@url \@url }%
\providecommand \@url [1]{\endgroup\@href {#1}{\urlprefix }}%
\providecommand \urlprefix  [0]{URL }%
\providecommand \Eprint [0]{\href }%
\providecommand \doibase [0]{https://doi.org/}%
\providecommand \selectlanguage [0]{\@gobble}%
\providecommand \bibinfo  [0]{\@secondoftwo}%
\providecommand \bibfield  [0]{\@secondoftwo}%
\providecommand \translation [1]{[#1]}%
\providecommand \BibitemOpen [0]{}%
\providecommand \bibitemStop [0]{}%
\providecommand \bibitemNoStop [0]{.\EOS\space}%
\providecommand \EOS [0]{\spacefactor3000\relax}%
\providecommand \BibitemShut  [1]{\csname bibitem#1\endcsname}%
\let\auto@bib@innerbib\@empty
\bibitem [{\citenamefont {Anderson}(1958)}]{Anderson1958}%
  \BibitemOpen
  \bibfield  {author} {\bibinfo {author} {\bibfnamefont {P.~W.}\ \bibnamefont
  {Anderson}},\ }\bibfield  {title} {\bibinfo {title} {{Absence of diffusion in
  certain random lattices}},\ }\href {https://doi.org/10.1103/PhysRev.109.1492}
  {\bibfield  {journal} {\bibinfo  {journal} {Physical Review}\ }\textbf
  {\bibinfo {volume} {109}},\ \bibinfo {pages} {1492} (\bibinfo {year}
  {1958})}\BibitemShut {NoStop}%
\bibitem [{\citenamefont {Gornyi}\ \emph {et~al.}(2005)\citenamefont {Gornyi},
  \citenamefont {Mirlin},\ and\ \citenamefont {Polyakov}}]{Gornyi2005a}%
  \BibitemOpen
  \bibfield  {author} {\bibinfo {author} {\bibfnamefont {I.~V.}\ \bibnamefont
  {Gornyi}}, \bibinfo {author} {\bibfnamefont {A.~D.}\ \bibnamefont {Mirlin}},\
  and\ \bibinfo {author} {\bibfnamefont {D.~G.}\ \bibnamefont {Polyakov}},\
  }\bibfield  {title} {\bibinfo {title} {Interacting electrons in disordered
  wires: Anderson localization and low-$t$ transport},\ }\href
  {https://doi.org/10.1103/PhysRevLett.95.206603} {\bibfield  {journal}
  {\bibinfo  {journal} {Phys. Rev. Lett.}\ }\textbf {\bibinfo {volume} {95}},\
  \bibinfo {pages} {206603} (\bibinfo {year} {2005})}\BibitemShut {NoStop}%
\bibitem [{\citenamefont {Basko}\ \emph {et~al.}(2006)\citenamefont {Basko},
  \citenamefont {Aleiner},\ and\ \citenamefont {Altshuler}}]{Basko2006}%
  \BibitemOpen
  \bibfield  {author} {\bibinfo {author} {\bibfnamefont {D.~M.}\ \bibnamefont
  {Basko}}, \bibinfo {author} {\bibfnamefont {I.~L.}\ \bibnamefont {Aleiner}},\
  and\ \bibinfo {author} {\bibfnamefont {B.~L.}\ \bibnamefont {Altshuler}},\
  }\bibfield  {title} {\bibinfo {title} {{Metal-insulator transition in a
  weakly interacting many-electron system with localized single-particle
  states}},\ }\href {https://doi.org/10.1016/j.aop.2005.11.014} {\bibfield
  {journal} {\bibinfo  {journal} {Annals of Physics}\ }\textbf {\bibinfo
  {volume} {321}},\ \bibinfo {pages} {1126} (\bibinfo {year}
  {2006})}\BibitemShut {NoStop}%
\bibitem [{\citenamefont {Abanin}\ \emph {et~al.}(2019)\citenamefont {Abanin},
  \citenamefont {Altman}, \citenamefont {Bloch},\ and\ \citenamefont
  {Serbyn}}]{Abanin2019}%
  \BibitemOpen
  \bibfield  {author} {\bibinfo {author} {\bibfnamefont {D.~A.}\ \bibnamefont
  {Abanin}}, \bibinfo {author} {\bibfnamefont {E.}~\bibnamefont {Altman}},
  \bibinfo {author} {\bibfnamefont {I.}~\bibnamefont {Bloch}},\ and\ \bibinfo
  {author} {\bibfnamefont {M.}~\bibnamefont {Serbyn}},\ }\bibfield  {title}
  {\bibinfo {title} {Colloquium: Many-body localization, thermalization, and
  entanglement},\ }\href
  {https://link.aps.org/doi/10.1103/RevModPhys.91.021001} {\bibfield  {journal}
  {\bibinfo  {journal} {Rev. Mod. Phys.}\ }\textbf {\bibinfo {volume} {91}},\
  \bibinfo {pages} {021001} (\bibinfo {year} {2019})}\BibitemShut {NoStop}%
\bibitem [{\citenamefont {Bardarson}\ \emph {et~al.}(2012)\citenamefont
  {Bardarson}, \citenamefont {Pollmann},\ and\ \citenamefont
  {Moore}}]{Bardarson2012}%
  \BibitemOpen
  \bibfield  {author} {\bibinfo {author} {\bibfnamefont {J.~H.}\ \bibnamefont
  {Bardarson}}, \bibinfo {author} {\bibfnamefont {F.}~\bibnamefont
  {Pollmann}},\ and\ \bibinfo {author} {\bibfnamefont {J.~E.}\ \bibnamefont
  {Moore}},\ }\bibfield  {title} {\bibinfo {title} {Unbounded growth of
  entanglement in models of many-body localization},\ }\href
  {https://doi.org/10.1103/PhysRevLett.109.017202} {\bibfield  {journal}
  {\bibinfo  {journal} {Phys. Rev. Lett.}\ }\textbf {\bibinfo {volume} {109}},\
  \bibinfo {pages} {017202} (\bibinfo {year} {2012})}\BibitemShut {NoStop}%
\bibitem [{\citenamefont {Serbyn}\ \emph
  {et~al.}(2013{\natexlab{a}})\citenamefont {Serbyn}, \citenamefont
  {Papi{\'{c}}},\ and\ \citenamefont {Abanin}}]{Serbyn2013a}%
  \BibitemOpen
  \bibfield  {author} {\bibinfo {author} {\bibfnamefont {M.}~\bibnamefont
  {Serbyn}}, \bibinfo {author} {\bibfnamefont {Z.}~\bibnamefont
  {Papi{\'{c}}}},\ and\ \bibinfo {author} {\bibfnamefont {D.~A.}\ \bibnamefont
  {Abanin}},\ }\bibfield  {title} {\bibinfo {title} {Universal slow growth of
  entanglement in interacting strongly disordered systems},\ }\href
  {https://doi.org/10.1103/PhysRevLett.110.260601} {\bibfield  {journal}
  {\bibinfo  {journal} {Phys. Rev. Lett.}\ }\textbf {\bibinfo {volume} {110}},\
  \bibinfo {pages} {260601} (\bibinfo {year} {2013}{\natexlab{a}})}\BibitemShut
  {NoStop}%
\bibitem [{\citenamefont {Serbyn}\ \emph
  {et~al.}(2013{\natexlab{b}})\citenamefont {Serbyn}, \citenamefont
  {Papi\'{c}},\ and\ \citenamefont {Abanin}}]{Serbyn2013}%
  \BibitemOpen
  \bibfield  {author} {\bibinfo {author} {\bibfnamefont {M.}~\bibnamefont
  {Serbyn}}, \bibinfo {author} {\bibfnamefont {Z.}~\bibnamefont {Papi\'{c}}},\
  and\ \bibinfo {author} {\bibfnamefont {D.~A.}\ \bibnamefont {Abanin}},\
  }\bibfield  {title} {\bibinfo {title} {Local conservation laws and the
  structure of the many-body localized states},\ }\href
  {https://link.aps.org/doi/10.1103/PhysRevLett.111.127201} {\bibfield
  {journal} {\bibinfo  {journal} {Phys. Rev. Lett.}\ }\textbf {\bibinfo
  {volume} {111}},\ \bibinfo {pages} {127201} (\bibinfo {year}
  {2013}{\natexlab{b}})}\BibitemShut {NoStop}%
\bibitem [{\citenamefont {Huse}\ \emph {et~al.}(2014)\citenamefont {Huse},
  \citenamefont {Nandkishore},\ and\ \citenamefont {Oganesyan}}]{Huse2014}%
  \BibitemOpen
  \bibfield  {author} {\bibinfo {author} {\bibfnamefont {D.~A.}\ \bibnamefont
  {Huse}}, \bibinfo {author} {\bibfnamefont {R.}~\bibnamefont {Nandkishore}},\
  and\ \bibinfo {author} {\bibfnamefont {V.}~\bibnamefont {Oganesyan}},\
  }\bibfield  {title} {\bibinfo {title} {Phenomenology of fully
  many-body-localized systems},\ }\href
  {https://link.aps.org/doi/10.1103/PhysRevB.90.174202} {\bibfield  {journal}
  {\bibinfo  {journal} {Phys. Rev. B}\ }\textbf {\bibinfo {volume} {90}},\
  \bibinfo {pages} {174202} (\bibinfo {year} {2014})}\BibitemShut {NoStop}%
\bibitem [{\citenamefont {Imbrie}(2016)}]{Imbrie2016}%
  \BibitemOpen
  \bibfield  {author} {\bibinfo {author} {\bibfnamefont {J.~Z.}\ \bibnamefont
  {Imbrie}},\ }\bibfield  {title} {\bibinfo {title} {{On Many-Body Localization
  for Quantum Spin Chains}},\ }\href
  {https://doi.org/10.1007/s10955-016-1508-x} {\bibfield  {journal} {\bibinfo
  {journal} {Journal of Statistical Physics}\ }\textbf {\bibinfo {volume}
  {163}},\ \bibinfo {pages} {998} (\bibinfo {year} {2016})}\BibitemShut
  {NoStop}%
\bibitem [{\citenamefont {Schreiber}\ \emph {et~al.}(2015)\citenamefont
  {Schreiber}, \citenamefont {Hodgman}, \citenamefont {Bordia}, \citenamefont
  {L{\"u}schen}, \citenamefont {Fischer}, \citenamefont {Vosk}, \citenamefont
  {Altman}, \citenamefont {Schneider},\ and\ \citenamefont
  {Bloch}}]{Schreiber2015}%
  \BibitemOpen
  \bibfield  {author} {\bibinfo {author} {\bibfnamefont {M.}~\bibnamefont
  {Schreiber}}, \bibinfo {author} {\bibfnamefont {S.~S.}\ \bibnamefont
  {Hodgman}}, \bibinfo {author} {\bibfnamefont {P.}~\bibnamefont {Bordia}},
  \bibinfo {author} {\bibfnamefont {H.~P.}\ \bibnamefont {L{\"u}schen}},
  \bibinfo {author} {\bibfnamefont {M.~H.}\ \bibnamefont {Fischer}}, \bibinfo
  {author} {\bibfnamefont {R.}~\bibnamefont {Vosk}}, \bibinfo {author}
  {\bibfnamefont {E.}~\bibnamefont {Altman}}, \bibinfo {author} {\bibfnamefont
  {U.}~\bibnamefont {Schneider}},\ and\ \bibinfo {author} {\bibfnamefont
  {I.}~\bibnamefont {Bloch}},\ }\bibfield  {title} {\bibinfo {title}
  {Observation of many-body localization of interacting fermions in a
  quasirandom optical lattice},\ }\href
  {https://doi.org/10.1126/science.aaa7432} {\bibfield  {journal} {\bibinfo
  {journal} {Science}\ }\textbf {\bibinfo {volume} {349}},\ \bibinfo {pages}
  {842} (\bibinfo {year} {2015})}\BibitemShut {NoStop}%
\bibitem [{\citenamefont {L\"uschen}\ \emph
  {et~al.}(2017{\natexlab{a}})\citenamefont {L\"uschen}, \citenamefont
  {Bordia}, \citenamefont {Scherg}, \citenamefont {Alet}, \citenamefont
  {Altman}, \citenamefont {Schneider},\ and\ \citenamefont
  {Bloch}}]{Bloch2017a}%
  \BibitemOpen
  \bibfield  {author} {\bibinfo {author} {\bibfnamefont {H.~P.}\ \bibnamefont
  {L\"uschen}}, \bibinfo {author} {\bibfnamefont {P.}~\bibnamefont {Bordia}},
  \bibinfo {author} {\bibfnamefont {S.}~\bibnamefont {Scherg}}, \bibinfo
  {author} {\bibfnamefont {F.}~\bibnamefont {Alet}}, \bibinfo {author}
  {\bibfnamefont {E.}~\bibnamefont {Altman}}, \bibinfo {author} {\bibfnamefont
  {U.}~\bibnamefont {Schneider}},\ and\ \bibinfo {author} {\bibfnamefont
  {I.}~\bibnamefont {Bloch}},\ }\bibfield  {title} {\bibinfo {title}
  {Observation of slow dynamics near the many-body localization transition in
  one-dimensional quasiperiodic systems},\ }\href
  {https://doi.org/10.1103/PhysRevLett.119.260401} {\bibfield  {journal}
  {\bibinfo  {journal} {Phys. Rev. Lett.}\ }\textbf {\bibinfo {volume} {119}},\
  \bibinfo {pages} {260401} (\bibinfo {year} {2017}{\natexlab{a}})}\BibitemShut
  {NoStop}%
\bibitem [{\citenamefont {Lukin}\ \emph
  {et~al.}(2019{\natexlab{a}})\citenamefont {Lukin}, \citenamefont {Rispoli},
  \citenamefont {Schittko}, \citenamefont {Tai}, \citenamefont {Kaufman},
  \citenamefont {Choi}, \citenamefont {Khemani}, \citenamefont {L{\'e}onard},\
  and\ \citenamefont {Greiner}}]{Greiner2019a}%
  \BibitemOpen
  \bibfield  {author} {\bibinfo {author} {\bibfnamefont {A.}~\bibnamefont
  {Lukin}}, \bibinfo {author} {\bibfnamefont {M.}~\bibnamefont {Rispoli}},
  \bibinfo {author} {\bibfnamefont {R.}~\bibnamefont {Schittko}}, \bibinfo
  {author} {\bibfnamefont {M.~E.}\ \bibnamefont {Tai}}, \bibinfo {author}
  {\bibfnamefont {A.~M.}\ \bibnamefont {Kaufman}}, \bibinfo {author}
  {\bibfnamefont {S.}~\bibnamefont {Choi}}, \bibinfo {author} {\bibfnamefont
  {V.}~\bibnamefont {Khemani}}, \bibinfo {author} {\bibfnamefont
  {J.}~\bibnamefont {L{\'e}onard}},\ and\ \bibinfo {author} {\bibfnamefont
  {M.}~\bibnamefont {Greiner}},\ }\bibfield  {title} {\bibinfo {title} {Probing
  entanglement in a many-body--localized system},\ }\href
  {https://doi.org/10.1126/science.aau0818} {\bibfield  {journal} {\bibinfo
  {journal} {Science}\ }\textbf {\bibinfo {volume} {364}},\ \bibinfo {pages}
  {256} (\bibinfo {year} {2019}{\natexlab{a}})}\BibitemShut {NoStop}%
\bibitem [{\citenamefont {Rispoli}\ \emph {et~al.}(2019)\citenamefont
  {Rispoli}, \citenamefont {Lukin}, \citenamefont {Schittko}, \citenamefont
  {Sooshin}, \citenamefont {Tai}, \citenamefont {L{\'e}onard},\ and\
  \citenamefont {Greiner}}]{Greiner2019b}%
  \BibitemOpen
  \bibfield  {author} {\bibinfo {author} {\bibfnamefont {M.}~\bibnamefont
  {Rispoli}}, \bibinfo {author} {\bibfnamefont {A.}~\bibnamefont {Lukin}},
  \bibinfo {author} {\bibfnamefont {R.}~\bibnamefont {Schittko}}, \bibinfo
  {author} {\bibfnamefont {K.}~\bibnamefont {Sooshin}}, \bibinfo {author}
  {\bibfnamefont {M.~E.}\ \bibnamefont {Tai}}, \bibinfo {author} {\bibfnamefont
  {J.}~\bibnamefont {L{\'e}onard}},\ and\ \bibinfo {author} {\bibfnamefont
  {M.}~\bibnamefont {Greiner}},\ }\bibfield  {title} {\bibinfo {title} {Quantum
  critical behaviour at the many-body localization transition},\ }\href
  {https://doi.org/10.1038/s41586-019-1527-2} {\bibfield  {journal} {\bibinfo
  {journal} {Nature}\ }\textbf {\bibinfo {volume} {573}},\ \bibinfo {pages}
  {385} (\bibinfo {year} {2019})}\BibitemShut {NoStop}%
\bibitem [{\citenamefont {Choi}\ \emph {et~al.}(2016)\citenamefont {Choi},
  \citenamefont {Hild}, \citenamefont {Zeiher}, \citenamefont {Schau{\ss}},
  \citenamefont {Rubio-Abadal}, \citenamefont {Yefsah}, \citenamefont
  {Khemani}, \citenamefont {Huse}, \citenamefont {Bloch},\ and\ \citenamefont
  {Gross}}]{Choi2016}%
  \BibitemOpen
  \bibfield  {author} {\bibinfo {author} {\bibfnamefont {J.-y.}\ \bibnamefont
  {Choi}}, \bibinfo {author} {\bibfnamefont {S.}~\bibnamefont {Hild}}, \bibinfo
  {author} {\bibfnamefont {J.}~\bibnamefont {Zeiher}}, \bibinfo {author}
  {\bibfnamefont {P.}~\bibnamefont {Schau{\ss}}}, \bibinfo {author}
  {\bibfnamefont {A.}~\bibnamefont {Rubio-Abadal}}, \bibinfo {author}
  {\bibfnamefont {T.}~\bibnamefont {Yefsah}}, \bibinfo {author} {\bibfnamefont
  {V.}~\bibnamefont {Khemani}}, \bibinfo {author} {\bibfnamefont {D.~A.}\
  \bibnamefont {Huse}}, \bibinfo {author} {\bibfnamefont {I.}~\bibnamefont
  {Bloch}},\ and\ \bibinfo {author} {\bibfnamefont {C.}~\bibnamefont {Gross}},\
  }\bibfield  {title} {\bibinfo {title} {Exploring the many-body localization
  transition in two dimensions},\ }\href
  {https://doi.org/10.1126/science.aaf8834} {\bibfield  {journal} {\bibinfo
  {journal} {Science}\ }\textbf {\bibinfo {volume} {352}},\ \bibinfo {pages}
  {1547} (\bibinfo {year} {2016})}\BibitemShut {NoStop}%
\bibitem [{\citenamefont {Bordia}\ \emph {et~al.}(2017)\citenamefont {Bordia},
  \citenamefont {L\"uschen}, \citenamefont {Scherg}, \citenamefont
  {Gopalakrishnan}, \citenamefont {Knap}, \citenamefont {Schneider},\ and\
  \citenamefont {Bloch}}]{Bloch2017c}%
  \BibitemOpen
  \bibfield  {author} {\bibinfo {author} {\bibfnamefont {P.}~\bibnamefont
  {Bordia}}, \bibinfo {author} {\bibfnamefont {H.}~\bibnamefont {L\"uschen}},
  \bibinfo {author} {\bibfnamefont {S.}~\bibnamefont {Scherg}}, \bibinfo
  {author} {\bibfnamefont {S.}~\bibnamefont {Gopalakrishnan}}, \bibinfo
  {author} {\bibfnamefont {M.}~\bibnamefont {Knap}}, \bibinfo {author}
  {\bibfnamefont {U.}~\bibnamefont {Schneider}},\ and\ \bibinfo {author}
  {\bibfnamefont {I.}~\bibnamefont {Bloch}},\ }\bibfield  {title} {\bibinfo
  {title} {Probing slow relaxation and many-body localization in
  two-dimensional quasiperiodic systems},\ }\href
  {https://doi.org/10.1103/PhysRevX.7.041047} {\bibfield  {journal} {\bibinfo
  {journal} {Phys. Rev. X}\ }\textbf {\bibinfo {volume} {7}},\ \bibinfo {pages}
  {041047} (\bibinfo {year} {2017})}\BibitemShut {NoStop}%
\bibitem [{\citenamefont {{Guo}}\ \emph {et~al.}(2019)\citenamefont {{Guo}},
  \citenamefont {{Cheng}}, \citenamefont {{Sun}}, \citenamefont {{Song}},
  \citenamefont {{Li}}, \citenamefont {{Wang}}, \citenamefont {{Ren}},
  \citenamefont {{Dong}}, \citenamefont {{Zheng}}, \citenamefont {{Zhang}},
  \citenamefont {{Mondaini}}, \citenamefont {{Fan}},\ and\ \citenamefont
  {{Wang}}}]{Guo2019}%
  \BibitemOpen
  \bibfield  {author} {\bibinfo {author} {\bibfnamefont {Q.}~\bibnamefont
  {{Guo}}}, \bibinfo {author} {\bibfnamefont {C.}~\bibnamefont {{Cheng}}},
  \bibinfo {author} {\bibfnamefont {Z.-H.}\ \bibnamefont {{Sun}}}, \bibinfo
  {author} {\bibfnamefont {Z.}~\bibnamefont {{Song}}}, \bibinfo {author}
  {\bibfnamefont {H.}~\bibnamefont {{Li}}}, \bibinfo {author} {\bibfnamefont
  {Z.}~\bibnamefont {{Wang}}}, \bibinfo {author} {\bibfnamefont
  {W.}~\bibnamefont {{Ren}}}, \bibinfo {author} {\bibfnamefont
  {H.}~\bibnamefont {{Dong}}}, \bibinfo {author} {\bibfnamefont
  {D.}~\bibnamefont {{Zheng}}}, \bibinfo {author} {\bibfnamefont {Y.-R.}\
  \bibnamefont {{Zhang}}}, \bibinfo {author} {\bibfnamefont {R.}~\bibnamefont
  {{Mondaini}}}, \bibinfo {author} {\bibfnamefont {H.}~\bibnamefont {{Fan}}},\
  and\ \bibinfo {author} {\bibfnamefont {H.}~\bibnamefont {{Wang}}},\
  }\bibfield  {title} {\bibinfo {title} {{Observation of energy resolved
  many-body localization}},\ }\href@noop {} {\bibfield  {journal} {\bibinfo
  {journal} {arXiv e-prints}\ ,\ \bibinfo {eid} {arXiv:1912.02818}} (\bibinfo
  {year} {2019})},\ \Eprint {https://arxiv.org/abs/1912.02818}
  {arXiv:1912.02818 [quant-ph]} \BibitemShut {NoStop}%
\bibitem [{\citenamefont {De~Roeck}\ \emph {et~al.}(2016)\citenamefont
  {De~Roeck}, \citenamefont {Huveneers}, \citenamefont {M\"uller},\ and\
  \citenamefont {Schiulaz}}]{DeRoeck2016}%
  \BibitemOpen
  \bibfield  {author} {\bibinfo {author} {\bibfnamefont {W.}~\bibnamefont
  {De~Roeck}}, \bibinfo {author} {\bibfnamefont {F.}~\bibnamefont {Huveneers}},
  \bibinfo {author} {\bibfnamefont {M.}~\bibnamefont {M\"uller}},\ and\
  \bibinfo {author} {\bibfnamefont {M.}~\bibnamefont {Schiulaz}},\ }\bibfield
  {title} {\bibinfo {title} {Absence of many-body mobility edges},\ }\href
  {https://link.aps.org/doi/10.1103/PhysRevB.93.014203} {\bibfield  {journal}
  {\bibinfo  {journal} {Phys. Rev. B}\ }\textbf {\bibinfo {volume} {93}},\
  \bibinfo {pages} {014203} (\bibinfo {year} {2016})}\BibitemShut {NoStop}%
\bibitem [{\citenamefont {Luitz}\ \emph {et~al.}(2015)\citenamefont {Luitz},
  \citenamefont {Laflorencie},\ and\ \citenamefont {Alet}}]{Luitz2015}%
  \BibitemOpen
  \bibfield  {author} {\bibinfo {author} {\bibfnamefont {D.~J.}\ \bibnamefont
  {Luitz}}, \bibinfo {author} {\bibfnamefont {N.}~\bibnamefont {Laflorencie}},\
  and\ \bibinfo {author} {\bibfnamefont {F.}~\bibnamefont {Alet}},\ }\bibfield
  {title} {\bibinfo {title} {Many-body localization edge in the random-field
  heisenberg chain},\ }\href
  {https://link.aps.org/doi/10.1103/PhysRevB.91.081103} {\bibfield  {journal}
  {\bibinfo  {journal} {Phys. Rev. B}\ }\textbf {\bibinfo {volume} {91}},\
  \bibinfo {pages} {081103} (\bibinfo {year} {2015})}\BibitemShut {NoStop}%
\bibitem [{\citenamefont {Brighi}\ \emph {et~al.}(2020)\citenamefont {Brighi},
  \citenamefont {Abanin},\ and\ \citenamefont {Serbyn}}]{Brighi2020}%
  \BibitemOpen
  \bibfield  {author} {\bibinfo {author} {\bibfnamefont {P.}~\bibnamefont
  {Brighi}}, \bibinfo {author} {\bibfnamefont {D.~A.}\ \bibnamefont {Abanin}},\
  and\ \bibinfo {author} {\bibfnamefont {M.}~\bibnamefont {Serbyn}},\
  }\bibfield  {title} {\bibinfo {title} {Stability of mobility edges in
  disordered interacting systems},\ }\href
  {https://doi.org/10.1103/PhysRevB.102.060202} {\bibfield  {journal} {\bibinfo
   {journal} {Phys. Rev. B}\ }\textbf {\bibinfo {volume} {102}},\ \bibinfo
  {pages} {060202} (\bibinfo {year} {2020})}\BibitemShut {NoStop}%
\bibitem [{\citenamefont {Yao}\ and\ \citenamefont {Zakrzewski}(2020)}]{Yao20}%
  \BibitemOpen
  \bibfield  {author} {\bibinfo {author} {\bibfnamefont {R.}~\bibnamefont
  {Yao}}\ and\ \bibinfo {author} {\bibfnamefont {J.}~\bibnamefont
  {Zakrzewski}},\ }\bibfield  {title} {\bibinfo {title} {Many-body localization
  in the bose-hubbard model: Evidence for mobility edge},\ }\href
  {https://doi.org/10.1103/PhysRevB.102.014310} {\bibfield  {journal} {\bibinfo
   {journal} {Phys. Rev. B}\ }\textbf {\bibinfo {volume} {102}},\ \bibinfo
  {pages} {014310} (\bibinfo {year} {2020})}\BibitemShut {NoStop}%
\bibitem [{\citenamefont {{Pomata}}\ \emph {et~al.}(2020)\citenamefont
  {{Pomata}}, \citenamefont {{Ganeshan}},\ and\ \citenamefont
  {{Wei}}}]{Pomata}%
  \BibitemOpen
  \bibfield  {author} {\bibinfo {author} {\bibfnamefont {N.}~\bibnamefont
  {{Pomata}}}, \bibinfo {author} {\bibfnamefont {S.}~\bibnamefont
  {{Ganeshan}}},\ and\ \bibinfo {author} {\bibfnamefont {T.-C.}\ \bibnamefont
  {{Wei}}},\ }\bibfield  {title} {\bibinfo {title} {{In search of a many-body
  mobility edge with matrix product states in a Generalized Aubry-Andr{\'e}
  model with interactions}},\ }\href@noop {} {\bibfield  {journal} {\bibinfo
  {journal} {arXiv e-prints}\ ,\ \bibinfo {eid} {arXiv:2012.09853}} (\bibinfo
  {year} {2020})},\ \Eprint {https://arxiv.org/abs/2012.09853}
  {arXiv:2012.09853 [cond-mat.dis-nn]} \BibitemShut {NoStop}%
\bibitem [{\citenamefont {De~Roeck}\ and\ \citenamefont
  {Huveneers}(2017)}]{DeRoeck2017}%
  \BibitemOpen
  \bibfield  {author} {\bibinfo {author} {\bibfnamefont {W.}~\bibnamefont
  {De~Roeck}}\ and\ \bibinfo {author} {\bibfnamefont {F.}~\bibnamefont
  {Huveneers}},\ }\bibfield  {title} {\bibinfo {title} {Stability and
  instability towards delocalization in many-body localization systems},\
  }\href {https://link.aps.org/doi/10.1103/PhysRevB.95.155129} {\bibfield
  {journal} {\bibinfo  {journal} {Phys. Rev. B}\ }\textbf {\bibinfo {volume}
  {95}},\ \bibinfo {pages} {155129} (\bibinfo {year} {2017})}\BibitemShut
  {NoStop}%
\bibitem [{\citenamefont {Potirniche}\ \emph {et~al.}(2019)\citenamefont
  {Potirniche}, \citenamefont {Banerjee},\ and\ \citenamefont
  {Altman}}]{Altman2019}%
  \BibitemOpen
  \bibfield  {author} {\bibinfo {author} {\bibfnamefont {I.-D.}\ \bibnamefont
  {Potirniche}}, \bibinfo {author} {\bibfnamefont {S.}~\bibnamefont
  {Banerjee}},\ and\ \bibinfo {author} {\bibfnamefont {E.}~\bibnamefont
  {Altman}},\ }\bibfield  {title} {\bibinfo {title} {Exploration of the
  stability of many-body localization in $d>1$},\ }\href
  {https://doi.org/10.1103/PhysRevB.99.205149} {\bibfield  {journal} {\bibinfo
  {journal} {Phys. Rev. B}\ }\textbf {\bibinfo {volume} {99}},\ \bibinfo
  {pages} {205149} (\bibinfo {year} {2019})}\BibitemShut {NoStop}%
\bibitem [{\citenamefont {Wahl}\ \emph {et~al.}(2019)\citenamefont {Wahl},
  \citenamefont {Pal},\ and\ \citenamefont {Simon}}]{Pal2019}%
  \BibitemOpen
  \bibfield  {author} {\bibinfo {author} {\bibfnamefont {T.~B.}\ \bibnamefont
  {Wahl}}, \bibinfo {author} {\bibfnamefont {A.}~\bibnamefont {Pal}},\ and\
  \bibinfo {author} {\bibfnamefont {S.~H.}\ \bibnamefont {Simon}},\ }\bibfield
  {title} {\bibinfo {title} {Signatures of the many-body localized regime in
  two dimensions},\ }\href {https://doi.org/10.1038/s41567-018-0339-x}
  {\bibfield  {journal} {\bibinfo  {journal} {Nature Phys}\ }\textbf {\bibinfo
  {volume} {15}},\ \bibinfo {pages} {164} (\bibinfo {year} {2019})}\BibitemShut
  {NoStop}%
\bibitem [{\citenamefont {Doggen}\ \emph {et~al.}(2020)\citenamefont {Doggen},
  \citenamefont {Gornyi}, \citenamefont {Mirlin},\ and\ \citenamefont
  {Polyakov}}]{Mirlin2020}%
  \BibitemOpen
  \bibfield  {author} {\bibinfo {author} {\bibfnamefont {E.~V.~H.}\
  \bibnamefont {Doggen}}, \bibinfo {author} {\bibfnamefont {I.~V.}\
  \bibnamefont {Gornyi}}, \bibinfo {author} {\bibfnamefont {A.~D.}\
  \bibnamefont {Mirlin}},\ and\ \bibinfo {author} {\bibfnamefont {D.~G.}\
  \bibnamefont {Polyakov}},\ }\bibfield  {title} {\bibinfo {title} {Slow
  many-body delocalization beyond one dimension},\ }\href
  {https://doi.org/10.1103/PhysRevLett.125.155701} {\bibfield  {journal}
  {\bibinfo  {journal} {Phys. Rev. Lett.}\ }\textbf {\bibinfo {volume} {125}},\
  \bibinfo {pages} {155701} (\bibinfo {year} {2020})}\BibitemShut {NoStop}%
\bibitem [{\citenamefont {Kshetrimayum}\ \emph {et~al.}(2020)\citenamefont
  {Kshetrimayum}, \citenamefont {Goihl},\ and\ \citenamefont
  {Eisert}}]{Eisert2020}%
  \BibitemOpen
  \bibfield  {author} {\bibinfo {author} {\bibfnamefont {A.}~\bibnamefont
  {Kshetrimayum}}, \bibinfo {author} {\bibfnamefont {M.}~\bibnamefont
  {Goihl}},\ and\ \bibinfo {author} {\bibfnamefont {J.}~\bibnamefont
  {Eisert}},\ }\bibfield  {title} {\bibinfo {title} {Time evolution of
  many-body localized systems in two spatial dimensions},\ }\href
  {https://doi.org/10.1103/PhysRevB.102.235132} {\bibfield  {journal} {\bibinfo
   {journal} {Phys. Rev. B}\ }\textbf {\bibinfo {volume} {102}},\ \bibinfo
  {pages} {235132} (\bibinfo {year} {2020})}\BibitemShut {NoStop}%
\bibitem [{\citenamefont {Nandkishore}(2015)}]{Nandkishore2015a}%
  \BibitemOpen
  \bibfield  {author} {\bibinfo {author} {\bibfnamefont {R.}~\bibnamefont
  {Nandkishore}},\ }\bibfield  {title} {\bibinfo {title} {Many-body
  localization proximity effect},\ }\href
  {https://doi.org/10.1103/PhysRevB.92.245141} {\bibfield  {journal} {\bibinfo
  {journal} {Phys. Rev. B}\ }\textbf {\bibinfo {volume} {92}},\ \bibinfo
  {pages} {245141} (\bibinfo {year} {2015})}\BibitemShut {NoStop}%
\bibitem [{\citenamefont {Luitz}\ \emph {et~al.}(2017)\citenamefont {Luitz},
  \citenamefont {Huveneers},\ and\ \citenamefont {De~Roeck}}]{Luitz2017}%
  \BibitemOpen
  \bibfield  {author} {\bibinfo {author} {\bibfnamefont {D.~J.}\ \bibnamefont
  {Luitz}}, \bibinfo {author} {\bibfnamefont {F.}~\bibnamefont {Huveneers}},\
  and\ \bibinfo {author} {\bibfnamefont {W.}~\bibnamefont {De~Roeck}},\
  }\bibfield  {title} {\bibinfo {title} {How a small quantum bath can
  thermalize long localized chains},\ }\href
  {https://link.aps.org/doi/10.1103/PhysRevLett.119.150602} {\bibfield
  {journal} {\bibinfo  {journal} {Phys. Rev. Lett.}\ }\textbf {\bibinfo
  {volume} {119}},\ \bibinfo {pages} {150602} (\bibinfo {year}
  {2017})}\BibitemShut {NoStop}%
\bibitem [{\citenamefont {Huse}\ \emph {et~al.}(2015)\citenamefont {Huse},
  \citenamefont {Nandkishore}, \citenamefont {Pietracaprina}, \citenamefont
  {Ros},\ and\ \citenamefont {Scardicchio}}]{Huse2015}%
  \BibitemOpen
  \bibfield  {author} {\bibinfo {author} {\bibfnamefont {D.~A.}\ \bibnamefont
  {Huse}}, \bibinfo {author} {\bibfnamefont {R.}~\bibnamefont {Nandkishore}},
  \bibinfo {author} {\bibfnamefont {F.}~\bibnamefont {Pietracaprina}}, \bibinfo
  {author} {\bibfnamefont {V.}~\bibnamefont {Ros}},\ and\ \bibinfo {author}
  {\bibfnamefont {A.}~\bibnamefont {Scardicchio}},\ }\bibfield  {title}
  {\bibinfo {title} {Localized systems coupled to small baths: From anderson to
  zeno},\ }\href {https://doi.org/10.1103/PhysRevB.92.014203} {\bibfield
  {journal} {\bibinfo  {journal} {Phys. Rev. B}\ }\textbf {\bibinfo {volume}
  {92}},\ \bibinfo {pages} {014203} (\bibinfo {year} {2015})}\BibitemShut
  {NoStop}%
\bibitem [{\citenamefont {Wybo}\ \emph {et~al.}(2020)\citenamefont {Wybo},
  \citenamefont {Knap},\ and\ \citenamefont {Pollmann}}]{Pollmann2019}%
  \BibitemOpen
  \bibfield  {author} {\bibinfo {author} {\bibfnamefont {E.}~\bibnamefont
  {Wybo}}, \bibinfo {author} {\bibfnamefont {M.}~\bibnamefont {Knap}},\ and\
  \bibinfo {author} {\bibfnamefont {F.}~\bibnamefont {Pollmann}},\ }\bibfield
  {title} {\bibinfo {title} {Entanglement dynamics of a many-body localized
  system coupled to a bath},\ }\href
  {https://doi.org/10.1103/PhysRevB.102.064304} {\bibfield  {journal} {\bibinfo
   {journal} {Phys. Rev. B}\ }\textbf {\bibinfo {volume} {102}},\ \bibinfo
  {pages} {064304} (\bibinfo {year} {2020})}\BibitemShut {NoStop}%
\bibitem [{\citenamefont {Hyatt}\ \emph {et~al.}(2017)\citenamefont {Hyatt},
  \citenamefont {Garrison}, \citenamefont {Potter},\ and\ \citenamefont
  {Bauer}}]{Hyatt2017}%
  \BibitemOpen
  \bibfield  {author} {\bibinfo {author} {\bibfnamefont {K.}~\bibnamefont
  {Hyatt}}, \bibinfo {author} {\bibfnamefont {J.~R.}\ \bibnamefont {Garrison}},
  \bibinfo {author} {\bibfnamefont {A.~C.}\ \bibnamefont {Potter}},\ and\
  \bibinfo {author} {\bibfnamefont {B.}~\bibnamefont {Bauer}},\ }\bibfield
  {title} {\bibinfo {title} {Many-body localization in the presence of a small
  bath},\ }\href {https://doi.org/10.1103/PhysRevB.95.035132} {\bibfield
  {journal} {\bibinfo  {journal} {Phys. Rev. B}\ }\textbf {\bibinfo {volume}
  {95}},\ \bibinfo {pages} {035132} (\bibinfo {year} {2017})}\BibitemShut
  {NoStop}%
\bibitem [{\citenamefont {Goihl}\ \emph {et~al.}(2019)\citenamefont {Goihl},
  \citenamefont {Eisert},\ and\ \citenamefont {Krumnow}}]{Goihl2019}%
  \BibitemOpen
  \bibfield  {author} {\bibinfo {author} {\bibfnamefont {M.}~\bibnamefont
  {Goihl}}, \bibinfo {author} {\bibfnamefont {J.}~\bibnamefont {Eisert}},\ and\
  \bibinfo {author} {\bibfnamefont {C.}~\bibnamefont {Krumnow}},\ }\bibfield
  {title} {\bibinfo {title} {Exploration of the stability of many-body
  localized systems in the presence of a small bath},\ }\href
  {https://link.aps.org/doi/10.1103/PhysRevB.99.195145} {\bibfield  {journal}
  {\bibinfo  {journal} {Phys. Rev. B}\ }\textbf {\bibinfo {volume} {99}},\
  \bibinfo {pages} {195145} (\bibinfo {year} {2019})}\BibitemShut {NoStop}%
\bibitem [{\citenamefont {Rubio-Abadal}\ \emph {et~al.}(2019)\citenamefont
  {Rubio-Abadal}, \citenamefont {Choi}, \citenamefont {Zeiher}, \citenamefont
  {Hollerith}, \citenamefont {Rui}, \citenamefont {Bloch},\ and\ \citenamefont
  {Gross}}]{Rubio-Abadal2019}%
  \BibitemOpen
  \bibfield  {author} {\bibinfo {author} {\bibfnamefont {A.}~\bibnamefont
  {Rubio-Abadal}}, \bibinfo {author} {\bibfnamefont {J.-y.}\ \bibnamefont
  {Choi}}, \bibinfo {author} {\bibfnamefont {J.}~\bibnamefont {Zeiher}},
  \bibinfo {author} {\bibfnamefont {S.}~\bibnamefont {Hollerith}}, \bibinfo
  {author} {\bibfnamefont {J.}~\bibnamefont {Rui}}, \bibinfo {author}
  {\bibfnamefont {I.}~\bibnamefont {Bloch}},\ and\ \bibinfo {author}
  {\bibfnamefont {C.}~\bibnamefont {Gross}},\ }\bibfield  {title} {\bibinfo
  {title} {Many-body delocalization in the presence of a quantum bath},\ }\href
  {https://link.aps.org/doi/10.1103/PhysRevX.9.041014} {\bibfield  {journal}
  {\bibinfo  {journal} {Phys. Rev. X}\ }\textbf {\bibinfo {volume} {9}},\
  \bibinfo {pages} {041014} (\bibinfo {year} {2019})}\BibitemShut {NoStop}%
\bibitem [{\citenamefont {{L{\'e}onard}}\ \emph {et~al.}(2020)\citenamefont
  {{L{\'e}onard}}, \citenamefont {{Rispoli}}, \citenamefont {{Lukin}},
  \citenamefont {{Schittko}}, \citenamefont {{Kim}}, \citenamefont {{Kwan}},
  \citenamefont {{Sels}}, \citenamefont {{Demler}},\ and\ \citenamefont
  {{Greiner}}}]{Leonard2020}%
  \BibitemOpen
  \bibfield  {author} {\bibinfo {author} {\bibfnamefont {J.}~\bibnamefont
  {{L{\'e}onard}}}, \bibinfo {author} {\bibfnamefont {M.}~\bibnamefont
  {{Rispoli}}}, \bibinfo {author} {\bibfnamefont {A.}~\bibnamefont {{Lukin}}},
  \bibinfo {author} {\bibfnamefont {R.}~\bibnamefont {{Schittko}}}, \bibinfo
  {author} {\bibfnamefont {S.}~\bibnamefont {{Kim}}}, \bibinfo {author}
  {\bibfnamefont {J.}~\bibnamefont {{Kwan}}}, \bibinfo {author} {\bibfnamefont
  {D.}~\bibnamefont {{Sels}}}, \bibinfo {author} {\bibfnamefont
  {E.}~\bibnamefont {{Demler}}},\ and\ \bibinfo {author} {\bibfnamefont
  {M.}~\bibnamefont {{Greiner}}},\ }\bibfield  {title} {\bibinfo {title}
  {{Signatures of bath-induced quantum avalanches in a many-body--localized
  system}},\ }\href@noop {} {\bibfield  {journal} {\bibinfo  {journal} {arXiv
  e-prints}\ ,\ \bibinfo {eid} {arXiv:2012.15270}} (\bibinfo {year} {2020})},\
  \Eprint {https://arxiv.org/abs/2012.15270} {arXiv:2012.15270
  [cond-mat.quant-gas]} \BibitemShut {NoStop}%
\bibitem [{\citenamefont {Krause}\ \emph {et~al.}(2021)\citenamefont {Krause},
  \citenamefont {Pellegrin}, \citenamefont {Brouwer}, \citenamefont {Abanin},\
  and\ \citenamefont {Filippone}}]{Krause2021}%
  \BibitemOpen
  \bibfield  {author} {\bibinfo {author} {\bibfnamefont {U.}~\bibnamefont
  {Krause}}, \bibinfo {author} {\bibfnamefont {T.}~\bibnamefont {Pellegrin}},
  \bibinfo {author} {\bibfnamefont {P.~W.}\ \bibnamefont {Brouwer}}, \bibinfo
  {author} {\bibfnamefont {D.~A.}\ \bibnamefont {Abanin}},\ and\ \bibinfo
  {author} {\bibfnamefont {M.}~\bibnamefont {Filippone}},\ }\bibfield  {title}
  {\bibinfo {title} {Nucleation of ergodicity by a single mobile impurity in
  supercooled insulators},\ }\href
  {https://link.aps.org/doi/10.1103/PhysRevLett.126.030603} {\bibfield
  {journal} {\bibinfo  {journal} {Phys. Rev. Lett.}\ }\textbf {\bibinfo
  {volume} {126}},\ \bibinfo {pages} {030603} (\bibinfo {year}
  {2021})}\BibitemShut {NoStop}%
\bibitem [{\citenamefont {{Brighi}}\ \emph {et~al.}(2021)\citenamefont
  {{Brighi}}, \citenamefont {{Michailidis}}, \citenamefont {{Abanin}},\ and\
  \citenamefont {{Serbyn}}}]{Brighi2021a}%
  \BibitemOpen
  \bibfield  {author} {\bibinfo {author} {\bibfnamefont {P.}~\bibnamefont
  {{Brighi}}}, \bibinfo {author} {\bibfnamefont {A.~A.}\ \bibnamefont
  {{Michailidis}}}, \bibinfo {author} {\bibfnamefont {D.~A.}\ \bibnamefont
  {{Abanin}}},\ and\ \bibinfo {author} {\bibfnamefont {M.}~\bibnamefont
  {{Serbyn}}},\ }\bibfield  {title} {\bibinfo {title} {{Propagation of
  Many-body Localization in an Anderson Insulator}},\ }\href@noop {} {\bibfield
   {journal} {\bibinfo  {journal} {arXiv e-prints}\ ,\ \bibinfo {eid}
  {arXiv:2109.07332}} (\bibinfo {year} {2021})},\ \Eprint
  {https://arxiv.org/abs/2109.07332} {arXiv:2109.07332 [cond-mat.dis-nn]}
  \BibitemShut {NoStop}%
\bibitem [{\citenamefont {Verstraete}\ and\ \citenamefont
  {Cirac}(2006)}]{Verstraete2006}%
  \BibitemOpen
  \bibfield  {author} {\bibinfo {author} {\bibfnamefont {F.}~\bibnamefont
  {Verstraete}}\ and\ \bibinfo {author} {\bibfnamefont {J.~I.}\ \bibnamefont
  {Cirac}},\ }\bibfield  {title} {\bibinfo {title} {Matrix product states
  represent ground states faithfully},\ }\href
  {https://doi.org/10.1103/PhysRevB.73.094423} {\bibfield  {journal} {\bibinfo
  {journal} {Phys. Rev. B}\ }\textbf {\bibinfo {volume} {73}},\ \bibinfo
  {pages} {094423} (\bibinfo {year} {2006})}\BibitemShut {NoStop}%
\bibitem [{\citenamefont {Serbyn}\ \emph {et~al.}(2016)\citenamefont {Serbyn},
  \citenamefont {Michailidis}, \citenamefont {Abanin},\ and\ \citenamefont
  {Papi\ifmmode~\acute{c}\else \'{c}\fi{}}}]{Serbyn2016}%
  \BibitemOpen
  \bibfield  {author} {\bibinfo {author} {\bibfnamefont {M.}~\bibnamefont
  {Serbyn}}, \bibinfo {author} {\bibfnamefont {A.~A.}\ \bibnamefont
  {Michailidis}}, \bibinfo {author} {\bibfnamefont {D.~A.}\ \bibnamefont
  {Abanin}},\ and\ \bibinfo {author} {\bibfnamefont {Z.}~\bibnamefont
  {Papi\ifmmode~\acute{c}\else \'{c}\fi{}}},\ }\bibfield  {title} {\bibinfo
  {title} {Power-law entanglement spectrum in many-body localized phases},\
  }\href {https://doi.org/10.1103/PhysRevLett.117.160601} {\bibfield  {journal}
  {\bibinfo  {journal} {Phys. Rev. Lett.}\ }\textbf {\bibinfo {volume} {117}},\
  \bibinfo {pages} {160601} (\bibinfo {year} {2016})}\BibitemShut {NoStop}%
\bibitem [{\citenamefont {Yu}\ \emph {et~al.}(2017)\citenamefont {Yu},
  \citenamefont {Pekker},\ and\ \citenamefont {Clark}}]{DMRG-X}%
  \BibitemOpen
  \bibfield  {author} {\bibinfo {author} {\bibfnamefont {X.}~\bibnamefont
  {Yu}}, \bibinfo {author} {\bibfnamefont {D.}~\bibnamefont {Pekker}},\ and\
  \bibinfo {author} {\bibfnamefont {B.~K.}\ \bibnamefont {Clark}},\ }\bibfield
  {title} {\bibinfo {title} {Finding matrix product state representations of
  highly excited eigenstates of many-body localized hamiltonians},\ }\href
  {https://doi.org/10.1103/PhysRevLett.118.017201} {\bibfield  {journal}
  {\bibinfo  {journal} {Phys. Rev. Lett.}\ }\textbf {\bibinfo {volume} {118}},\
  \bibinfo {pages} {017201} (\bibinfo {year} {2017})}\BibitemShut {NoStop}%
\bibitem [{\citenamefont {Khemani}\ \emph {et~al.}(2016)\citenamefont
  {Khemani}, \citenamefont {Pollmann},\ and\ \citenamefont
  {Sondhi}}]{Pollman2016}%
  \BibitemOpen
  \bibfield  {author} {\bibinfo {author} {\bibfnamefont {V.}~\bibnamefont
  {Khemani}}, \bibinfo {author} {\bibfnamefont {F.}~\bibnamefont {Pollmann}},\
  and\ \bibinfo {author} {\bibfnamefont {S.~L.}\ \bibnamefont {Sondhi}},\
  }\bibfield  {title} {\bibinfo {title} {Obtaining highly excited eigenstates
  of many-body localized hamiltonians by the density matrix renormalization
  group approach},\ }\href {https://doi.org/10.1103/PhysRevLett.116.247204}
  {\bibfield  {journal} {\bibinfo  {journal} {Phys. Rev. Lett.}\ }\textbf
  {\bibinfo {volume} {116}},\ \bibinfo {pages} {247204} (\bibinfo {year}
  {2016})}\BibitemShut {NoStop}%
\bibitem [{\citenamefont {L\"uschen}\ \emph
  {et~al.}(2017{\natexlab{b}})\citenamefont {L\"uschen}, \citenamefont
  {Bordia}, \citenamefont {Hodgman}, \citenamefont {Schreiber}, \citenamefont
  {Sarkar}, \citenamefont {Daley}, \citenamefont {Fischer}, \citenamefont
  {Altman}, \citenamefont {Bloch},\ and\ \citenamefont
  {Schneider}}]{Bloch2017b}%
  \BibitemOpen
  \bibfield  {author} {\bibinfo {author} {\bibfnamefont {H.~P.}\ \bibnamefont
  {L\"uschen}}, \bibinfo {author} {\bibfnamefont {P.}~\bibnamefont {Bordia}},
  \bibinfo {author} {\bibfnamefont {S.~S.}\ \bibnamefont {Hodgman}}, \bibinfo
  {author} {\bibfnamefont {M.}~\bibnamefont {Schreiber}}, \bibinfo {author}
  {\bibfnamefont {S.}~\bibnamefont {Sarkar}}, \bibinfo {author} {\bibfnamefont
  {A.~J.}\ \bibnamefont {Daley}}, \bibinfo {author} {\bibfnamefont {M.~H.}\
  \bibnamefont {Fischer}}, \bibinfo {author} {\bibfnamefont {E.}~\bibnamefont
  {Altman}}, \bibinfo {author} {\bibfnamefont {I.}~\bibnamefont {Bloch}},\ and\
  \bibinfo {author} {\bibfnamefont {U.}~\bibnamefont {Schneider}},\ }\bibfield
  {title} {\bibinfo {title} {Signatures of many-body localization in a
  controlled open quantum system},\ }\href
  {https://doi.org/10.1103/PhysRevX.7.011034} {\bibfield  {journal} {\bibinfo
  {journal} {Phys. Rev. X}\ }\textbf {\bibinfo {volume} {7}},\ \bibinfo {pages}
  {011034} (\bibinfo {year} {2017}{\natexlab{b}})}\BibitemShut {NoStop}%
\bibitem [{\citenamefont {Thouless}(1972)}]{Thouless1972}%
  \BibitemOpen
  \bibfield  {author} {\bibinfo {author} {\bibfnamefont {D.~J.}\ \bibnamefont
  {Thouless}},\ }\bibfield  {title} {\bibinfo {title} {A relation between the
  density of states and range of localization for one dimensional random
  systems},\ }\href {https://doi.org/10.1088/0022-3719/5/1/010} {\bibfield
  {journal} {\bibinfo  {journal} {Journal of Physics C: Solid State Physics}\
  }\textbf {\bibinfo {volume} {5}},\ \bibinfo {pages} {77} (\bibinfo {year}
  {1972})}\BibitemShut {NoStop}%
\bibitem [{\citenamefont {Nandy}\ \emph {et~al.}(2021)\citenamefont {Nandy},
  \citenamefont {Evers},\ and\ \citenamefont {Bera}}]{Bera2021}%
  \BibitemOpen
  \bibfield  {author} {\bibinfo {author} {\bibfnamefont {S.}~\bibnamefont
  {Nandy}}, \bibinfo {author} {\bibfnamefont {F.}~\bibnamefont {Evers}},\ and\
  \bibinfo {author} {\bibfnamefont {S.}~\bibnamefont {Bera}},\ }\bibfield
  {title} {\bibinfo {title} {Dephasing in strongly disordered interacting
  quantum wires},\ }\href {https://doi.org/10.1103/PhysRevB.103.085105}
  {\bibfield  {journal} {\bibinfo  {journal} {Phys. Rev. B}\ }\textbf {\bibinfo
  {volume} {103}},\ \bibinfo {pages} {085105} (\bibinfo {year}
  {2021})}\BibitemShut {NoStop}%
\bibitem [{\citenamefont {Gopalakrishnan}\ \emph {et~al.}(2017)\citenamefont
  {Gopalakrishnan}, \citenamefont {Islam},\ and\ \citenamefont
  {Knap}}]{Knap2017}%
  \BibitemOpen
  \bibfield  {author} {\bibinfo {author} {\bibfnamefont {S.}~\bibnamefont
  {Gopalakrishnan}}, \bibinfo {author} {\bibfnamefont {K.~R.}\ \bibnamefont
  {Islam}},\ and\ \bibinfo {author} {\bibfnamefont {M.}~\bibnamefont {Knap}},\
  }\bibfield  {title} {\bibinfo {title} {Noise-induced subdiffusion in strongly
  localized quantum systems},\ }\href
  {https://doi.org/10.1103/PhysRevLett.119.046601} {\bibfield  {journal}
  {\bibinfo  {journal} {Phys. Rev. Lett.}\ }\textbf {\bibinfo {volume} {119}},\
  \bibinfo {pages} {046601} (\bibinfo {year} {2017})}\BibitemShut {NoStop}%
\bibitem [{\citenamefont {Lorenzo}\ \emph {et~al.}(2018)\citenamefont
  {Lorenzo}, \citenamefont {Apollaro}, \citenamefont {Palma}, \citenamefont
  {Nandkishore}, \citenamefont {Silva},\ and\ \citenamefont
  {Marino}}]{Marino2018}%
  \BibitemOpen
  \bibfield  {author} {\bibinfo {author} {\bibfnamefont {S.}~\bibnamefont
  {Lorenzo}}, \bibinfo {author} {\bibfnamefont {T.}~\bibnamefont {Apollaro}},
  \bibinfo {author} {\bibfnamefont {G.~M.}\ \bibnamefont {Palma}}, \bibinfo
  {author} {\bibfnamefont {R.}~\bibnamefont {Nandkishore}}, \bibinfo {author}
  {\bibfnamefont {A.}~\bibnamefont {Silva}},\ and\ \bibinfo {author}
  {\bibfnamefont {J.}~\bibnamefont {Marino}},\ }\bibfield  {title} {\bibinfo
  {title} {Remnants of anderson localization in prethermalization induced by
  white noise},\ }\href {https://doi.org/10.1103/PhysRevB.98.054302} {\bibfield
   {journal} {\bibinfo  {journal} {Phys. Rev. B}\ }\textbf {\bibinfo {volume}
  {98}},\ \bibinfo {pages} {054302} (\bibinfo {year} {2018})}\BibitemShut
  {NoStop}%
\bibitem [{\citenamefont {Gefen}\ and\ \citenamefont
  {Sch\"on}(1984)}]{Gefen1984}%
  \BibitemOpen
  \bibfield  {author} {\bibinfo {author} {\bibfnamefont {Y.}~\bibnamefont
  {Gefen}}\ and\ \bibinfo {author} {\bibfnamefont {G.}~\bibnamefont
  {Sch\"on}},\ }\bibfield  {title} {\bibinfo {title} {Effect of inelastic
  processes on localization in one dimension},\ }\href
  {https://doi.org/10.1103/PhysRevB.30.7323} {\bibfield  {journal} {\bibinfo
  {journal} {Phys. Rev. B}\ }\textbf {\bibinfo {volume} {30}},\ \bibinfo
  {pages} {7323} (\bibinfo {year} {1984})}\BibitemShut {NoStop}%
\bibitem [{\citenamefont {Logan}\ and\ \citenamefont
  {Wolynes}(1987)}]{Logan1987}%
  \BibitemOpen
  \bibfield  {author} {\bibinfo {author} {\bibfnamefont {D.~E.}\ \bibnamefont
  {Logan}}\ and\ \bibinfo {author} {\bibfnamefont {P.~G.}\ \bibnamefont
  {Wolynes}},\ }\bibfield  {title} {\bibinfo {title} {Dephasing and anderson
  localization in topologically disordered systems},\ }\href
  {https://doi.org/10.1103/PhysRevB.36.4135} {\bibfield  {journal} {\bibinfo
  {journal} {Phys. Rev. B}\ }\textbf {\bibinfo {volume} {36}},\ \bibinfo
  {pages} {4135} (\bibinfo {year} {1987})}\BibitemShut {NoStop}%
\bibitem [{\citenamefont {Taylor}\ and\ \citenamefont
  {Scardicchio}(2021)}]{Scardicchio2021}%
  \BibitemOpen
  \bibfield  {author} {\bibinfo {author} {\bibfnamefont {S.~R.}\ \bibnamefont
  {Taylor}}\ and\ \bibinfo {author} {\bibfnamefont {A.}~\bibnamefont
  {Scardicchio}},\ }\bibfield  {title} {\bibinfo {title} {Subdiffusion in a
  one-dimensional anderson insulator with random dephasing: Finite-size
  scaling, griffiths effects, and possible implications for many-body
  localization},\ }\href {https://doi.org/10.1103/PhysRevB.103.184202}
  {\bibfield  {journal} {\bibinfo  {journal} {Phys. Rev. B}\ }\textbf {\bibinfo
  {volume} {103}},\ \bibinfo {pages} {184202} (\bibinfo {year}
  {2021})}\BibitemShut {NoStop}%
\bibitem [{\citenamefont {{Lezama}}\ and\ \citenamefont {{Bar
  Lev}}(2021)}]{BarLev2021}%
  \BibitemOpen
  \bibfield  {author} {\bibinfo {author} {\bibfnamefont {T.~L.~M.}\
  \bibnamefont {{Lezama}}}\ and\ \bibinfo {author} {\bibfnamefont
  {Y.}~\bibnamefont {{Bar Lev}}},\ }\bibfield  {title} {\bibinfo {title}
  {{Logarithmic, noise-induced dynamics in the Anderson insulator}},\
  }\href@noop {} {\bibfield  {journal} {\bibinfo  {journal} {arXiv e-prints}\
  ,\ \bibinfo {eid} {arXiv:2108.08861}} (\bibinfo {year} {2021})},\ \Eprint
  {https://arxiv.org/abs/2108.08861} {arXiv:2108.08861 [cond-mat.dis-nn]}
  \BibitemShut {NoStop}%
\bibitem [{\citenamefont {Vidal}(2003)}]{Vidal2003}%
  \BibitemOpen
  \bibfield  {author} {\bibinfo {author} {\bibfnamefont {G.}~\bibnamefont
  {Vidal}},\ }\bibfield  {title} {\bibinfo {title} {Efficient classical
  simulation of slightly entangled quantum computations},\ }\href
  {https://doi.org/10.1103/PhysRevLett.91.147902} {\bibfield  {journal}
  {\bibinfo  {journal} {Phys. Rev. Lett.}\ }\textbf {\bibinfo {volume} {91}},\
  \bibinfo {pages} {147902} (\bibinfo {year} {2003})}\BibitemShut {NoStop}%
\bibitem [{\citenamefont {Lukin}\ \emph
  {et~al.}(2019{\natexlab{b}})\citenamefont {Lukin}, \citenamefont {Rispoli},
  \citenamefont {Schittko}, \citenamefont {Tai}, \citenamefont {Kaufman},
  \citenamefont {Choi}, \citenamefont {Khemani}, \citenamefont {L{\'e}onard},\
  and\ \citenamefont {Greiner}}]{Lukin2019}%
  \BibitemOpen
  \bibfield  {author} {\bibinfo {author} {\bibfnamefont {A.}~\bibnamefont
  {Lukin}}, \bibinfo {author} {\bibfnamefont {M.}~\bibnamefont {Rispoli}},
  \bibinfo {author} {\bibfnamefont {R.}~\bibnamefont {Schittko}}, \bibinfo
  {author} {\bibfnamefont {M.~E.}\ \bibnamefont {Tai}}, \bibinfo {author}
  {\bibfnamefont {A.~M.}\ \bibnamefont {Kaufman}}, \bibinfo {author}
  {\bibfnamefont {S.}~\bibnamefont {Choi}}, \bibinfo {author} {\bibfnamefont
  {V.}~\bibnamefont {Khemani}}, \bibinfo {author} {\bibfnamefont
  {J.}~\bibnamefont {L{\'e}onard}},\ and\ \bibinfo {author} {\bibfnamefont
  {M.}~\bibnamefont {Greiner}},\ }\bibfield  {title} {\bibinfo {title} {Probing
  entanglement in a many-body{\textendash}localized system},\ }\href
  {https://doi.org/10.1126/science.aau0818} {\bibfield  {journal} {\bibinfo
  {journal} {Science}\ }\textbf {\bibinfo {volume} {364}},\ \bibinfo {pages}
  {256} (\bibinfo {year} {2019}{\natexlab{b}})}\BibitemShut {NoStop}%
\bibitem [{\citenamefont {Kiefer-Emmanouilidis}\ \emph
  {et~al.}(2020)\citenamefont {Kiefer-Emmanouilidis}, \citenamefont {Unanyan},
  \citenamefont {Fleischhauer},\ and\ \citenamefont
  {Sirker}}]{Kiefer-Emmanouilidis2020}%
  \BibitemOpen
  \bibfield  {author} {\bibinfo {author} {\bibfnamefont {M.}~\bibnamefont
  {Kiefer-Emmanouilidis}}, \bibinfo {author} {\bibfnamefont {R.}~\bibnamefont
  {Unanyan}}, \bibinfo {author} {\bibfnamefont {M.}~\bibnamefont
  {Fleischhauer}},\ and\ \bibinfo {author} {\bibfnamefont {J.}~\bibnamefont
  {Sirker}},\ }\bibfield  {title} {\bibinfo {title} {Evidence for unbounded
  growth of the number entropy in many-body localized phases},\ }\href
  {https://link.aps.org/doi/10.1103/PhysRevLett.124.243601} {\bibfield
  {journal} {\bibinfo  {journal} {Phys. Rev. Lett.}\ }\textbf {\bibinfo
  {volume} {124}},\ \bibinfo {pages} {243601} (\bibinfo {year}
  {2020})}\BibitemShut {NoStop}%
\bibitem [{\citenamefont {Kiefer-Emmanouilidis}\ \emph
  {et~al.}(2021)\citenamefont {Kiefer-Emmanouilidis}, \citenamefont {Unanyan},
  \citenamefont {Fleischhauer},\ and\ \citenamefont
  {Sirker}}]{Kiefer-Emmanouilidis2021}%
  \BibitemOpen
  \bibfield  {author} {\bibinfo {author} {\bibfnamefont {M.}~\bibnamefont
  {Kiefer-Emmanouilidis}}, \bibinfo {author} {\bibfnamefont {R.}~\bibnamefont
  {Unanyan}}, \bibinfo {author} {\bibfnamefont {M.}~\bibnamefont
  {Fleischhauer}},\ and\ \bibinfo {author} {\bibfnamefont {J.}~\bibnamefont
  {Sirker}},\ }\bibfield  {title} {\bibinfo {title} {Slow delocalization of
  particles in many-body localized phases},\ }\href
  {https://doi.org/10.1103/PhysRevB.103.024203} {\bibfield  {journal} {\bibinfo
   {journal} {Phys. Rev. B}\ }\textbf {\bibinfo {volume} {103}},\ \bibinfo
  {pages} {024203} (\bibinfo {year} {2021})}\BibitemShut {NoStop}%
\bibitem [{\citenamefont {Luitz}\ and\ \citenamefont {Lev}(2020)}]{Luitz2020}%
  \BibitemOpen
  \bibfield  {author} {\bibinfo {author} {\bibfnamefont {D.~J.}\ \bibnamefont
  {Luitz}}\ and\ \bibinfo {author} {\bibfnamefont {Y.~B.}\ \bibnamefont
  {Lev}},\ }\bibfield  {title} {\bibinfo {title} {Absence of slow particle
  transport in the many-body localized phase},\ }\href
  {https://doi.org/10.1103/PhysRevB.102.100202} {\bibfield  {journal} {\bibinfo
   {journal} {Phys. Rev. B}\ }\textbf {\bibinfo {volume} {102}},\ \bibinfo
  {pages} {100202} (\bibinfo {year} {2020})}\BibitemShut {NoStop}%
\bibitem [{\citenamefont {Bera}\ \emph {et~al.}(2015)\citenamefont {Bera},
  \citenamefont {Schomerus}, \citenamefont {Heidrich-Meisner},\ and\
  \citenamefont {Bardarson}}]{Bardarson2015}%
  \BibitemOpen
  \bibfield  {author} {\bibinfo {author} {\bibfnamefont {S.}~\bibnamefont
  {Bera}}, \bibinfo {author} {\bibfnamefont {H.}~\bibnamefont {Schomerus}},
  \bibinfo {author} {\bibfnamefont {F.}~\bibnamefont {Heidrich-Meisner}},\ and\
  \bibinfo {author} {\bibfnamefont {J.~H.}\ \bibnamefont {Bardarson}},\
  }\bibfield  {title} {\bibinfo {title} {Many-body localization characterized
  from a one-particle perspective},\ }\href
  {https://doi.org/10.1103/PhysRevLett.115.046603} {\bibfield  {journal}
  {\bibinfo  {journal} {Phys. Rev. Lett.}\ }\textbf {\bibinfo {volume} {115}},\
  \bibinfo {pages} {046603} (\bibinfo {year} {2015})}\BibitemShut {NoStop}%
\bibitem [{\citenamefont {Lezama}\ \emph {et~al.}(2017)\citenamefont {Lezama},
  \citenamefont {Bera}, \citenamefont {Schomerus}, \citenamefont
  {Heidrich-Meisner},\ and\ \citenamefont {Bardarson}}]{Bardarson2017}%
  \BibitemOpen
  \bibfield  {author} {\bibinfo {author} {\bibfnamefont {T.~L.~M.}\
  \bibnamefont {Lezama}}, \bibinfo {author} {\bibfnamefont {S.}~\bibnamefont
  {Bera}}, \bibinfo {author} {\bibfnamefont {H.}~\bibnamefont {Schomerus}},
  \bibinfo {author} {\bibfnamefont {F.}~\bibnamefont {Heidrich-Meisner}},\ and\
  \bibinfo {author} {\bibfnamefont {J.~H.}\ \bibnamefont {Bardarson}},\
  }\bibfield  {title} {\bibinfo {title} {One-particle density matrix occupation
  spectrum of many-body localized states after a global quench},\ }\href
  {https://doi.org/10.1103/PhysRevB.96.060202} {\bibfield  {journal} {\bibinfo
  {journal} {Phys. Rev. B}\ }\textbf {\bibinfo {volume} {96}},\ \bibinfo
  {pages} {060202} (\bibinfo {year} {2017})}\BibitemShut {NoStop}%
\bibitem [{\citenamefont {Bauer}\ and\ \citenamefont
  {Nayak}(2013)}]{Bauer2013}%
  \BibitemOpen
  \bibfield  {author} {\bibinfo {author} {\bibfnamefont {B.}~\bibnamefont
  {Bauer}}\ and\ \bibinfo {author} {\bibfnamefont {C.}~\bibnamefont {Nayak}},\
  }\bibfield  {title} {\bibinfo {title} {Area laws in a many-body localized
  state and its implications for topological order},\ }\href
  {https://doi.org/10.1088/1742-5468/2013/09/p09005} {\bibfield  {journal}
  {\bibinfo  {journal} {J. Stat. Mech.}\ }\textbf {\bibinfo {volume} {2013}},\
  \bibinfo {pages} {P09005} (\bibinfo {year} {2013})}\BibitemShut {NoStop}%
\bibitem [{\citenamefont {Schollwöck}(2011)}]{SCHOLLWOCK201196}%
  \BibitemOpen
  \bibfield  {author} {\bibinfo {author} {\bibfnamefont {U.}~\bibnamefont
  {Schollwöck}},\ }\bibfield  {title} {\bibinfo {title} {The density-matrix
  renormalization group in the age of matrix product states},\ }\href
  {https://doi.org/https://doi.org/10.1016/j.aop.2010.09.012} {\bibfield
  {journal} {\bibinfo  {journal} {Annals of Physics}\ }\textbf {\bibinfo
  {volume} {326}},\ \bibinfo {pages} {96} (\bibinfo {year} {2011})},\ \bibinfo
  {note} {January 2011 Special Issue}\BibitemShut {NoStop}%
\end{thebibliography}

%

\end{document}